\documentclass[fleqn,usenatbib]{mnras}

\usepackage[utf8]{inputenc}
\DeclareUnicodeCharacter{2248}{\approx}
\usepackage{newtxtext,newtxmath}
\usepackage[T1]{fontenc}

\DeclareRobustCommand{\VAN}[3]{#2}
\let\VANthebibliography\thebibliography
\def\thebibliography{\DeclareRobustCommand{\VAN}[3]{##3}\VANthebibliography}

\usepackage{graphicx}
\usepackage{amsmath}

\usepackage{amssymb}
\usepackage{color}
\usepackage{fix-cm}
\usepackage{newtxtext,newtxmath}

\def\HI{\ion{H}{I}}
\def\V{{\mathcal V}}
\def\del{\partial}

\title[Measuring the \ion{H}{I} power spectrum at low-$z$ with MIGHTEE]{Measurements of the \ion{H}{I} intensity mapping power spectrum at low redshifts with MIGHTEE data: comparison with detected \ion{H}{I} galaxies}

\author[J. Townsend et al.]{
Junaid Townsend,$^{1}$\thanks{E-mail: junaidtownsend@gmail.com }
Mario G. Santos,$^{1,2}$
Suman Chatterjee,$^{1}$
Zhaoting Chen,$^{3,4}$
Sourabh Paul,$^{4}$
\newauthor
Aishrila Mazumder,$^{4}$
Laura Wolz,$^{4}$
Matt J. Jarvis$^{1,5}$ and Bradley S. Frank$^{6,7}$
\\
$^{1}$Department of Physics and Astronomy, University of the Western Cape, Robert Sobukwe Road, Bellville, 7535, South Africa\\
$^{2}$South African Radio Astronomy Observatory (SARAO), Cape Town, 7700, South Africa\\
$^{3}$Institute for Astronomy, The University of Edinburgh, Royal Observatory, Edinburgh EH9 3HJ, UK\\
$^{4}$Jodrell Bank Centre for Astrophysics, Department of Physics and Astronomy, The University of Manchester, Manchester M13 9PL, UK\\
$^{5}$Astrophysics, Department of Physics, University of Oxford, Keble Road,
Oxford, OX1 3RH, UK\\
$^{6}$STFC UK Astronomy Technology Centre, Royal Observatory, Edinburgh, Blackford Hill, Edinburgh, EH9 3HJ\\
$^{7}$Department of Astronomy, University of Cape Town, Private Bag X3, Rondebosch 7701, South Africa}
\date{Accepted XXX. Received YYY; in original form ZZZ}

\pubyear{2026}

\begin{document}

\definecolor{forestgreen(web)}{rgb}{0.13, 0.55, 0.13}

\label{firstpage}
\pagerange{\pageref{firstpage}--\pageref{lastpage}}
\maketitle


\begin{abstract}
Line intensity mapping provides a statistical approach to tracing the large-scale distribution of matter in the Universe. We apply the \ion{H}{I} intensity mapping technique to interferometric data from the MeerKAT International GHz-Tiered Extragalactic Explorations (MIGHTEE) Survey, analysing 17.5 hours of a single pointing in the COSMOS field, using a 60 MHz sub-band in the frequency range 1332 - 1392 MHz ($0.02 \lesssim z \lesssim 0.07$). Using a delay-spectrum-based estimator, we measure the \ion{H}{I} power spectrum on sub-megaparsec scales and compare it directly to the power spectrum inferred from a catalogue of individually detected \ion{H}{I} galaxies in the same field. After mitigating low-level broadband contamination through conservative outlier flagging in the three-dimensional power spectrum, cross-correlation of time-split visibilities yields a statistically significant detection on scales $3 \lesssim k \lesssim 20 \, \mathrm{Mpc}^{-1}$ with a total signal-to-noise ratio of $\sim 13$. Over this range, the power spectra obtained from visibilities and detected galaxies are consistent within uncertainties and have comparable amplitudes of order $10^{-2}$ – $10^{-1}$ $\mathrm{mK}^2 \mathrm{Mpc}^3$. End-to-end validation is performed by propagating detected galaxies through the power spectrum estimator via both direct intensity-field construction and simulated visibilities, demonstrating agreement up to $k \sim 20 \ \mathrm{Mpc}^{-1}$, beyond which measurements become noise-dominated. A statistically significant correlation is also observed between the data and the simulated visibilities from the detected \ion{H}{I} galaxies, which should be free of systematics. These results provide a self-consistent validation of interferometric \ion{H}{I} intensity mapping at low redshift and demonstrate agreement with galaxy-based measurements within the same cosmological volume.
\end{abstract}



\section{Introduction}
\label{section:intro}

Intensity mapping (IM) with the 21 cm emission line of neutral hydrogen ($\ion{H}{I}$) has been proposed as a novel technique to explore the cosmological and astrophysical properties of the Universe \citep{Bharadwaj2001a, Bharadwaj2001, Battye2004, McQuinn2006a, Chang2008, wyithe2008mnras, Bull2015, Santos2015b, Kovetz2017}. In the era of precision cosmology in which the Cosmic Microwave Background (CMB) from Recombination at z $\sim$ 1100 has provided significant insights into our understanding of the Universe (e.g. \citealt{Planck_18_params}), IM with the 21-cm line would help make this picture more complete with its large area and deep volume coverage capabilities, allowing us to probe substantial portions of the Universe's cosmic history. In addition to its fast survey speeds, $\ion{H}{I}$ IM also provides excellent redshift information since it measures the 21 cm line, with a rest frame frequency of $\sim$ 1420 MHz, while radio telescopes typically provide sub-MHz frequency resolutions, without the need for special spectroscopic techniques required for optical surveys. Coupled with the abundance of \ion{H}{I} throughout the Universe's cosmic history, IM provides a means of probing cosmological information all the way up to $z\sim200$ \citep{2006PhRvD..74h3517S}, well into the Dark Ages. As such, it has the means of providing a comprehensive view of the Universe's evolution from the present day until the first stars and galaxies started forming and even beyond.
\par

First predicted by \cite{Hulst1945} and then detected by \cite{Ewen1951}, this 21 cm emission line arises from the spin-flip transition of the electron within the hydrogen atom, which then emits a photon of energy at a wavelength of about 21 cm. Despite this transition being highly forbidden, we can observe this emission line with radio telescopes sensitive to centimetre-wavelength radiation due to hydrogen being the most abundant form of baryonic matter in the Universe. During the Epoch of Reionisation (EoR), the hydrogen in the intergalactic medium went from neutral to fully ionised due to the ultraviolet photons and X-ray radiation from the first luminous objects, such as stars within early galaxies \citep{2006PhR...433..181F}. In the post-reionisation Universe, we expect to find most of the \ion{H}{I} within so-called damped Lyman-alpha (Ly-$\alpha$) systems \citep{Pritchard:2011xb, Villaescusa-Navarro2014, Spinelli2020} that provide shielding from those sources that would otherwise potentially ionise all the \ion{H}{I}. Due to this, we observe most of the \ion{H}{I} after reionisation within galaxies, galaxy groups, and galaxy clusters. This makes it an excellent tracer of the underlying dark matter density field, which gravitationally binds the Baryonic matter and, therefore, the large-scale structure of the Universe and structure formation history across time. IM with \ion{H}{I} also allows us to study the gas content of galaxies in the post-reionisation era \citep{10.1093/mnras/stx1388}, which serves as the fundamental fuel reservoir for star formation in galaxies \citep{2012ApJ...756..113H, 2015MNRAS.447.1610M}. It helps probe galaxy evolution across cosmic time, which can be compared against state-of-the-art \ion{H}{I} galaxy simulations such as SIMBA \citep{Dave_2019}. These comparisons will enable us to understand better the astrophysics of \ion{H}{I} gas and the galaxies or clusters of galaxies that host it.
\par

Measurements of the \ion{H}{I} IM signal are done in practice using statistical measurements via the power spectrum, a 2-point Fourier statistic, which tells us how much \ion{H}{I} clustering there is on a range of cosmological scales \citep{Peacock2003, Dodelson2003}. This can be achieved by surveying large volumes of the Universe with radio telescopes sensitive to the 21 cm line emission, making sky maps across frequency, which can be used to calculate the power spectrum statistic in either cross-correlation with other tracers of dark matter, such as galaxy surveys, or in auto-correlation of the \ion{H}{I} intensity maps. There have been numerous experiments designed to do \ion{H}{I} IM, with some of them having made detections of the \ion{H}{I} power spectrum at various epochs in cross-correlation with galaxy surveys \citep{Chang2010, Masui2013, Anderson2018, Wolz2021, Cunnington2022, Amiri_2023} or provided upper limits on the auto-correlations \citep{Switzer2013, 2025MNRAS.537.3632M} (also see \citealt{Liu2020} for a review). A detection of the \ion{H}{I} auto-power spectrum on small scales at redshifts $z\sim 0.32$ and $z\sim 0.44$ has been reported in \cite{Paul2023}, based on 96 hours of MeerKAT data on a single-pointing \citep{Mauch2020}. Generally, two experimental configurations exist for \ion{H}{I} intensity mapping. These include single-dish and interferometric instruments. In the post-reionisation era ($z < 6$), single-dish experiments focus on probing larger scales ($k \lesssim  0.1 \ \text{Mpc}^{-1}$) by rapidly surveying large sky areas with low angular resolution \citep{2015aska.confE..19S}, while the interferometric experiments probe smaller, linear to non-linear scales ($k \gtrsim 0.1 \ \text{Mpc}^{-1}$) due to tightly-packed baselines between individual antennas that result in better angular resolution \citep{2016SPIE.9906E..5XN, 2022ApJS..261...29C}. In this manner, the two survey types are complementary on angular scales, $k_{\perp}$, while being sensitive to the same range of line-of-sight scales, $k_{\parallel}$, set by the bandwidth and frequency resolution of the observing instrument \citep{Bull2015}.
\par

A direct detection of the \ion{H}{I} intensity mapping power spectrum presents several challenges due to a combination of instrumental systematics and foregrounds several orders of magnitude stronger than the signal \citep{10.1093/mnras/stw1380, 10.1093/mnras/stab3064, Chen2022}. Cross-correlations with external galaxy surveys can confirm the presence of an \ion{H}{I} signal but do not guarantee that the measured auto-power spectrum is free from systematics. Residual contamination such as low-level broadband radio frequency interference (RFI) can contribute power to the measurement. While null tests can increase the confidence in a detection, they cannot completely rule out subtle, undetected systematics. Ultimately, establishing this technique as a robust cosmological probe will require a combination of well-tested data reduction pipelines and multiple, independent measurements.
\par

In this paper, we aim to address these issues by applying our signal extraction methodology to a self-contained test case, where we directly compare the measured intensity mapping power spectrum to that derived from \ion{H}{I} galaxies detected in the same field. We use Early Science visibility data from the MeerKAT International GHz-Tiered Extragalactic Explorations (MIGHTEE) Survey \citep{Jarvis2016}, employing the delay-spectrum technique \citep{Morales2004, Parsons2012a, Parsons2012, Thyagarajan2013, Parsons2014, Thyagarajan2015, Paul:2016blh, Morales2019} to measure the \ion{H}{I} power spectrum on sub-Megaparsec scales at low redshifts. This technique has primarily been applied to studies of the Epoch of Reionisation (EoR), where it has been used to set upper limits on the 21 cm power spectrum with instruments such as PAPER, the Murchison Widefield Array (MWA), and the Hydrogen Epoch of Reionisation Array (HERA) \citep[e.g.,][]{Parsons2012a, Pober2013, Parsons2014, Deboer2017, TheHERACollaboration2021a}. More recently, applications of this method in the post-EoR regime have shown promising results, employing foreground avoidance and using statistical measures such as spherical and cylindrical power spectra to analyse both simulated and observational data \citep{Paul2021, Chen2021, Chen2022, Chen2023, Paul2023, 2025MNRAS.541..476M}. Comparing the intensity mapping power spectrum to the galaxy power spectrum derived from directly-detected \ion{H}{I} galaxies in the same MIGHTEE dataset allows us to test the robustness of the intensity mapping approach in a regime where traditional \ion{H}{I} galaxy surveys are still feasible, given the relatively low redshift. This comparison further allows us to evaluate whether either the intensity mapping technique or direct detections using radio interferometric data systematically miss part of the \ion{H}{I} content and how consistent the two power spectrum measurements are. Our analysis builds upon and tests the methodology developed in \cite{Paul2023}. It also serves as a follow-up to the MIGHTEE-COSMOS analysis presented in \cite{2025MNRAS.541..476M}, where data from multiple COSMOS pointings were incoherently combined to obtain upper limits on the \ion{H}{I} power spectrum at $z \sim 0.44$. That study was itself a comparative test of the tentative detection reported in \cite{Paul2023}. The MIGHTEE survey, leveraging the wide field-of-view and high sensitivity of MeerKAT, is particularly well-suited for these tests, providing a good sample of \ion{H}{I} galaxies at low redshifts.
\par

The rest of the paper is structured as follows. Section \ref{section:data} outlines the data used in this work. Section \ref{section:delay_pspec} focuses on the 'delay'-spectrum methodology and pipeline developed to measure the $\ion{H}{I}$ power spectrum from the visibility data, while also showcasing the visibility power spectrum results. Section \ref{section:hi_gal_pspec} discusses validations employed using a simulated cosmological volume which contains detected \ion{H}{I} galaxies from MIGHTEE. Further validations of the power spectrum results using visibilities simulated on these \ion{H}{I} detections are discussed in Section \ref{section:pipeline_validation}, followed by the summary and conclusions in Section \ref{section:conclusion}.
\par

Throughout this paper, a flat ($\Omega_{\text{k}} =0$) $\Lambda$CDM cosmology is assumed, with cosmological parameters [$h, \Omega_{\text{M}}, \Omega_{\text{b}}, n_{\text{S}}, \sigma_8$] = [0.674, 0.311, 0.049, 0.965, 0.811], corresponding with those from the latest Planck results \citep{Planck_18_params}.
\par


\section{MIGHTEE Data}
\label{section:data}

In this work, data from the MeerKAT telescope's MIGHTEE survey \citep{Jarvis2016} is used to make a statistical measurement of the \ion{H}{I} signal at low $z$ and on small scales (smaller than about 1 Mpc) via the \ion{H}{I} power spectrum. The details of the specific MIGHTEE survey data are discussed below.
\par

\subsection{The MIGHTEE Survey}
\label{subsection:mightee_survey}

The MIGHTEE Survey\footnote{\url{https://www.mighteesurvey.org/}} is one of the Large Survey Projects (LSPs) conducted using the MeerKAT telescope. It covers four well-studied extragalactic fields, namely COSMOS\footnote{\url{https://cosmos.astro.caltech.edu/}}, XMM-LSS, CDFS, and ELAIS-S1, with a total observed area of about 20 $\deg^2$. These various fields also have the potential to be used for multi-wavelength studies, which is a good trade-off for the fact that it is shallower in comparison to the complementary, deeper LADUMA Survey \citep{2018AAS...23123107B}, which will go to much higher redshifts despite covering a much smaller patch of the sky. Importantly for the context of this study, the full MIGHTEE survey will reach similar depths ($\sim$ 1 $\mu$Jy) as the planned SKA Observatory, thus making it a useful survey with which assess the level of sensitivity achievable for IM. To achieve its scientific goals, MIGHTEE conducts its survey by simultaneously making continuum \citep{2025MNRAS.536.2187H}, polarimetry (MIGHTEE-pol) \citep{2024MNRAS.528.2511T}, and spectral-line (MIGHTEE-\ion{H}{I}) \citep{2024MNRAS.534...76H} measurements. Observations with the MIGHTEE survey have concluded, taking approximately 1000 hours of data in L-band over the four fields described above. In this work, the focus is on the use of various "Early-Science" data products, particularly the COSMOS central pointing (RA $\sim$ 150.12 $\deg$, Dec. $\sim$ 2.2 $\deg$) visibility data and MIGHTEE-\ion{H}{I} galaxy detections obtained using said visibility data, over the low-redshift band (0 < $z$ < 0.09, corresponding to the frequency range of 1310 MHz < $\nu$ < 1420 MHz).
\par

\subsection{Visibility data}
\label{subsection:mightee_hi_vis_data}

The data used comprises the interferometric visibilities observed with the MeerKAT telescope on the central pointing of the COSMOS field, as well as a catalogue of detected \ion{H}{I} galaxies created using these interferometric observations \citep{Ponomareva2023}. The visibility data consist of three measurement sets (or data blocks) with a total on-source observation time of 17.45 hours. For the exact visibility data specifications, refer to Table \ref{tab:mightee_vis_data_params}. Although there are other blocks available over the COSMOS field, they correspond to different pointings and cannot be easily coherently averaged for the visibility-based power spectrum analysis. The initial two data blocks were observed with 4K (4096) channels over the L-band, with the last observed using the 32K (32768 channels) mode configuration of the telescope. The time resolution is 8 seconds. These data blocks are each divided into scans. The length of each of these scans in block 1524147354 and block 1525613583 is $\sim$ 10 minutes, amounting to 37 and 31 scans for each, respectively. Block 1524147354, therefore, has a total observation time of 6.1 hours, while block 1525613583 has an observation time of 5.1 hours. The 32K block, block 1587911796, has longer scans of $\sim$ 55 minutes, and a total observation time of 6.25 hours spread over 7 scans.
\par

\begin{figure}
    \centering
    \includegraphics[width=\columnwidth]{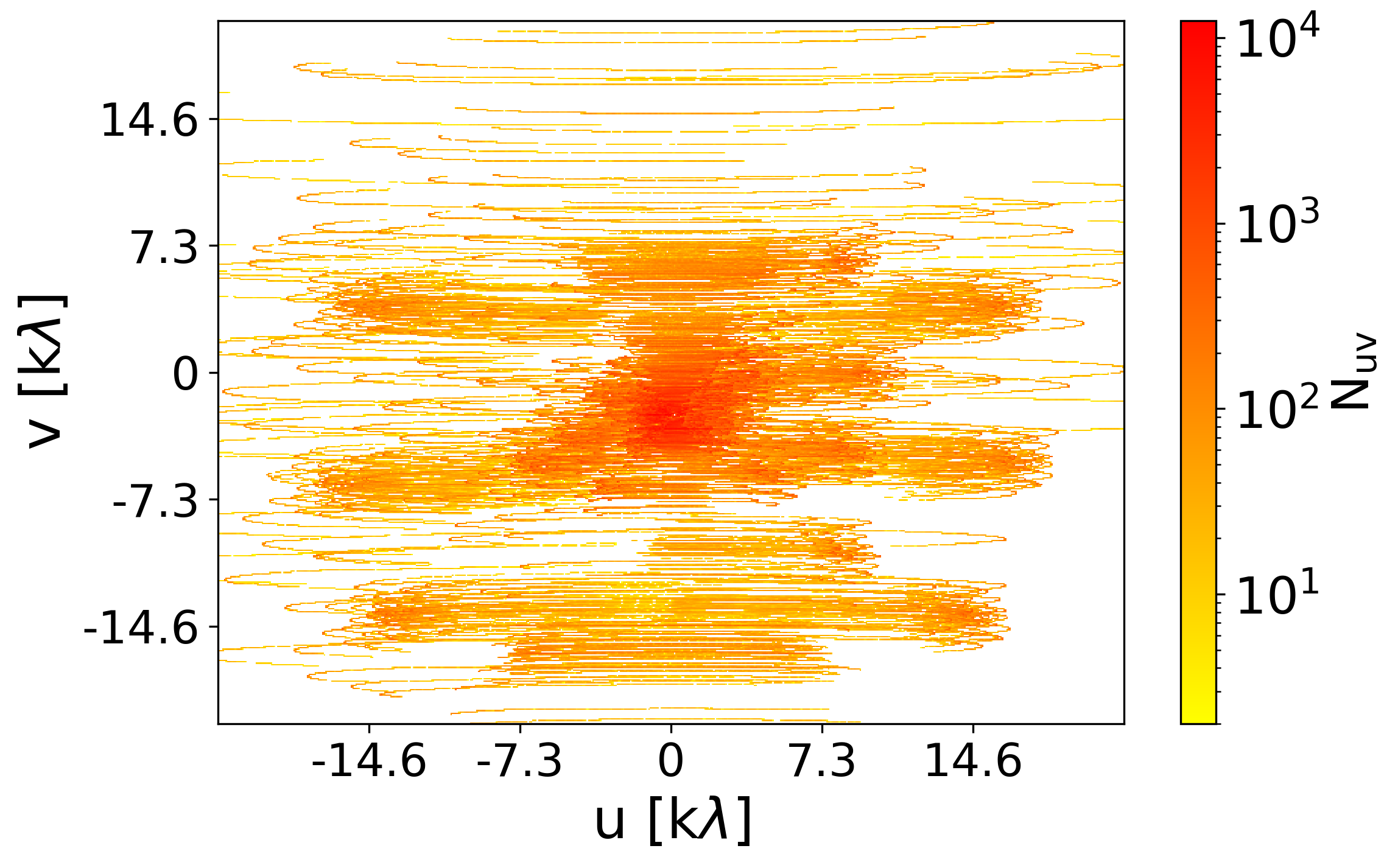}
    \caption{The $uv$ distribution of the combined COSMOS field data used in this study at the central frequency of 1362.425 MHz. This distribution is shown in 2D, with the number of $uv$ points displayed in the colour bar. The grid size used here is $\Delta u = \Delta v = 73\lambda$, where $\lambda$ is the central wavelength of the band used in this study.}
    \label{fig:cosmos_data_uvdist}
\end{figure}

Being a core-heavy instrument, data observed with MeerKAT will be most sensitive to the larger range of cosmological ($k$) scales we are sensitive to. The core-concentration of MeerKAT is demonstrated by the $uv$ distribution of the combined data blocks shown in Figure \ref{fig:cosmos_data_uvdist}. For this study, we use the calibrated, flagged, and reduced measurement sets. The reduction process for this visibility data is described in \cite{Maddox2021} for the 4K data blocks, and in \cite{Heywood2021} for the 32K data block. We do not repeat such processes before putting the visibility data through the gridding, averaging, and power spectrum estimation pipeline discussed in Section \ref{section:delay_pspec}. We make use of the frequency channels stored in the observed measurement sets over the range 1310 - 1420 MHz. From this selection of data blocks, we choose a relatively RFI-free region spanning approximately 1332.64 - 1392.20 MHz ($0.020 < z < 0.066$), resulting in a bandwidth of about 60 MHz. This range was chosen to match the redshift region where we have a complete catalogue of \ion{H}{I} detected galaxies \citep{Ponomareva2023}. In 4K resolution, this results in 284 channels. To ensure data uniformity, the 32K data block, 1587911796, was reduced to 4K by averaging over every 8 frequency channels, thus allowing us to coherently combine and average the data from all three data blocks for the power spectrum estimation.
\par

\begin{table*}
  \begin{tabular}{cccccccccc}
    \hline
    Block ID & Date & R.A. & Dec. & $\text{t}_{\text{obs}}$ (on-source) & $\text{N}_{\text{chan}}$ & $\text{N}_{\text{ant}}$ & Primary & Secondary \\
     & (UT, J2000) & & & (h) & & & calibrator & calibrator \\
    \hline
    1524147354 & 2018-04-19 & 10$^{\text{h}}$00$^{\text{m}}$28$^{\text{s}}$.6s & +02$\degr$12$\arcmin$21$\arcsec$ & 6.1 & 4096 & 64 & J0408-6545 & 3C237\\
    
    1525613583 & 2018-05-06 & 10$^{\text{h}}$00$^{\text{m}}$28$^{\text{s}}$.6s & +02$\degr$12$\arcmin$21$\arcsec$ & 5.1 & 4096 & 62 & J0408-6545 & 3C237\\
    
    1587911796 & 2020-04-26 & 10$^{\text{h}}$00$^{\text{m}}$28$^{\text{s}}$.6s & +02$\degr$12$\arcmin$21$\arcsec$ & 6.25 & 32768 & 59 & J0408-6545 & 3C237\\
    \hline
  \end{tabular}
  \caption{The MIGHTEE Early Science COSMOS central pointing visibility data sets used in this work. The table is adapted from \protect\cite{Heywood2021}.}
  \label{tab:mightee_vis_data_params}
\end{table*}

\subsection{\texorpdfstring{\ion{H}{I}}{HI} galaxy detections}
\label{subsection:mightee_hi_catalogue_data}

In addition to the visibility data, a catalogue of detected \ion{H}{I} galaxies from the MIGHTEE data is used in this study. The catalogue forms part of the outputs of the Early Science data processing for MIGHTEE-\ion{H}{I} described in \cite{Ponomareva2023}. It comprises \ion{H}{I} detections on both the single COSMOS pointing and three overlapping pointings on XMM-LSS.
\par

These detections were made, in the first instance, visually, with no guidance, on imaged cubes using the Cube Analysis and Rendering Tool for Astronomy (CARTA\footnote{\url{https://cartavis.github.io/}}) software suite. Once the positions and frequencies of each source are known, cubelets of each galaxy are extracted around the sources from the main cube. These cubelets are then processed by the source-finding and parametrisation tool SOFIA \citep{2015MNRAS.448.1922S}. Masks are created around the emission from the frequency channels that correspond to the galaxy sources. After masking, the flux is summed, the central frequency of individual sources is calculated, and this \ion{H}{I} flux, S, is then subsequently converted to \ion{H}{I}
mass, M$_{\ion{H}{I}}$, using the relation from \cite{2017PASA...34...52M}:

\begin{equation} \label{eq:flux_to_mass}
\left( \frac{M_{\ion{H}{I}}}{M_{\odot}} \right) = \frac{2.356 \times 10^{5}}{1+z} \left( \frac{D_L}{\text{Mpc}} \right)^2 \left( \frac{S}{\text{Jy km/s}} \right),
\end{equation}
\par

where $D_L$ is the luminosity distance to the source at redshift $z$, and $S$ is the flux calculated from the individual cubelets for each source.
\par

On the COSMOS pointing specifically, there are 80 detections over a redshift range of $0.02 < z < 0.09$, and spanning a \ion{H}{I} mass range of $4 \times 10^{7} M_{\odot} \lesssim M_{\ion{H}{I}} \lesssim 10^{10} M_{\odot}$. The catalogue includes the measured velocity width corresponding to the point of the global \ion{H}{I} profile at 50\% peak flux density, also known as $W_{50}$, values. This $W_{50}$ represents the maximum rotational velocity measured from a resolved rotation curve of a given galaxy, i.e., $V_{\text{max}} = W_{50}/2 \sin(i)$, where $i$ is the inclination angle of said galaxy \citep{2021MNRAS.508.1195P}. The details of the construction of the overall catalogue are described in \cite{Maddox2021}, with additional details on how the $W_{50}$ values were measured found in \cite{2021MNRAS.508.1195P}. Further, detailed \ion{H}{I} studies making use of the data in this catalogue of detections can be found, for instance, in \cite{Rajohnson2022, Tudorache2022, Pan2022, Ponomareva2023}.
\par

\begin{figure}
    \centering
    \includegraphics[width=\columnwidth]{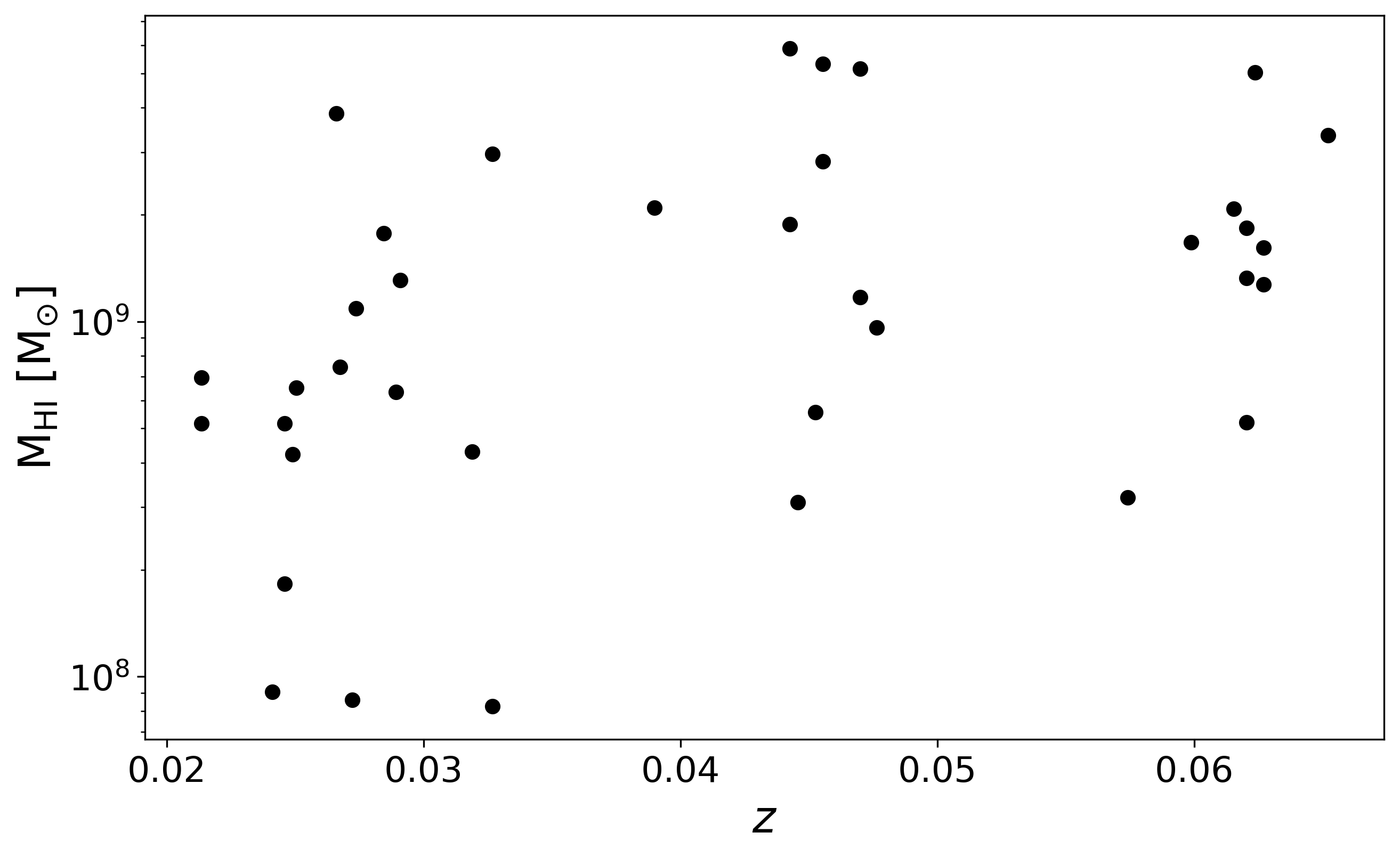}
    \caption{The \ion{H}{I} masses, $M_{\ion{H}{I}}$, for the selected galaxies from the full MIGHTEE-\ion{H}{I} Early Science catalogue plotted at their respective redshifts, $z$. The sample shown contains galaxies on the central COSMOS pointing only and is sourced over a redshift range that aligns with the frequency range chosen for the visibility data.}
    \label{fig:mhi_z_gal_plot}
\end{figure}

\begin{figure}
    \centering
    \includegraphics[width=\columnwidth]{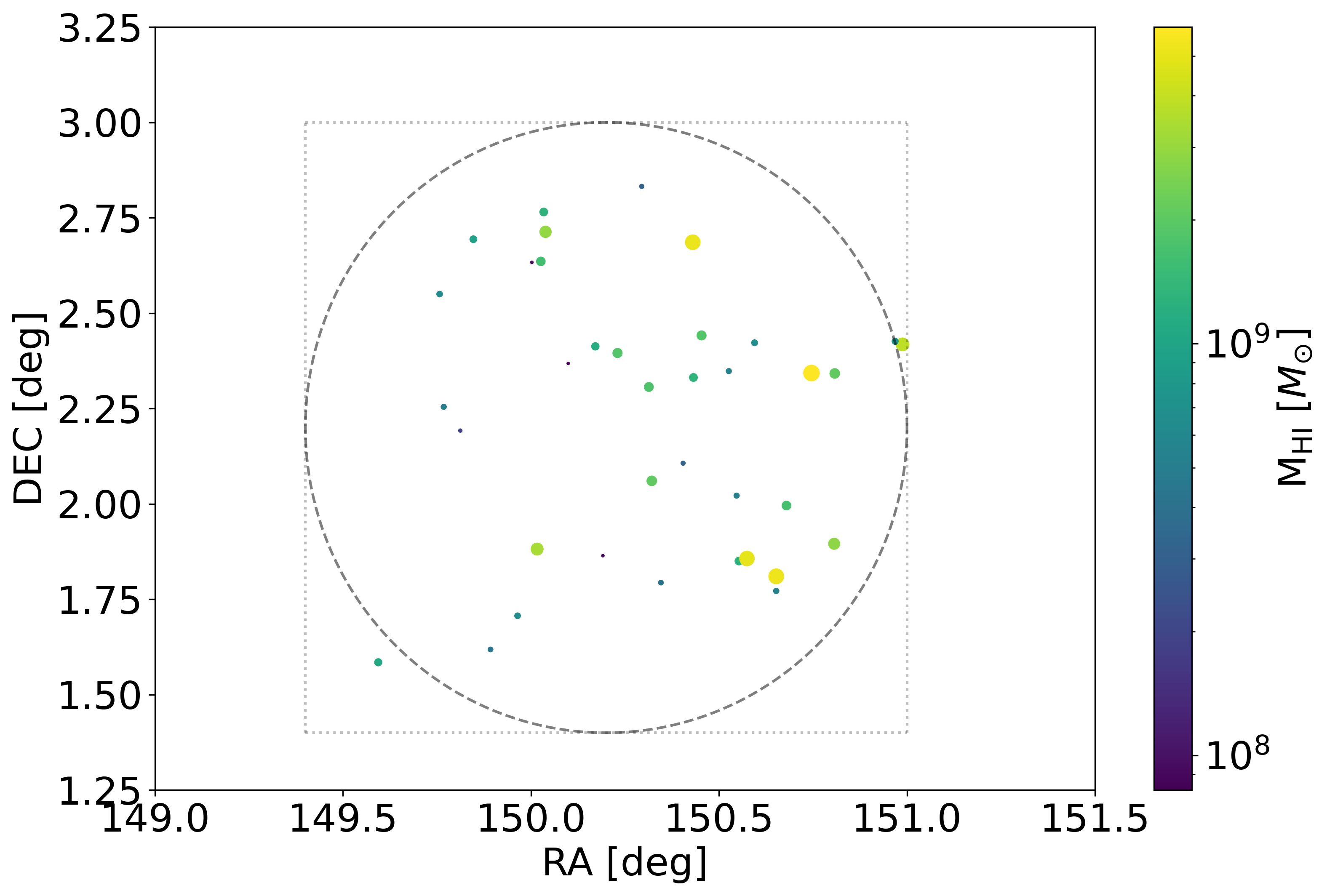}
    \caption{The \ion{H}{I} masses, $M_{\ion{H}{I}}$, and positions for each galaxy selected in the sub-band from the catalogue in R.A. and Dec. The dashed circle corresponds to the main lobe of the MeerKAT primary beam at 1362.425 MHz, while the dotted square denotes the edges along the Right Ascension and Declination maximum and minimum limits of the box in which the galaxy masses are placed (see Section \ref{section:hi_gal_pspec}).}
    \label{fig:mhi_gal_field}
\end{figure}

Out of the 80 galaxies detected on the COSMOS field, 37 galaxies are in the 0.02 - 0.066 redshift range of interest. They are used to compare with the power spectrum estimation from the visibility data (see Section \ref{section:hi_gal_pspec}). One caveat worth mentioning here is that there may be galaxies that lie just beyond the frequency range that may leak a few channels into this range because of their width. Despite this, the effect of this will be negligible overall when comparing power spectra. The measured \ion{H}{I} masses for each of these galaxies along redshift are shown in Figure \ref{fig:mhi_z_gal_plot}, with an accompanying plot for R.A. and Dec. shown in Figure \ref{fig:mhi_gal_field}. Given the redshift range, as well as the area along R.A. and Dec., we estimate a cosmological volume of $\sim$ (389 Mpc/h)$^3$, or 1254 Mpc$^3$. For 37 galaxies, this corresponds to a galaxy density per Mpc$^3$, $n_{\text{gal}}$, of $\sim$ 0.03 Mpc$^{-3}$.
\par


\section{\texorpdfstring{\ion{H}{I}}{HI} Power spectrum from the MIGHTEE visibility data}
\label{section:delay_pspec}

This section discusses the methodology used to estimate the \ion{H}{I} power spectrum on the MIGHTEE calibrated and processed visibility data. While the details discussed here are comprehensive, some useful complementary references to the technical details about the pipelines discussed below can be found in \cite{Paul2021, Chen2022, Paul2023}.
\par

\subsection{Delay spectrum pipeline and power spectrum estimator}
\label{subsection:delay_vis_pipeline}

\begin{figure*}
    \includegraphics[width=0.975\columnwidth]{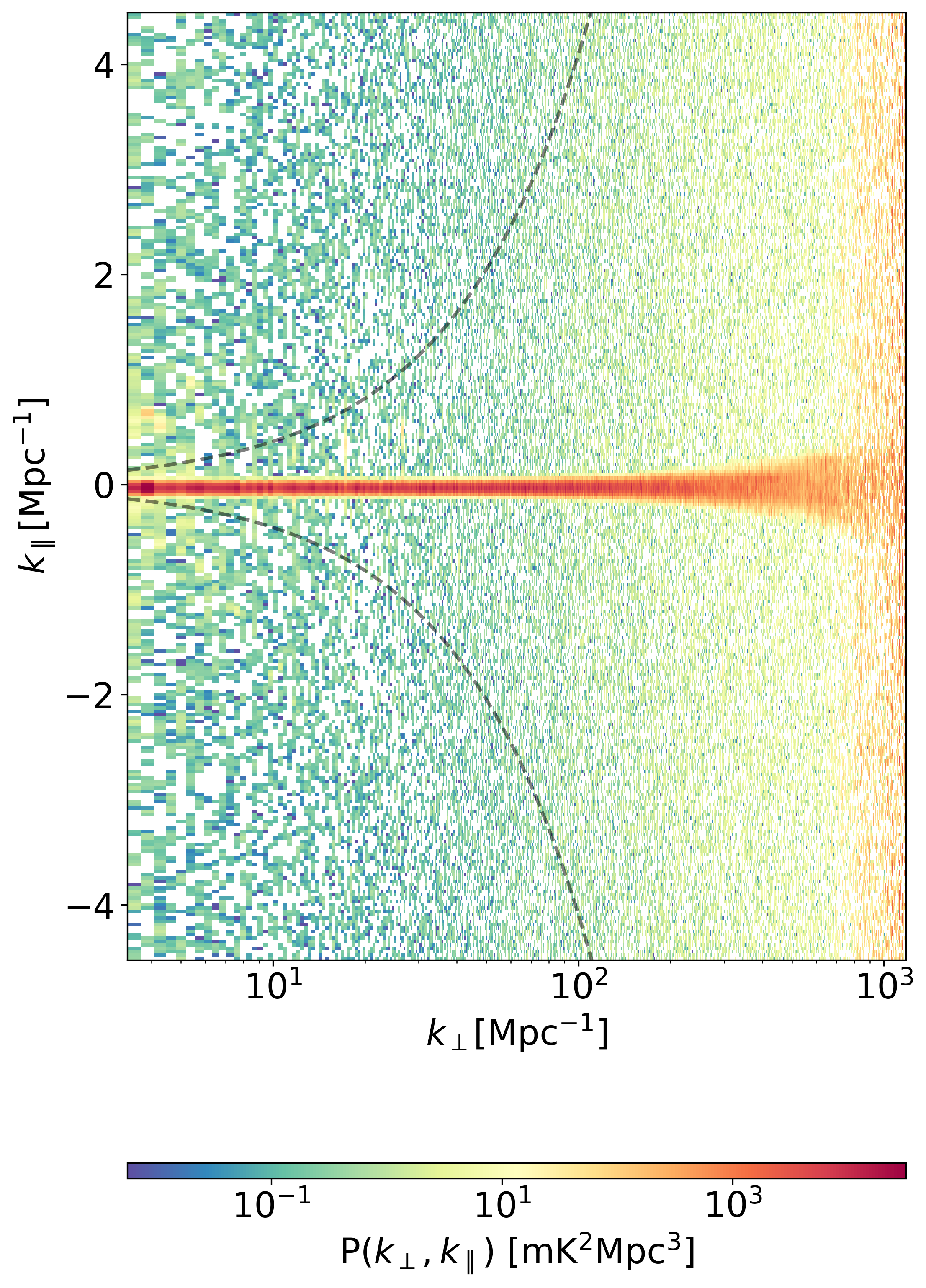}
    \includegraphics[width=0.975\columnwidth]{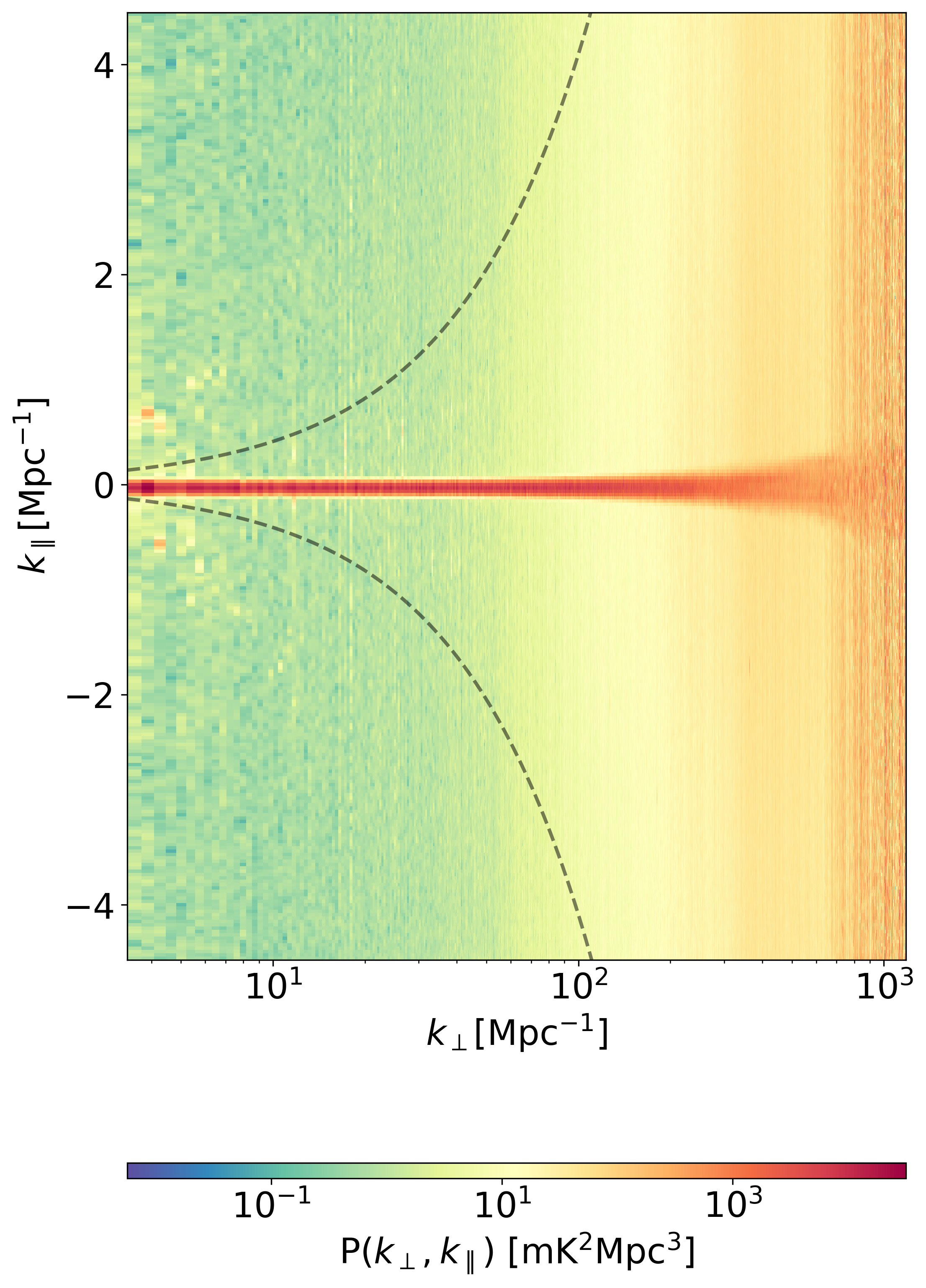}
    \caption{The Stokes I cylindrical cross-correlation (left) and auto-correlation (right) power spectra of the combined 17.45h of MIGHTEE COSMOS Early Science visibility data before any contamination removal is applied to the 3D power spectrum cube, $P(\boldsymbol{k}_{\perp}, k_{\parallel})$. Also shown is the horizon limit, $k_{\parallel}^{\text{horizon}}(k_{\perp})$ as a dashed grey line for $\pm k_{\parallel}$.}
    \label{fig:stokesI_2d_pk_before_flagging}
\end{figure*}

Here, we describe the methodology employed to produce the  3D power spectrum cube on the interferometric data observed using MeerKAT. This methodology was first developed in \cite{Paul2021}, and has undergone improvements since then \citep{Paul2023}. From this 3D power spectrum, a methodology is described to calculate the subsequent cylindrical-averaged (2D) and spherically-averaged (1D) power spectra. The paper discussed the possibility of estimating the \ion{H}{I} power spectrum by producing simulations of MIGHTEE observations of the COSMOS field informed by the observational parameters and first measurement sets from the survey itself. Since then, the proposed methodology has also been used in the analysis of $\sim$ 96 hours of MeerKAT data observed on a single pointing on the DEEP2 field \citep{Mauch2020} in \cite{Paul2023} to make a tentative detection of the \ion{H}{I} signal in auto-correlation. 
\par

In this work, applying the techniques and lessons learnt so far to a set of interferometric observations which can be directly compared to \ion{H}{I} detections made on the data, we can further validate the methods employed to obtain the \cite{Paul2023} results, in addition to being a means of providing an estimate of the \ion{H}{I} power spectrum on the MIGHTEE data at low redshift ($z < 0.1$).
\par

The first step in the pipeline is to average all the visibilities into a grid in the $(u,v)$ plane at each frequency channel, $\nu$. This is done for both the Stokes I visibility data and Stokes V, as described later. We choose 1500 grid points and select $\Delta u = \Delta v = 73\lambda$, which approximately corresponds to the width of the MeerKAT primary beam in $k_{\perp}$ space at the central frequency ($\sim$ 1362 MHz) of the band. We consider two scenarios:

\begin{enumerate}

    \item 
    A single grid on which all the visibility data is gridded and averaged. Power spectra estimations on this are designated as auto-correlation power spectra, or "Visibilities (total)", to denote power spectra derived from this single-grid, combined MIGHTEE visibility data in some comparative power-spectrum results.
    \\

    \item 
    In the cross-correlation case, instead of combining all the visibility data on one grid, each consecutive 8-second time stamp of data is placed on one of two visibility grids. We denote these visibility grids as "odd" and "even".  The power spectra estimated on the cross-correlation of these grids are labelled as "Visibilities Cross" or "Visibilities (odd $\times$ even)".
    
\end{enumerate}

Since the noise in each grid is not expected to correlate, this cross-correlation eliminates the noise bias. Hence, where there is predominantly noise in the data, we expect the power spectrum to be consistent with zero. At the same time, where we have sufficient sensitivity, our estimator should yield the unbiased \ion{H}{I} signal power spectrum amplitude. In the auto-correlation case, to remove the thermal noise contribution, it has to be modelled through simulations based on the observational parameters from the data (this is discussed further in Section \ref{subsection:thermal_noise_sims}).
\par

The gridded visibilities are then Fourier transformed along the frequency axis and multiplied by their conjugate (or by the conjugate of the second grid in the case of cross-correlation) to obtain the 3D power spectrum in units of Jy$^{2}$ Hz$^{2}$. This $uv\tau$ cube is then multiplied by normalisation factors to obtain the power spectrum in cosmological units of Mpc$^{3}$ mK$^{2}$. Mathematically, the delay power spectrum can be written as \citep{Morales2004, McQuinn2006a, Parsons2012, Pober2013, Thyagarajan2013, Parsons2014, Thyagarajan2015, Paul:2016blh}

\begin{equation}
\label{eq:pspec_delay_norm_final}
    \hat{P}_{\text{D}}(\boldsymbol{k}_{\perp}, k_{\parallel}) = \left( \frac{\lambda^2}{2 k_{\text{B}}} \right)^2 \frac{A_{\text{e}}}{\lambda^2 B} \frac{r_{\nu}^2 \Delta r_{\nu}}{B} \Re\{ \tilde{V}_{i}(\boldsymbol{u},\tau) \tilde{V}_{j}^{*}(\boldsymbol{u},\tau)  \},
\end{equation}

\noindent
where $\boldsymbol{k}_{\perp}$ is composed of the $k_x$ and $k_y$ angular scales and is given by $\boldsymbol{k}_{\perp} = 2\pi \boldsymbol{u} / r_{\nu}$, with $\boldsymbol{u} = (u,v)$ and $r_{\nu}$ the comoving distance at the central redshift of the observation, $z$. The central wavelength of the observed bandwidth is $\lambda$, $k_{\text{B}}$ is the Boltzmann constant, $A_{\text{e}}$ is the effective area of the dish, $B$ is the bandwidth of the observation and $\Delta r_{\nu}$ is the comoving width corresponding to this bandwidth. For this power spectrum, we take the real part of the visibility correlation, $\Re \{ \tilde{V}_i \tilde{V}_j^* \}$, where $i$ and $j$ correspond to the two visibility grids generated for the cross-correlation case, and $i = j$ for the auto-correlation. The above result is valid in the regime where there is negligible redshift evolution along the light cone and also necessarily relies on the flat sky approximation \citep{Chen2021}. To estimate the 1D power spectrum from the 3D power spectrum cube given by $\hat{P}_{\text{D}}(\boldsymbol{k}_{\perp}, k_{\parallel})$, an inverse noise variance weighted spherical average is taken in logarithmic $k$ bins, $\Delta k_{i}$, which can be expressed as \citep{Paul2023}

\begin{equation} \label{eq:pk_1d_estimator}
    \hat{P}_{\text{D}}^{i}(k) = \sum_{j} P_{\text{D}}^j w_{j}  \big /  \sum_{j} w_{j},
\end{equation}

\noindent
where $w_j = 1/\sigma^2_{j, P_{\text{TN}}}$ are the thermal noise variance weights (discussed further in Section \ref{subsection:thermal_noise_sims}) and $P_{\text{D}}^j$ are the 3D power spectrum amplitude values, $j$, in the selected $k$ bin. Further, the error on this power spectrum estimate can be derived from the sampling variance \citep{Paul2023},

\begin{equation} \label{eq:pk_estimator_errbar}
    (\Delta \hat{P}_{\text{D}}^{i})^2 = \sum_{j} (P^{j}_{\text{D}} - \hat{P}^i_{\text{D}})^2 w_{j}^2 \ \big / \ \left(\sum_{j} w_{j} \right)^2.
\end{equation}

\noindent
The cylindrically-averaged (2D) power spectrum can also be estimated similarly by applying the same inverse noise variance weighting and binning cylindrically in transverse and line-of-sight $k$ modes, $(k_{\perp}, k_{\parallel})$, giving an estimator of the form

\begin{equation} \label{eq:pk_2d_estimator} 
    \hat{P}^i_{\text{D}}(k_{\perp}, k_{\parallel}) = \sum_j P^j_{\text{D}}(\boldsymbol{k}_{\perp}, k_{\parallel}) \ w_j(\boldsymbol{k}_{\perp}, k_{\parallel}) \big / \sum_j w_j(\boldsymbol{k}_{\perp}, k_{\parallel}),
\end{equation}

\noindent
which collapses the 2D $\boldsymbol{k}_{\perp}$ plane while performing the above averaging and noise weighting. To make use of foreground avoidance, modes are sampled above the horizon limit \citep{Datta2010, Parsons2012, Liu2014, Thyagarajan2013, Thyagarajan2015},

\begin{equation} \label{eq:horizon_lim}
    k_{\parallel} = \frac{r_{\nu}(z) E(z) H_0}{c(1+z)} \sin(\theta) \ k_{\perp},
\end{equation}

\noindent
where $H_0$ is the Hubble parameter at $z = 0$, $E(z) = H(z)/H_0$ and $c$ is the speed of light in a vacuum. In this work, we use $\sin(\theta) = 1$ as this corresponds to the maximum delay, e.g. largest $k_{\parallel}$ for the foreground wedge. Any spectrally smooth signal should be well below this boundary. This ensures that, for the power spectrum estimation, modes are sampled well above the region where we expect foregrounds to be located in the 3D power spectrum cube, giving minimal spillover into the \ion{H}{I} window. At $z \sim 0.042$, this limit reduces to $k_{\parallel} \simeq 0.041 \ k_{\perp}$. This is shown plotted on the cylindrical power spectra as a dashed curve throughout the paper for both $\pm k_{\parallel}$.
\par

Figure \ref{fig:stokesI_2d_pk_before_flagging} shows the Stokes I cylindrical power spectra estimated for both auto- and cross-correlation of delay-transformed visibility grids. Also shown is the horizon limit as a dashed line along $k_{\perp}$ for both $\pm k_{\parallel}$. The main foreground wedge is mainly confined below the horizon limit due to the relatively small size of the MeerKAT beam. The Early Science COSMOS data have been put through a comprehensive calibration and flagging process, and as a result, we have a clean \ion{H}{I} window over which to estimate power spectra. Despite this, some residual contamination appears to be present above the theoretical horizon limit when examining the cylindrical power spectra. This is seen in the combined data as well as in each data block. It is present in both auto- and cross-power spectra, although it is more obvious in the auto-power spectra, as can be seen from the figure. The strongest contamination is localised below $k_{\parallel} \sim 1 \ \text{Mpc}^{-1}$ and $k_{\perp} \sim 6 \ \text{Mpc}^{-1}$. This corresponds to the region in which the first two cylindrical $k$ bins of the estimator are selected, so we expect these bins to have some contamination if no extra flagging is applied. We will next discuss our approach to dealing with this contamination in the 3D power spectrum cube before making estimates of the 1D \ion{H}{I} power spectrum.
\par

\subsection{Removal of low-level contamination}
\label{subsection:5sigma_flag_methodology}

The contamination found in the cosmology window of our 2D power spectrum only affects a few pixels but has the potential to contribute to the bulk of the power in the $k$ bins over which we estimate the 1D power spectrum. At the same time, it is broadband and of a low enough level to evade standard flagging approaches. Therefore, we adopt the approach of flagging the contamination at the 3D power spectrum level. We have a total of $\sim 3 \times 10^{7}$ 3D $\boldsymbol{k}$ modes above the horizon limit we choose in our data, and we expect these to be predominantly noise-dominated. By comparing these 3D values to the expected noise distribution, we can identify and flag extreme outliers. If this number is small, the impact on signal loss should be negligible. Therefore, we create a flag mask, based on detailed thermal noise simulations, to be applied to the voxels in the 3D auto-power spectrum cube. This is done before performing any power spectrum averaging on the 3D cube to obtain 2D and 1D power spectrum estimates. The flag criterion is given by

\begin{equation} \label{eq:sigma_cut_condition}
    \left|P_{\text{D}}(\boldsymbol{k}_{\perp},k_{\parallel}) - \bar{P}_{\text{TN}}(\boldsymbol{k}_{\perp},k_{\parallel})\right| > N \times \sigma_{P_{\text{TN}}}(\boldsymbol{k}_{\perp},k_{\parallel}),
\end{equation}

\noindent
That is, we flag voxels when the difference between the delay power spectrum, $P_{\text{D}}$, and the average expected thermal noise power spectrum, $\bar{P}_{\text{TN}}$, at $(\boldsymbol{k}_{\perp},k_{\parallel})$ is greater than some integer multiple, $N$, of the standard deviation, $\sigma_{P_{\text{TN}}}$, of this set of thermal noise power spectra generated from the thermal noise visibility realisations (we discuss these realisations further in Section \ref{subsection:thermal_noise_sims}). This procedure is done on the 3D auto-power spectrum cube, $P_{\text{D}}$, of both the "odd" and "even" visibility grids, and then the masks generated on each are combined. If voxels are flagged in the "odd" or "even" case, then the final mask will have that voxel flagged (regardless of whether or not it is only flagged in one or the other of the odd/even cases). Further, we used $N = 5$, which should only pick up extreme outliers while eliminating most of the trace contamination that shows up in the Stokes I power spectrum.
\par

Guaranteeing that our data are indeed free of outliers might be subject to some discussion. In this case, visual inspection on the 2D power spectrum helped us identify some "hot" pixels that should not be present if the data were \ion{H}{I} or noise-dominated. We have a good understanding of how both the \ion{H}{I} and noise should be distributed at the 3D power level for our dataset. Any strong outliers from the expected distribution (e.g., by looking at the histograms) are almost certainly due to contamination. Of course, this will become more difficult if the contamination level is close to the noise level. Another way to spot contamination is to look at clear patterns in the flag fraction, as can be seen later in Section \ref{subsec:sigma_flag_vis_tests}. Ideally, this contamination should be picked earlier in the data processing, but wide-band, low-level RFI might very well only show up at the 3D power spectrum level, after some visibility averaging is done. Further cases for $N$ and the power spectrum results of this are discussed in Section \ref{subsec:sigma_flag_vis_tests}. Again, it is important to note that the \ion{H}{I} signal is highly anisotropic at these scales, so the impact of this flagging will depend on where it is applied, at least in terms of the 2D power spectrum. A comparison with any model should be done by applying the same flags to the model in the 2D space before any spherical averaging.
\par

With a means to grid, average, and transform the visibility data and then subsequently estimate power spectra with thermal noise weighting and contamination flagging, these methods can be applied to the Stokes I data. We next discuss the thermal noise simulations.
\par

\subsection{Thermal noise simulations}
\label{subsection:thermal_noise_sims}

\begin{figure}
    \centering
    \includegraphics[width=\columnwidth]{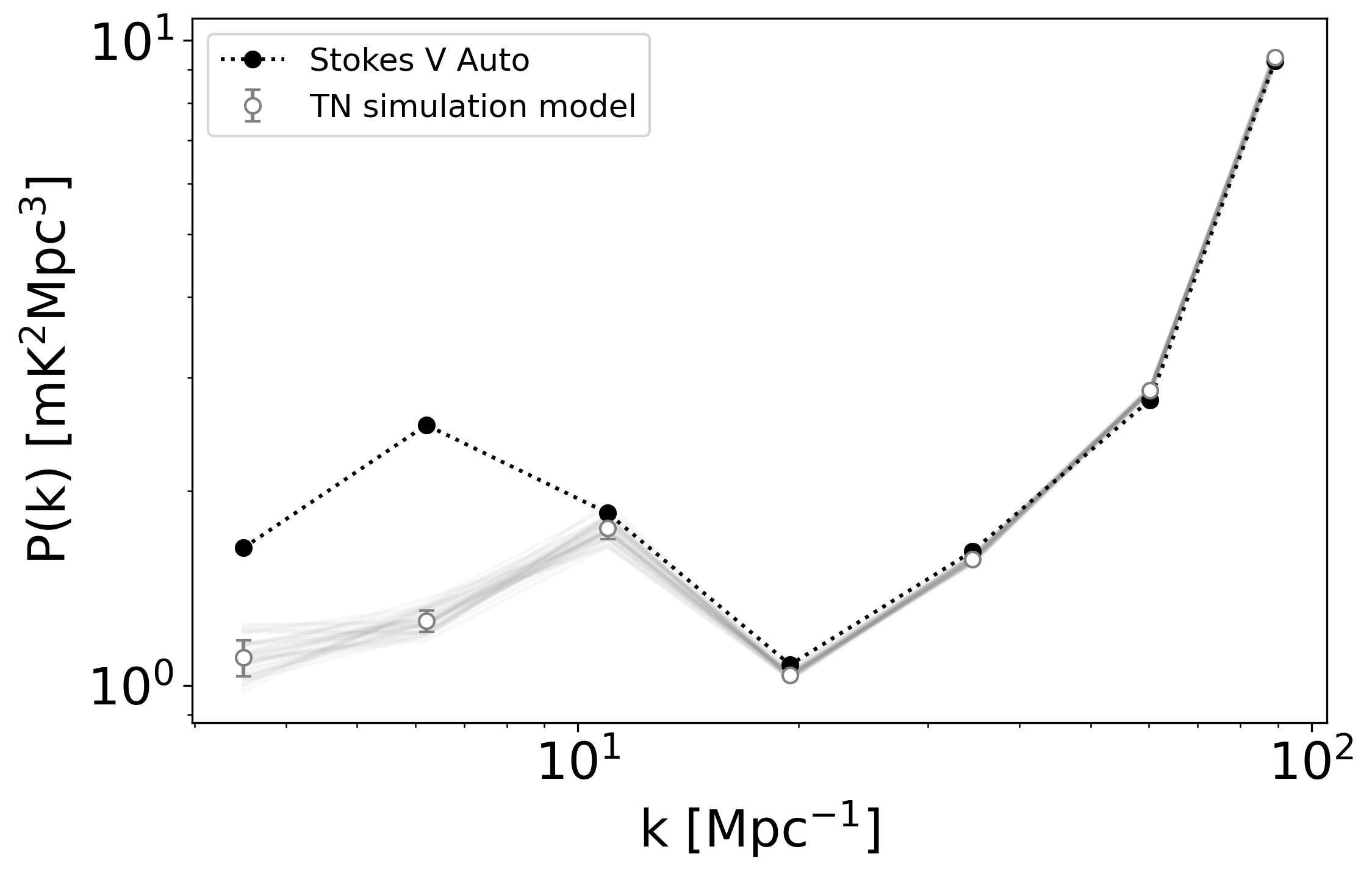}
    \caption{Comparison between the Stokes V and refitted simulated thermal noise 1D auto-power spectra. The increased amplitude we observe at $k \lesssim 10$ Mpc$^{-1}$ for Stokes V is primarily due to leakage from Stokes I, resulting in the discrepancy in the two power spectra. Beyond this, the power spectra are consistent across all $k$ scales.}
    \label{fig:stokes_v_tn_auto_pk1d_comp}
\end{figure}

Thermal noise simulations are required to effectively apply the estimator to the Stokes I (or V) data, as it relies on these simulations for the noise weighting, as well as removing contamination found in the \ion{H}{I} window of the power spectrum cube generated on the visibility data. This is done at the visibility level. For each baseline, time stamp, and frequency channel, the thermal noise is assigned randomly from a complex Gaussian distribution with a mean of zero, $\mu_{\text{TN}} = 0$, and a standard deviation given by the radiometer equation \citep{2002ASPC..278...81C, Morales2005}:

\begin{equation} \label{eq:sigma_tn}
    \sigma_{\text{TN}} = \frac{2k_{\text{B}}T_{\text{sys}}}{A_e \sqrt{\delta\nu \ \delta t}},
\end{equation}

\noindent
where $T_{\text{sys}}$ is the system temperature of the instrument, $A_e$ is the effective area of the dishes and $\delta\nu$ and $\delta t$ are the frequency and time resolutions, respectively. The quantity $A_e/T_{\text{sys}}$ is known as the natural sensitivity. It is particularly crucial for ensuring that the simulated thermal noise amplitude is accurate when comparing it to the Stokes I power spectrum amplitude.
\par

The reported\footnote{\url{https://skaafrica.atlassian.net/wiki/spaces/ESDKB/pages/277315585/MeerKAT+specifications}} MeerKAT natural sensitivity for the selected frequency range is 6.22 $\text{m}^2 / \text{K}$. Since the reported value from MeerKAT and that obtained from the data itself may be different due to differing system temperature, it needs to be corrected. To do this, we compare simulated thermal noise power spectra amplitudes based on the observation parameters of our data to the power spectrum obtained from the Stokes V data. At least at large $k$ values, we expect such data to be noise-dominated. The Stokes V data are put through the same gridding and averaging procedures as described for Stokes I. The 1D power spectrum is then compared to the 1D thermal noise power spectrum averaged over 10,000 realisations put through the same procedures described for Stokes I and Stokes V. The ratio of the 1D thermal noise power and Stokes V power at $k > 10$ Mpc$^{-1}$ is then used as a factor to rescale the initial value of $A_e/T_{\text{sys}}$. Doing so results in a value of 6.75 m$^2$/K, which is still consistent with what we expect for MeerKAT.
\par

Once a refitted value for the natural sensitivity is obtained, the 10,000 realisations of the thermal noise simulations are used as a model of the thermal noise present in the data. The spherically averaged noise power spectrum is shown in Figure \ref{fig:stokes_v_tn_auto_pk1d_comp} and compared to the 1D power spectrum from the Stokes V data. The noise increases for large $k_{\perp}$ due to the smaller number of long baselines. At $k \gtrsim 10 \ \text{Mpc}^{-1}$, there is good agreement between the power spectra as expected, since the Stokes V signal will be noise-dominated. Below that, we see more power from Stokes V which should be primarily leakage from Stokes I. We then use the variance over the set of thermal noise power spectrum realisations, $\sigma_{P_{\text{TN}}}^2$, as the weights, $w_j$, in the power spectrum estimator.
\par

\subsection{Power Spectrum Results}
\label{section:pspec_results}

\begin{figure*}
    \centering
    \includegraphics[width=0.975\columnwidth]{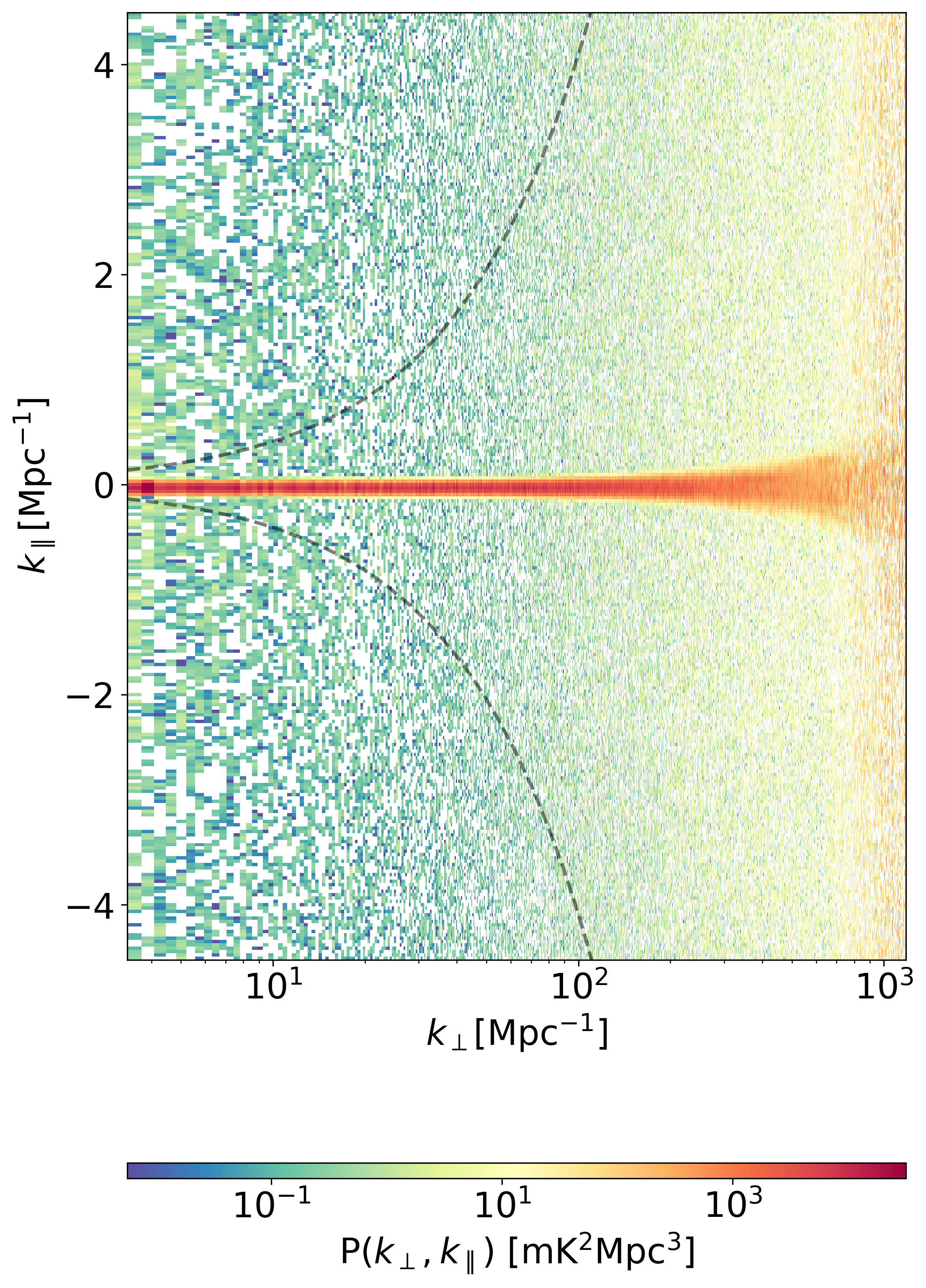}
    \includegraphics[width=0.975\columnwidth]{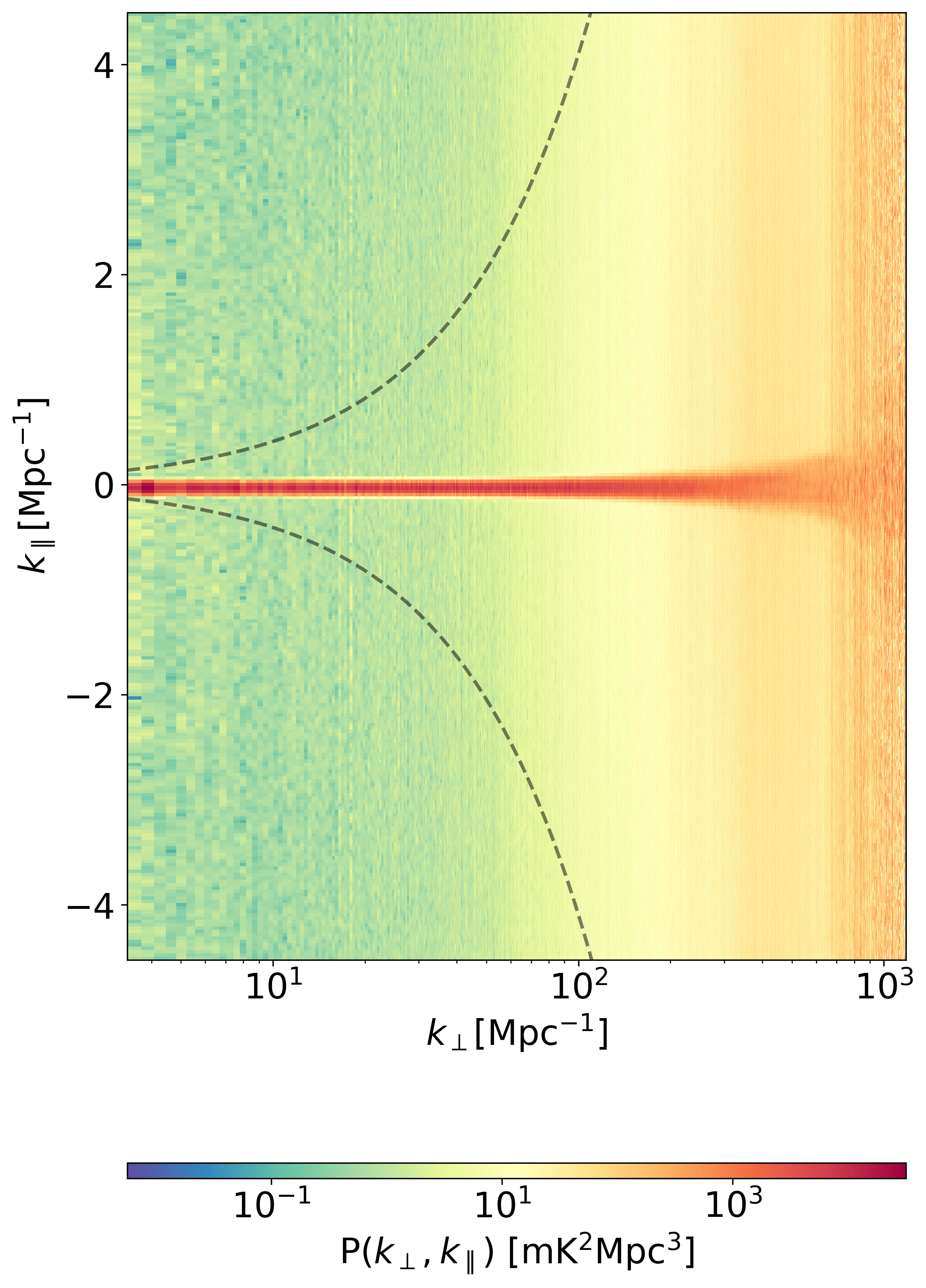}
    \caption{The Stokes I cylindrical power spectrum for the combined measurement set data after the $5\sigma$ cut has been applied for both cross- (left) and auto-correlation (right) cases. Also shown is the horizon limit, $k_{\parallel}^{\text{horizon}}(k_{\perp})$ as a dashed grey line for $\pm k_{\parallel}$. The contamination present in the same region above the horizon limit in the power spectrum shown in Figure \ref{fig:stokesI_2d_pk_before_flagging} was removed after applying this cut in either case (cross or auto).}
    \label{fig:2d_pk_5sigma_flag_auto_cross}
\end{figure*}

\begin{figure}
    \centering
    \includegraphics[width=\columnwidth]{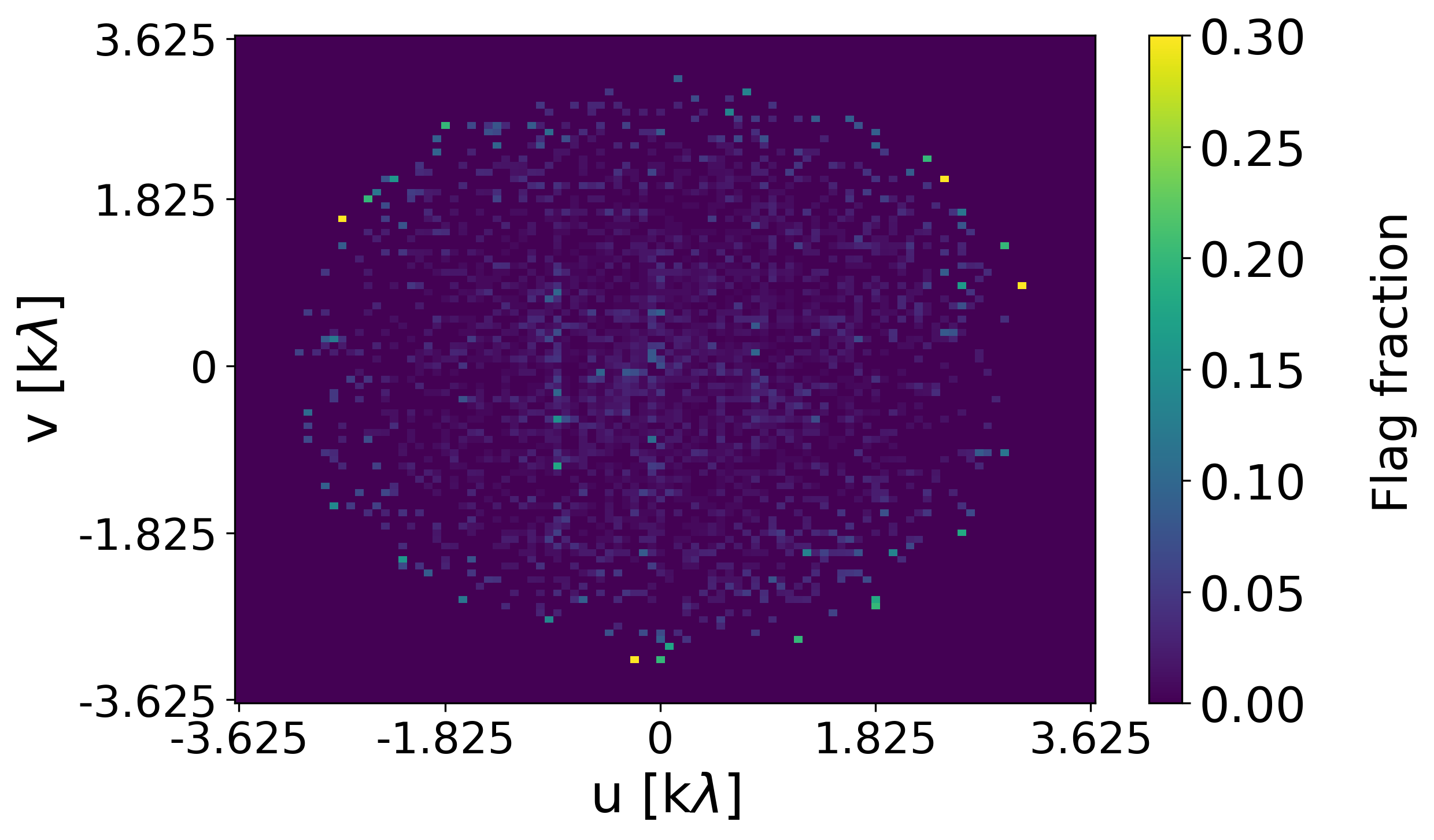}
    \caption{The $uv$-plane flagging fraction in the central portion of the $uv$ grid for the measurement set data used. The maximum fraction has been set to 30\% on the colour bar to clearly show lower level flagging applied by the $5\sigma$ cut, especially in the central region of the $uv$ distribution corresponding with small $k_{\perp}$.}
    \label{fig:5-sigma_flag_frac}
\end{figure}

We now turn to the power spectrum results using the pipeline described above. Figure \ref{fig:2d_pk_5sigma_flag_auto_cross} shows the cylindrical power spectra for auto-correlation and cross-correlation of the visibilities. Both cases are shown after 5$\sigma$ flags calculated on the data using Equation \ref{eq:sigma_cut_condition} have been applied to the 3D power spectrum cube. The flagging percentage in the region of $uv$ space corresponding to the region above the horizon line in $P(k_{\perp}, k_{\parallel})$ is shown in Figure \ref{fig:5-sigma_flag_frac}, demonstrating that the contamination that is flagged out by the criterion given in Equation \ref{eq:sigma_cut_condition} flags $\boldsymbol{k}$ modes in the cube that are localised at $\boldsymbol{k}_{\perp} \lesssim 20$ Mpc$^{-1}$, which corresponds to a radius in $uv$-space of $\sim$ 2700 $\lambda$. The percentage of $k$ modes flagged by this criterion in each $k$ bin is shown in Table \ref{tab:flag_frac_percentages}. For the visibility data, the flagging removes $\sim$ 5\% within the first $k$ bin and $\sim$ 2\% in all subsequent bins. This indicates that the flagging is not extensive and should not result in significant signal loss.

\begin{table}
    \centering
    Visibility power spectrum: odd $\times$ even \\
    \begin{tabular}{| c | c | c | c | c |}
     \hline
      $k$ bin [Mpc$^{-1}$] & Total & Sampled & Percentage Sampled \\
     \hline
     2.46   -- 4.34   & 1034 & 982 & 94.97\% \\
     4.34   -- 7.77   & 5062 & 4944 & 97.67\% \\
     7.77   -- 13.8   & 17900 & 17606 & 98.36\% \\
     13.8   -- 24.6   & 48548 & 47928 & 98.72\% \\
     24.6   -- 43.7   & 132944 & 131365 & 98.81\% \\
     \hline
    \end{tabular}    
    \caption{Table summarizing the $k$ bins, total $k$ modes in each bin, and the percentage of these that are sampled after applying the $5\sigma$ cut for the visibility data cross-power spectrum (odd $\times$ even).}
    \label{tab:flag_frac_percentages}
\end{table}

In addition to considering the power spectrum estimated on the cross-correlation of delay-transformed visibility cubes (even $\times$ odd), we also consider the auto-power of the combined visibility data placed in a single $uv\tau$ cube (odd $+$ even). To include this case in the comparison, the auto-power spectrum cube is also flagged using the 5$\sigma$ criteria to remove the contamination. After this, a thermal noise power spectrum model is subtracted from the 1D auto-power spectrum after applying the power spectrum estimator given in Equation \ref{eq:pk_1d_estimator}. The model that is subtracted from the auto-power spectrum is the mean of the 10,000 thermal noise realisations described in Section \ref{subsection:thermal_noise_sims}. Assuming that the estimated value of $A_e / T_{\text{sys}}$ accurately represents the noise in our data, this average will be a good approximation and can, therefore, be subtracted from the Stokes I auto-correlation power spectrum.
\par

\begin{figure}
    \includegraphics[width=0.975\columnwidth]{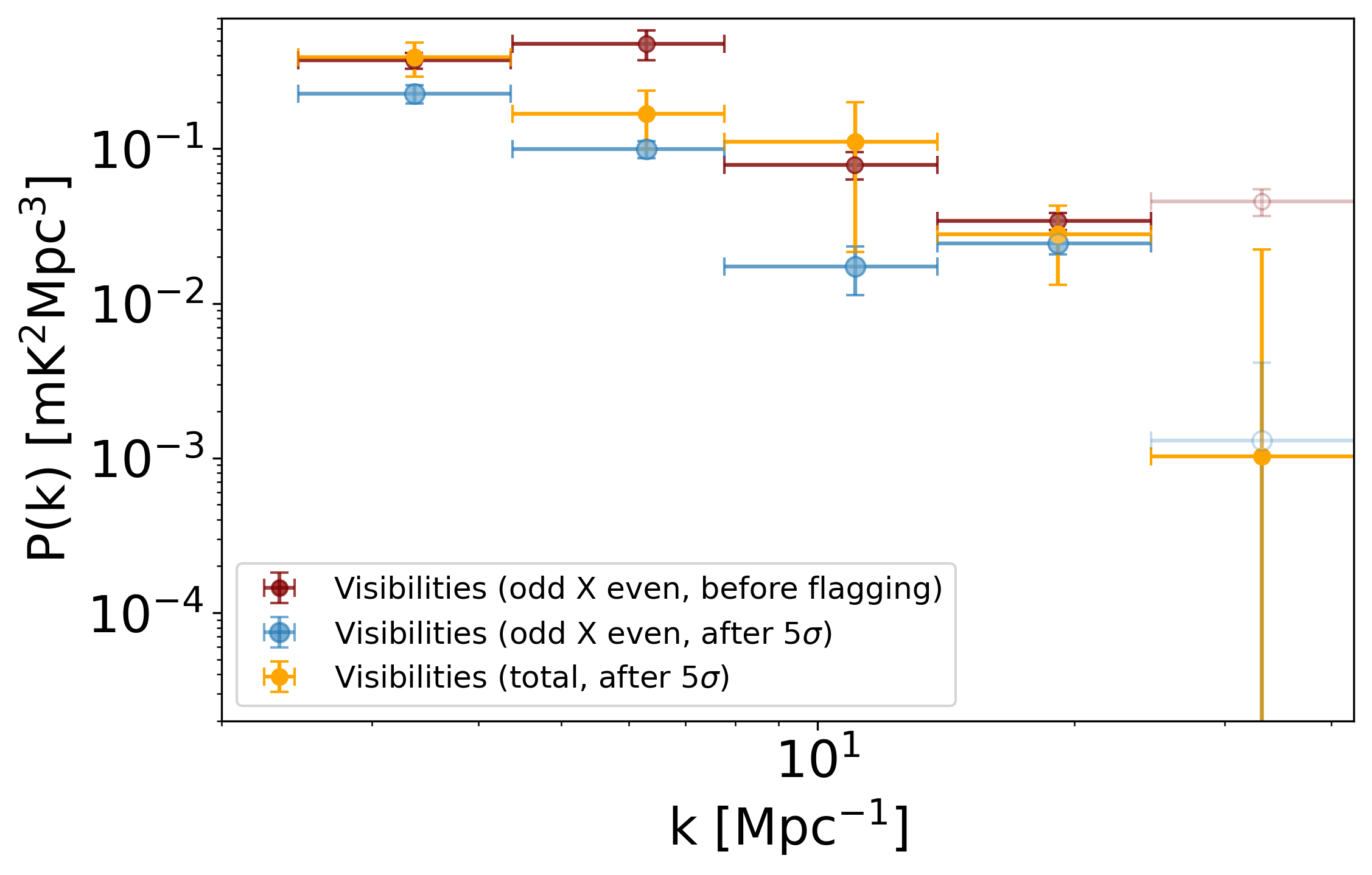}
    \caption{Comparison of the 1D power spectrum estimate from the visibility data before and after applying the $5\sigma$ flagging to the 3D power spectrum. Included is the auto-correlation power spectrum after doing a noise model subtraction and applying the same 5$\sigma$ flags.}
    \label{fig:1d_pk_auto_upper_5-sigma_w_no_flag_comp}
\end{figure}

Figure \ref{fig:1d_pk_auto_upper_5-sigma_w_no_flag_comp} shows a comparison of the 1D cross-power spectrum and auto-power spectrum estimated from the visibility data after applying a 5$\sigma$ cut. Included in the comparison is the power spectrum of the cross-correlated visibility data before the contamination flagging is performed. The shaded open circles show the negative power amplitudes. The auto-correlated visibility power spectrum should be an upper limit on the expected \ion{H}{I} power spectrum on these scales. It is interesting that, other than in the second and fourth $k$ bins, the thermal noise-subtracted auto-power is consistently higher than the power spectrum on the cross-correlation of visibilities before the 5$\sigma$ flagging is applied. This is likely due to small inaccuracies in the thermal noise power spectrum modelling at small $k$, despite its close correspondence with the Stokes V power spectrum at $k \gtrsim 20 \ \text{Mpc}^{-1}$. Additionally, it could also indicate that there is residual contamination present in the auto-correlation power spectrum, which is ultimately reduced when taking the cross-correlation. The cross power spectrum estimate after 5$\sigma$ flagging shows a significant drop in the second and third $k$ bins. The smaller drop in the first bin might indicate that there is still some contamination left after the 5$\sigma$ cut. The estimated power spectrum values after the 5$\sigma$ flagging are shown in Table \ref{tab:pk_1d_results} with an overall signal to noise of $\sim 13$. As we will see later, this measured power spectrum is consistent with the power from the detected \ion{H}{I} galaxies. Moreover, there is a clear non-zero signal when cross-correlating with the \ion{H}{I} galaxy data, as expected if the majority of the signal in the cleaned visibility dataset is from \ion{H}{I}. However, the power spectrum result presented here might still be biased, both due to residual contamination and possible signal loss. This is further discussed in the validation sections.
\par

\begin{table*}
    \centering
    Visibility power spectrum: odd $\times$ even (after $5\sigma$ cut) \\
    \begin{tabular}{| c | c | c | c |}
    \hline
    $k \ \left[ \text{Mpc}^{-1} \right]$ & $P(k) \ \left[ \text{mK}^2 \ \text{Mpc}^3 \right]$ & $\sigma_P (k) \ \left[ \text{mK}^2 \ \text{Mpc}^3 \right]$ & $P(k) / \sigma_P (k)$ \\
    \hline
    3.366 & 0.227 & 0.031 & 7.323 \\
    6.297 & 0.100 & 0.012 & 8.333 \\
    11.055 & 0.017 & 0.006 & 2.833 \\
    19.131 & 0.025 & 0.004 & 6.250 \\
    33.162 & -0.001 & 0.003 & -0.333 \\
    \hline
    \end{tabular}
    \caption{Estimated 1D power spectrum values. Shown are the effective $k$ at which the power spectrum amplitudes, $P(k)$, are quoted, as well as the errors, $\sigma_{P} (k)$, and the ratio of the amplitude to error, $P / \sigma_P$.}
    \label{tab:pk_1d_results}
\end{table*}


\section{Validations: \texorpdfstring{\ion{H}{I} Power spectrum from a box of detected \ion{H}{I} galaxies}{Validations: HI Power spectrum from a box of detected HI galaxies}}
\label{section:hi_gal_pspec}

In this section, we develop a pipeline to measure the \ion{H}{I} galaxy power spectrum, which will be used to compare to our visibility-based approach. The procedures in this section describe a method of going directly from the information provided in the galaxy catalogue described in Section \ref{subsection:mightee_hi_catalogue_data} such as \ion{H}{I} mass, M$_{\text{\ion{H}{I}}}$, RA, Dec., redshift ($z$) and $W_{50}$ to an \ion{H}{I} intensity cube and then the \ion{H}{I} power spectrum.
\par

\subsection{Temperature mesh assignment}
\label{subsection:pk_higal_pipeline}

The procedures applied to the catalogue to produce the temperature cube/mesh and the power spectrum estimation steps can be summarized as follows:

\begin{itemize}
    
    \item Galaxies within the redshift range (0.02 < $z$ < 0.066) are selected along with their R.A., Dec., and $M_{\text{\ion{H}{I}}}$. \\
    
    \item The galaxies are transformed from sky coordinates (R.A., Dec., $z$) to Cartesian coordinates $(L_x,L_y,L_z)$ in Mpc/h units. \\
     
    \item The Cartesian coordinates and the galaxy masses themselves are added to a mesh grid using the Nearest Grid Point (NGP) mass-assignment scheme. This scheme can be expressed as \citep{Jing2005, Cui2008, Colombi2009, 2024MNRAS.528.5586C}:
    
    \begin{equation}
        W_{\text{NGP}}(\textbf{x}) = 
        \begin{cases}
        1, \ &\text{for} \ |\textbf{x}| < \frac{1}{2} \\
        0, \ &\text{otherwise,}
        \end{cases}
    \end{equation}
    
    where \textbf{x} is the distance of the particle (in this case, a given galaxy from the catalogue) from the centre of the grid point, normalised by the cell (pixel) size. The pixel size selected for this is 0.05 Mpc/h to ensure that a pixel can realistically contain all the \ion{H}{I} for each galaxy. This choice is made based on the expected \ion{H}{I} sizes of the MIGHTEE-detected galaxies according to the \ion{H}{I} size-mass relation \citep{Rajohnson2022}. In our case, even the highest-mass galaxies have a $D_{\ion{H}{I}}$ that is smaller than the 0.05 Mpc/h pixel size, and so requires no further interpolation of \ion{H}{I} mass into neighbouring cells in the mesh grid, hence our use of the NGP mass-assignment scheme. This is illustrated in Figure \ref{fig:dhi_mhi_pixsize} via the expected \ion{H}{I} size calculated from the \ion{H}{I} masses of each galaxy in the catalogue based on the fit provided in \cite{Rajohnson2022}.\\

    \begin{figure}
        \centering
        \includegraphics[width=\columnwidth]{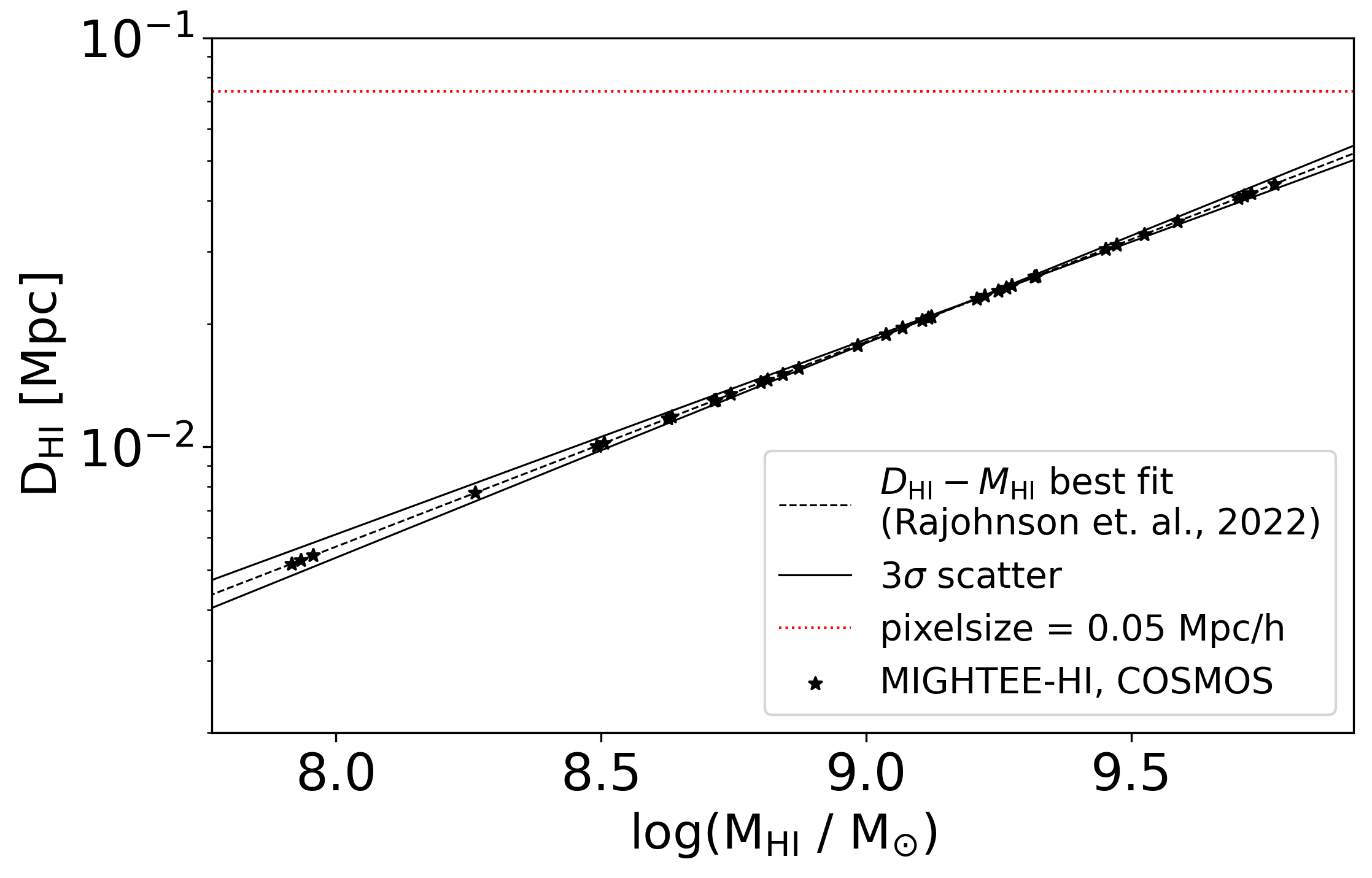}
        \caption{The best-fit \ion{H}{I} size-mass relation ($D_{\ion{H}{I}}$ - $M_{\ion{H}{I}}$) from \protect\cite{Rajohnson2022} with $3\sigma$ scatter, plotted with the predicted \ion{H}{I} sizes of the selected galaxies on the central COSMOS pointing according to their \ion{H}{I} masses. The red dotted line shows the pixel size selected for the mesh in Mpc.}
        \label{fig:dhi_mhi_pixsize}
    \end{figure}
    
    \item For completeness, higher-order mass assignment schemes are also tested without applying any aliasing corrections to see the deviation from the NGP assignment scheme case. In particular, the Cloud-in-Cell (CIC) and Triangular-Shaped Cloud (TSC) mass assignment schemes are tested.
    These are given in real space as

    \begin{equation}
        W_{\text{CIC}} (\textbf{x}) =
        \begin{cases}
            1 - |\textbf{x}|, & \text{for} \ \textbf{x} < 1 \\
            0, & \text{otherwise}
        \end{cases}
    \end{equation}

    \begin{equation}
        W_{\text{TSC}} (\textbf{x}) = 
        \begin{cases}
            \frac{3}{4} - |\textbf{x}|^2, & \text{for} \ |\textbf{x}| < \frac{1}{2} \\
            \frac{1}{2}\left( \frac{3}{2} - |\textbf{x}| \right)^2, & \text{for} \ \frac{1}{2} \leq |\textbf{x}| < \frac{3}{2}  \\
            0, & \text{otherwise}
        \end{cases}
    \end{equation}

    \begin{figure}
    \centering
    \includegraphics[width=\columnwidth, trim={0 3mm 0 0}, clip]{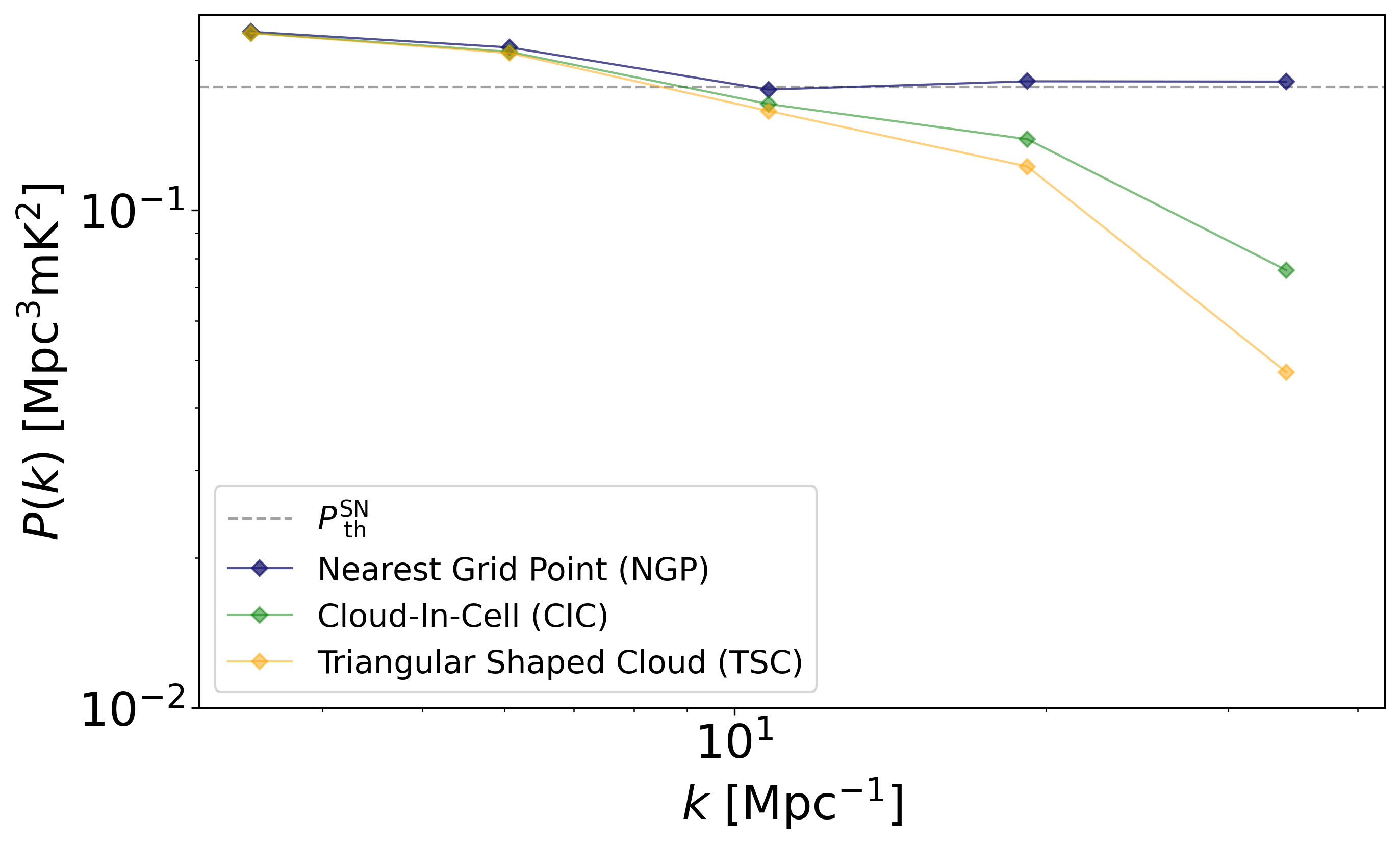}
    \caption{Comparison of the power spectrum using the standard mass-assignment schemes in \texttt{nbodykit}, using the standard power spectrum estimator on the \ion{H}{I} galaxy mesh. This does not include the spreading of mass along $L_z$, any $k$ mode-exclusion induced by the foreground avoidance, and the $5\sigma$ flagging or thermal noise weighting.}
    \label{fig:pk_mass_assign_comp}
\end{figure}

    The power spectrum results on the galaxy mesh grids comparing the various mass assignment schemes are shown in Figure \ref{fig:pk_mass_assign_comp}. This comparison is done without including the spreading of $M_{\ion{H}{I}}$ along $L_z$. This is to ensure that the comparison is done without including any window functions along $z$ which could induce suppression of the power spectra in addition to that imposed by the different mass assignment schemes.
    Also shown here is the \ion{H}{I} shot noise power spectrum model given by \citep{Chen2022}

    \begin{equation} \label{eq:p_hi_shotnoise_model}
        P_{\text{SN}} = \Bar{T}_{\ion{H}{I}}^2 V \sum_i \left( M_{\ion{H}{I}}^i \right)^2 / \ \left( \sum_i M_{\ion{H}{I}}^i \right)^2,
    \end{equation}
    
    \noindent
    where $\Bar{T}_{\ion{H}{I}}$ is the average \ion{H}{I} temperature, $V$ is the volume of the mesh, and the sum is over the \ion{H}{I} mass in the mesh. Since this model assumes point sources and does not include the effect of $W_{50}$ along the line-of-sight or the smoothing along $L_x$ and $L_y$ induced by mass assignment, we expect it to be consistent with the power spectrum on the \ion{H}{I} galaxies using the NGP scheme. We indeed see that the NGP power spectrum and shot noise model are consistent in Figure \ref{fig:pk_mass_assign_comp} for $k \gtrsim 10$ Mpc$^{-1}$. A derivation of this shot noise model can be found in Appendix \ref{appendix:A}.
    When including all the effects, such as the line width, and given the large cell size, we found that the NGP scheme is the optimal option for our purposes.
    \\
    
    \item Once the galaxy \ion{H}{I} masses have been placed in voxels, the mass is distributed along $L_z$ using the characteristic width given by the $W_{50}$ of each galaxy in the catalogue. Here, a choice for the shape is made as only a numerical value for the $W_{50}$ is provided in the catalogue. \cite{Li2024} has shown that the choice of the shape of the profile has minimal effects on the power spectrum estimated from the galaxies, hence we make the choice of using a Boxcar function which evenly distributes the \ion{H}{I} mass of each galaxy along $L_z$ over the pixels which correspond to the characteristic length scale given by the $W_{50}$. The relation between this length scale and the $W_{50}$ can be expressed as:

    \begin{equation} \label{eq:w50_to_d_gal}
        r_{\text{galaxy}} = \frac{W_{50}}{H(z)}(1+z),
    \end{equation}

    \noindent
    where $H$ is the Hubble factor at the redshift of the galaxy, $z$, in units of km/s/Mpc.
    \\

    \item The mesh now has the \ion{H}{I} mass of each galaxy selected from the catalogue in units of $M_{\odot}$. This is converted to \ion{H}{I} temperature using \citep{Chen2022}
    
    \begin{equation}
        T^{\text{pix}}_{\ion{H}{I}} = \frac{C_{\ion{H}{I}} (z) \ M^{\text{pix}}_{\ion{H}{I}}}{V_{\text{pix}}},
    \end{equation}
    where the conversion is done across the entire mesh at the central frequency of the mesh, $z_c$. \\

    \item So far, the steps described above make use of tools in the \texttt{nbodykit}\footnote{\url{https://nbodykit.readthedocs.io/en/latest/}} package \citep{Hand2018}, which is conveniently suited to dealing with the computations required for converting survey data into cosmological volumes ready for power spectrum estimation and other cosmological calculations. \\
    
    \item The \ion{H}{I} temperature mesh cube is then used to obtain an estimate of the power spectrum using equations \ref{eq:pk_1d_estimator} and \ref{eq:pk_estimator_errbar}, with the same foreground avoidance applied using Equation \ref{eq:horizon_lim}, and taking comparative cases in which the same thermal noise weights, $w_j$, are applied. For the standard estimator on the \ion{H}{I} temperature mesh, these weights are excluded, i.e., $w_j = 1, \ \forall \ j$. The thermal noise weighting and foreground avoidance help ensure that the estimated galaxy \ion{H}{I} power spectrum is equivalent to the estimate obtained from the visibility data. \\

    \item Finally, to be able to apply the same windows and weights to this 3D power spectrum, we interpolate and extrapolate the $(k_{\perp},k_{\parallel})$ galaxy cube to match exactly the one derived from the visibility data. The main issue is the perpendicular resolution of the galaxy box. In order to fix this, a simple extrapolation was made where higher $k_{\perp}$ values were matched to the galaxy box maximum $k_{\perp}$. This is enough because 1) the foreground cuts remove most of the high $k_{\perp}$ values, 2) we only show comparisons down to the galaxy box resolution, and 3) the shot noise will dominate at higher $k_{\perp}$.
    
\end{itemize}

\subsection{Impact of flags and weights} 
\label{subsection:pk_higal_additional_tools}

\begin{figure}
    \centering
    \includegraphics[width=\columnwidth, trim={0 3mm 0 0}, clip]{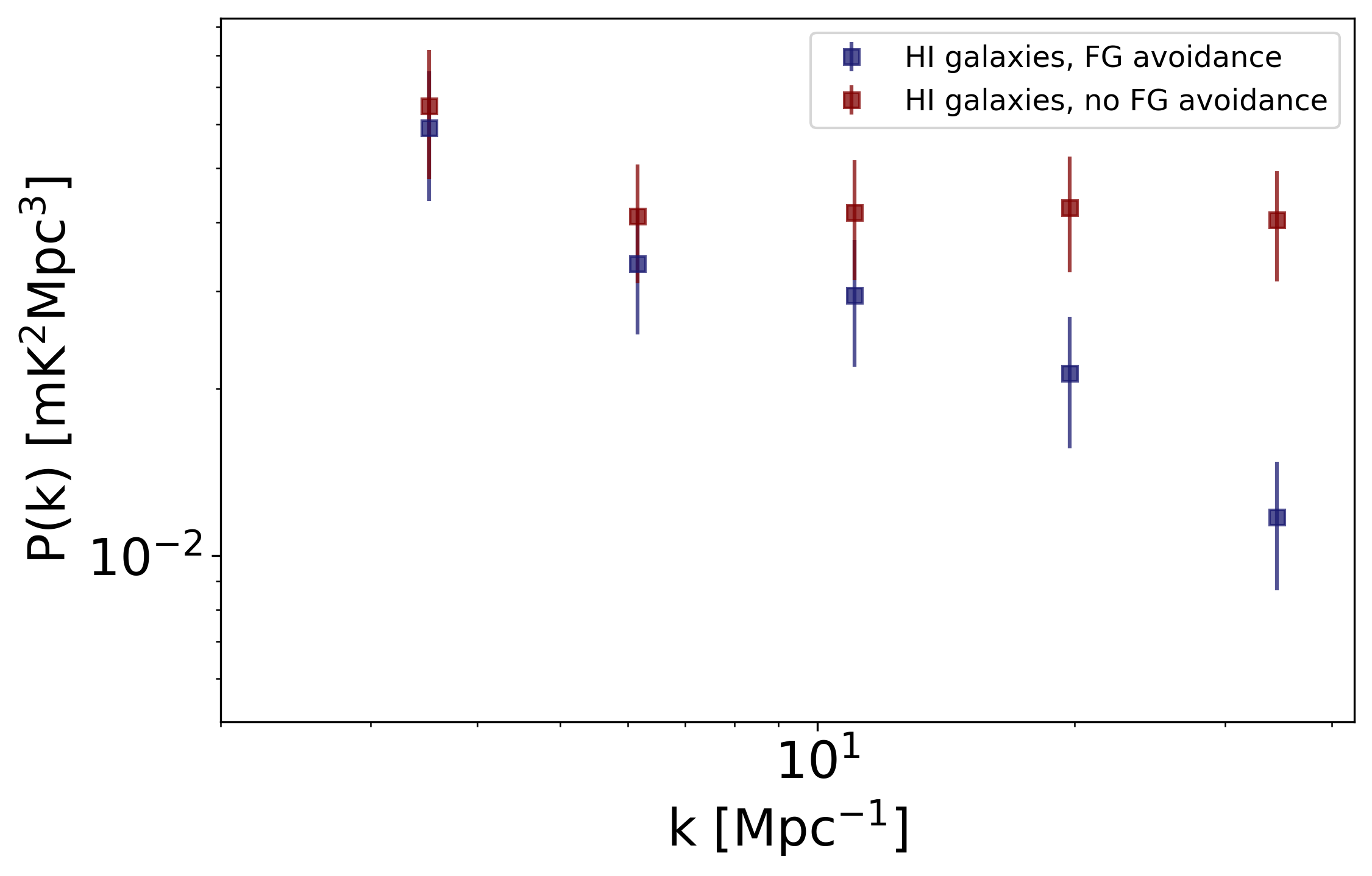}
    \caption{Comparing the \ion{H}{I} power spectrum from galaxies with and without foreground avoidance. This is done before applying the 5$\sigma$ flags or the thermal noise weights.}
    \label{fig:pk_1d_comparisons_reassignment}
\end{figure}

We now discuss the impact of foreground avoidance, noise weights, and power spectrum flags on the \ion{H}{I} galaxy power spectra. Since the \ion{H}{I} power spectrum is highly anisotropic, especially on the scales considered in this work, any change in the selected $({\bf k}_{\perp}, k_{\parallel})$ modes can impact the final 1D power spectrum. Figure \ref{fig:pk_1d_comparisons_reassignment} shows the effect of foreground avoidance on the mass spread ($W_{50}$) case. Note that for the case of no spread along $L_z$ in which each galaxy's mass is confined to a single voxel, we do not expect foreground avoidance to have a significant effect on the overall amplitude. On the other hand, for the $W_{50}$ case, the loss of low $k_{\parallel}$ modes leads to a loss of power which becomes more significant at large $k$.

\begin{figure}
    \centering
    \includegraphics[width=\columnwidth]{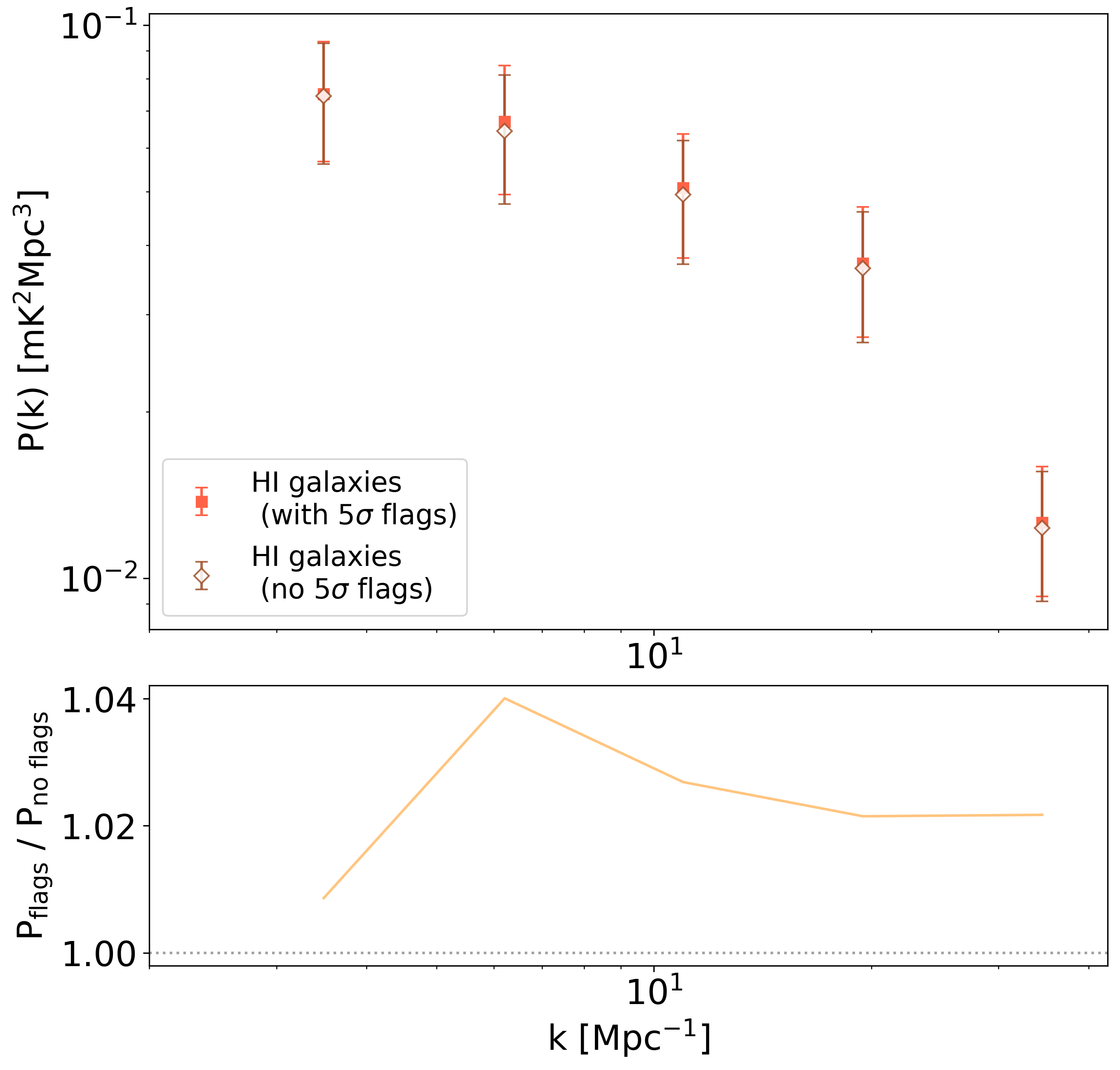}
    \caption{The \ion{H}{I} power spectrum estimated on the \ion{H}{I} galaxies comparing the cases of including and excluding the 5$\sigma$ flags (top), as well as the ratio of the two cases (bottom) over the full $k$ range.}
    \label{fig:p1d_higal_flag_comp}
\end{figure}

\begin{figure}
    \centering
    \includegraphics[width=\columnwidth]{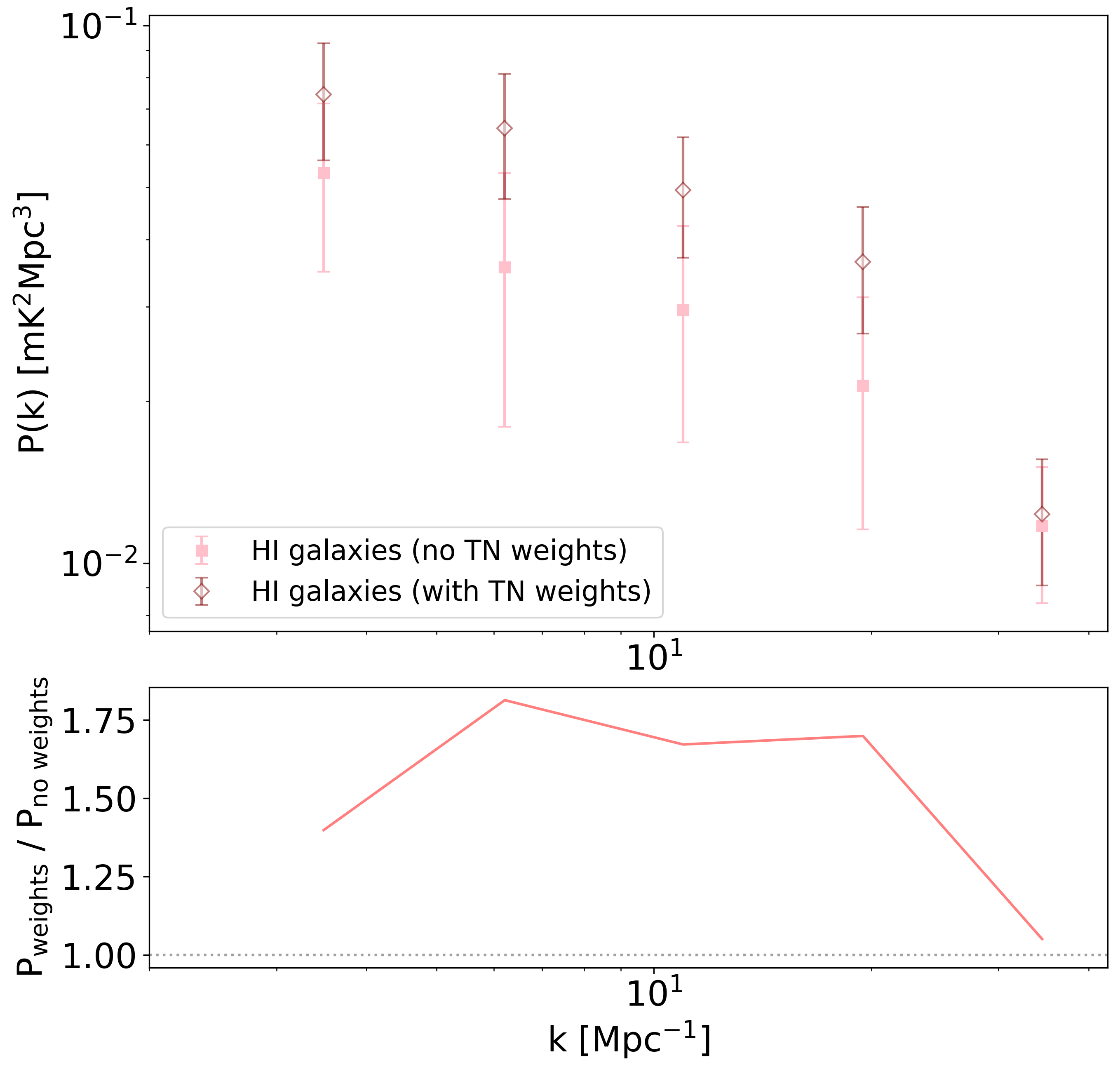}
    \caption{The \ion{H}{I} power spectrum on the \ion{H}{I} galaxies, comparing the cases of including and excluding the thermal noise weights when using the estimator.}
    \label{fig:p1d_higal_weight_comp}
\end{figure}

Figure \ref{fig:p1d_higal_flag_comp} shows the effect of applying the 5$\sigma$ flags. In this case, we flag the exact same 3D modes as the ones flagged on the visibility data. The difference between the flagged and unflagged cases is less than 5\%. This is somewhat at odds with what we have seen with the full visibility data, where the power spectrum shows a larger drop as we increase the $\sigma$ flagging. The comparison is non-trivial because the 2D power spectrum is highly anisotropic. This is why we can even have the signal amplitude increasing slightly, as seen in this figure. Our understanding is that the "real" signal has power more "spread out" and closer to the horizon cut in comparison to our \ion{H}{I} galaxy only signal. That is also where most of the $\sigma$ flagging is happening (Figure \ref{fig:sigma_flag_comp_pk_1d}). On the other hand, the thermal noise weights have a much more significant impact on the estimated power spectrum. Specifically, these weights raise the power spectrum amplitude at $k \lesssim 20$ Mpc$^{-1}$ as shown in Figure \ref{fig:p1d_higal_weight_comp}. This is because we have small $k_{\perp}$ values contributing to these modes, which have low noise due to the large number of short baselines and therefore get "up-weighted". Above $k \sim 20$ Mpc$^{-1}$, the noise will start dominating everywhere (Figure \ref{fig:stokes_v_tn_auto_pk1d_comp}). The impact of these weights on the final averaged power spectrum arises from the strong anisotropies in the 3D power spectrum. Therefore, the same weights, flags, and windows must be applied to any model power spectrum for a proper comparison.

\subsection{Estimating error bars based on \texorpdfstring{$M_{\ion{H}{I}}$}{MHI} uncertainties}
\label{subsection:pk_higal_uncertainties}

The final step is to estimate an error on the power spectrum from these galaxies. So far, the error bars shown on the power spectra have been estimated using a Jackknife approach. We mask pixels containing individual galaxies in the mesh and run the entire power spectrum code end-to-end, from assigning \ion{H}{I} mass to the mesh to the power spectrum estimation. Due to the small sample of galaxies, this method and the sampling variance in Equation \ref{eq:pk_estimator_errbar} do not capture the uncertainty in the power spectrum we are estimating, especially when we want to compare to the power spectrum estimated on the visibility data. The $M_{\ion{H}{I}}$ of these galaxies have an uncertainty associated with them according to the mass range in which they lie. These uncertainties are derived from the noise properties of the \ion{H}{I} cubes from which these galaxy masses are measured \citep{Maddox2021}, and a detailed description of these mass measurements is given in \cite{Ponomareva2023}. The exact levels are given at the 5\% level for $M_{\ion{H}{I}} > 10^{9} M_{\odot}$, 10\% for $10^{8} M_{\odot} < M_{\ion{H}{I}} < 10^{9} M_{\odot}$, and up to $\sim$ 20\% for $M_{\ion{H}{I}} < 10^{8} M_{\odot}$ \citep{2021MNRAS.508.1195P, Rajohnson2022, Ponomareva2023}. To capture this uncertainty in the power spectrum error bars, this percentage-based criterion is used to create a distribution for each $M_{\ion{H}{I}}$ selected from the catalogue with a standard deviation, $\sigma^{i}_{M_{\ion{H}{I}}}$ given as the percentage fraction of the individual \ion{H}{I} mass, $M_{\ion{H}{I}}^{i}$, with a mean, $\mu^i_{M_{\ion{H}{I}}} = M_{\ion{H}{I}}^{i}$. We produce 10,000 random samples for the distribution of each galaxy, and a catalogue is created by randomly selecting a sample \ion{H}{I} mass for each galaxy and running the pipeline end-to-end to estimate a power spectrum for each catalogue of samples. The mean and standard deviation of these power spectrum realisations are the final estimates we compare to the power spectrum estimated on the visibility data using Equation \ref{eq:pk_1d_estimator} with errors given by Equation \ref{eq:pk_estimator_errbar}. These errors are the only source of uncertainty used when showcasing and comparing the power spectrum estimated on the \ion{H}{I} galaxies in the rest of this study.

\subsection{Power Spectrum Comparison}
\label{subsection:pspec_comp}

\begin{figure}
    \includegraphics[width=\columnwidth]{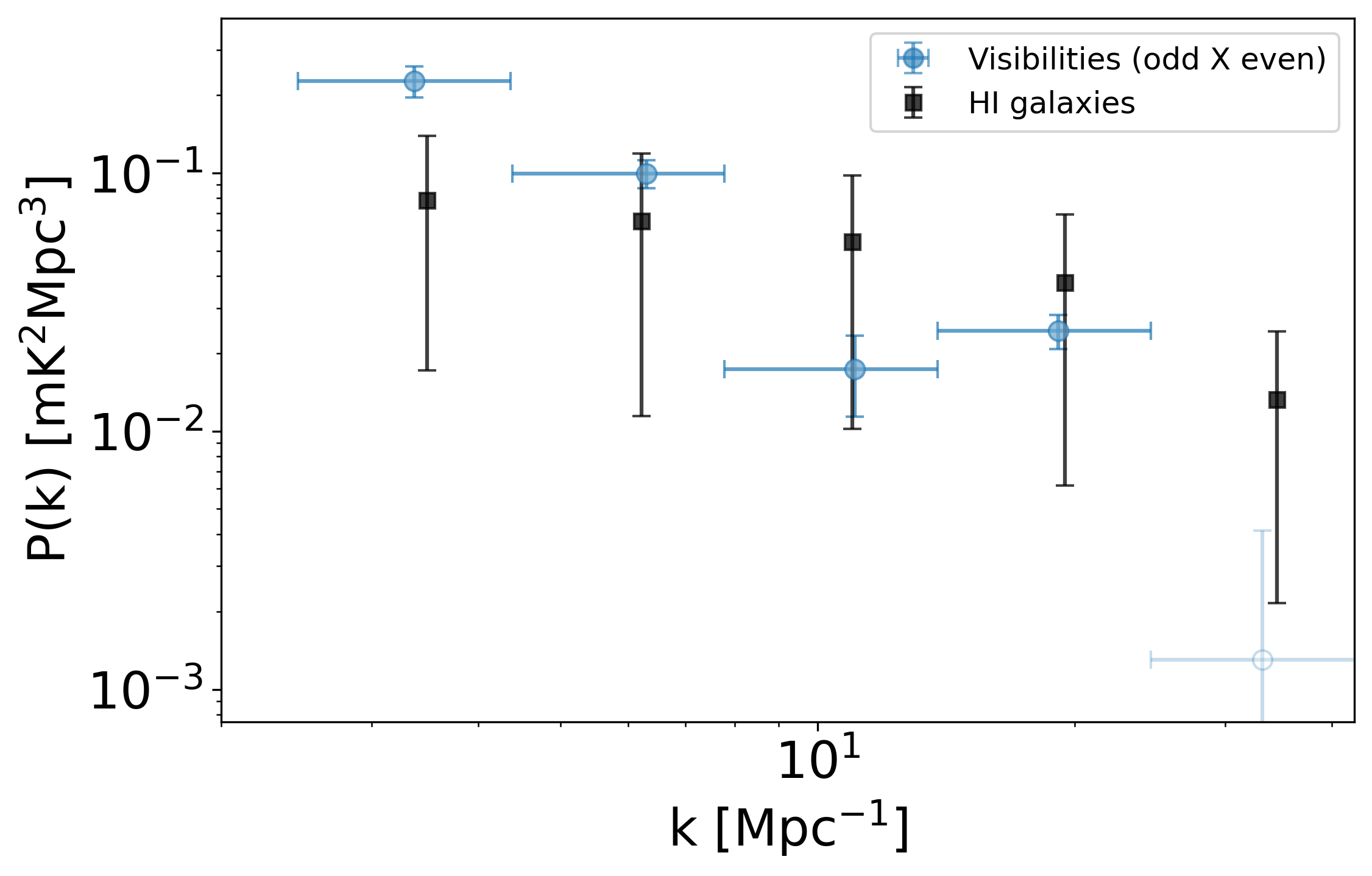}
    \caption{The comparison of the 1D power spectrum estimated on the "odd" $\times$ "even" power spectrum obtained from the MIGHTEE visibility data (blue circles) to the power spectrum estimated on the detected \ion{H}{I} galaxies (black squares). The 5$\sigma$ flagging criterion was applied to both 3D power spectrum cubes before performing the spherical average to obtain these 1D power spectrum estimates. Both also make use of the inverse thermal noise variance weighting. The errors estimated on the \ion{H}{I} galaxy case were generated using random sampling based on the uncertainties in the detected \ion{H}{I} mass, while the errors estimated in the visibility power spectrum case use Equation \ref{eq:pk_estimator_errbar}. Negative power amplitudes are shown as shaded open circles and are also included with the corresponding effective $k$ in Table \ref{tab:pk_1d_results}.}
    \label{fig:1d_pk_hi_gal_vis_5s_comparison}
\end{figure}

The key result in this section is the comparison of the power spectrum estimated on the cross-correlated combined COSMOS visibility data and the power spectrum estimated on the detected \ion{H}{I} galaxies using the same data set. This is shown in Figure \ref{fig:1d_pk_hi_gal_vis_5s_comparison}. Since these power spectra use different data products from the same set of MIGHTEE observations on a single pointing, the expectation is that the comparison should be consistent within the error bars, where the error in the power spectrum measured on the visibilities comes from sampling variance, weighted by thermal noise simulations (inverse noise variance weighting), while for the power spectrum from the \ion{H}{I} galaxies is sourced by the percentage uncertainties in the \ion{H}{I} masses in the catalogue. Any residual contamination is also expected to contribute to the error bars of the cross-correlation visibility power spectrum estimate and would be a source of bias in this comparison (as discussed in Section \ref{subsec:sigma_flag_vis_tests}). Furthermore, we do not expect cosmic variance or shot noise to account for the differences in the two power spectrum measurements, as they are both measured on the same field. However, uncertainties in the \ion{H}{I} mass estimation can lead to noticeable differences. For the Early Science MIGHTEE galaxy catalogue, the $W_{50}$ values are biased to smaller values and might not fully represent the velocity fields of all the galaxies on the COSMOS field accurately. Larger $W_{50}$ values for these galaxies would further suppress the estimated power spectrum, which could explain the differences between the result from the \ion{H}{I} galaxies and visibilities observed beyond $k = 10 \ \text{Mpc}^{-1}$. It is also likely the case that the error bars on the \ion{H}{I} galaxies are an overestimation since they are based on fixed percentage uncertainties in mass ranges, rather than uncertainties in the individual galaxy \ion{H}{I} masses themselves. Nonetheless, the comparison of the power spectra estimates are consistent, with the difference at about the 20\% level, which is a possible indication that the error bars on the \ion{H}{I} galaxy power spectrum might indeed be overestimated.
\par


\section{Validations: \texorpdfstring{\ion{H}{I} Power spectrum from the visibilities of detected \ion{H}{I} galaxies}{Validations: HI Power spectrum from the visibilities of detected HI galaxies}}
\label{section:pipeline_validation}

We further present tests by simulating a new visibility dataset from the detected MIGHTEE \ion{H}{I} galaxies and processing it using exactly the same pipeline as the one used on the original MIGHTEE visibility dataset (Section \ref{section:delay_pspec}). This would allow us to compare the signal levels and enable us to cross-correlate the two datasets, with the goal of testing that the visibilities and \ion{H}{I} galaxy detections do correlate in position and frequency. Both thermal noise and systematics will not correlate with the high signal-to-noise detected galaxies. Such an approach would require a visibility simulator. The details of this simulator and the validation tests are discussed in the subsections that follow.

\subsection{MeerKAT All-Sky Simulator (MASS)}
\label{subsection:mass_sims}

MASS is a simulator that estimates visibilities from a radio interferometric experiment and outputs them in the standard measurement set format for analysis. Considering the problem at hand, our aim here is to simulate visibilities given the \ion{H}{I} selected galaxy sky coordinates, redshifts, \ion{H}{I} mass, and $W_{50}$. For this work, we have 37 such galaxies that host the \ion{H}{I}. Here, we consider a simplified assumption that each galaxy subtends an angle that is much smaller than the resolution derived from the baseline range of our data. However, the velocity structure of the \ion{H}{I} inside the individual galaxies is well within the frequency resolution of our interest. We include the frequency spread of the \ion{H}{I} emission in the simulation by introducing a line profile $\phi_i(\nu = \nu_i)$ associated with each galaxy ($i$), where $\int \phi_i(\nu) d\nu = 1$. The line profile here has been defined with respect to the observed frequency in the rest frame of the telescope.
\par

Visibilities corresponding to the galaxy distribution are calculated using,
\begin{equation}\label{eq:vg}
\V(\mathbf{U_n},\nu_a)=  
\left( \frac{\del B}{\del T}\right)_{\nu_a} \frac{1}{ r^2 r'} 
 \sum_i T^{i}_{\HI}A(\mathbf{\theta}_i, \nu_i) \phi(\nu_a -\nu_i) 
e^{2\pi i \mathbf{U_n} \cdot \mathbf{\theta_{i}}} \,,
\end{equation}

\noindent
where $B = 2k_B T /\lambda^2$ is the Planck function in the Raleigh-Jeans limit, which is valid in the frequency range of our interest ($\sim$ 1330 - 1390 MHz), and $T^{i}_{\HI}$ is the brightness temperature
distribution from the $i^{\rm th}$ galaxy on the sky. The primary beam pattern for the telescope is given by $A(\mathbf{\theta}, \nu)$, $r$ is the 
comoving distance to the redshift ($z_c$) of the central frequency ($\nu_c$) channel and $r' =\mid dr/d\nu \mid_{\nu_c}$. For this work, we have chosen a simple line profile,

\begin{equation}
\label{eq:b3}
\phi(\nu-\nu_i) =
    \begin{cases}
\frac{1}{F_i\,\Delta \nu_c}, & \text{for} \mid \nu - \nu_i \mid \le \frac{F_i}{2} \times \Delta \nu_c \\
0, & \text{otherwise}.
    \end{cases}
\end{equation}

\noindent
Here, the value of $F_i$ has been estimated from the measured $W_{50}$ of each galaxy. Further details and results from using the \ion{H}{I} detections as the sky input to MASS are discussed in Section \ref{subsection:hi_pk_validation}.
\par

\begin{figure}
    \centering
    \includegraphics[width=\columnwidth]{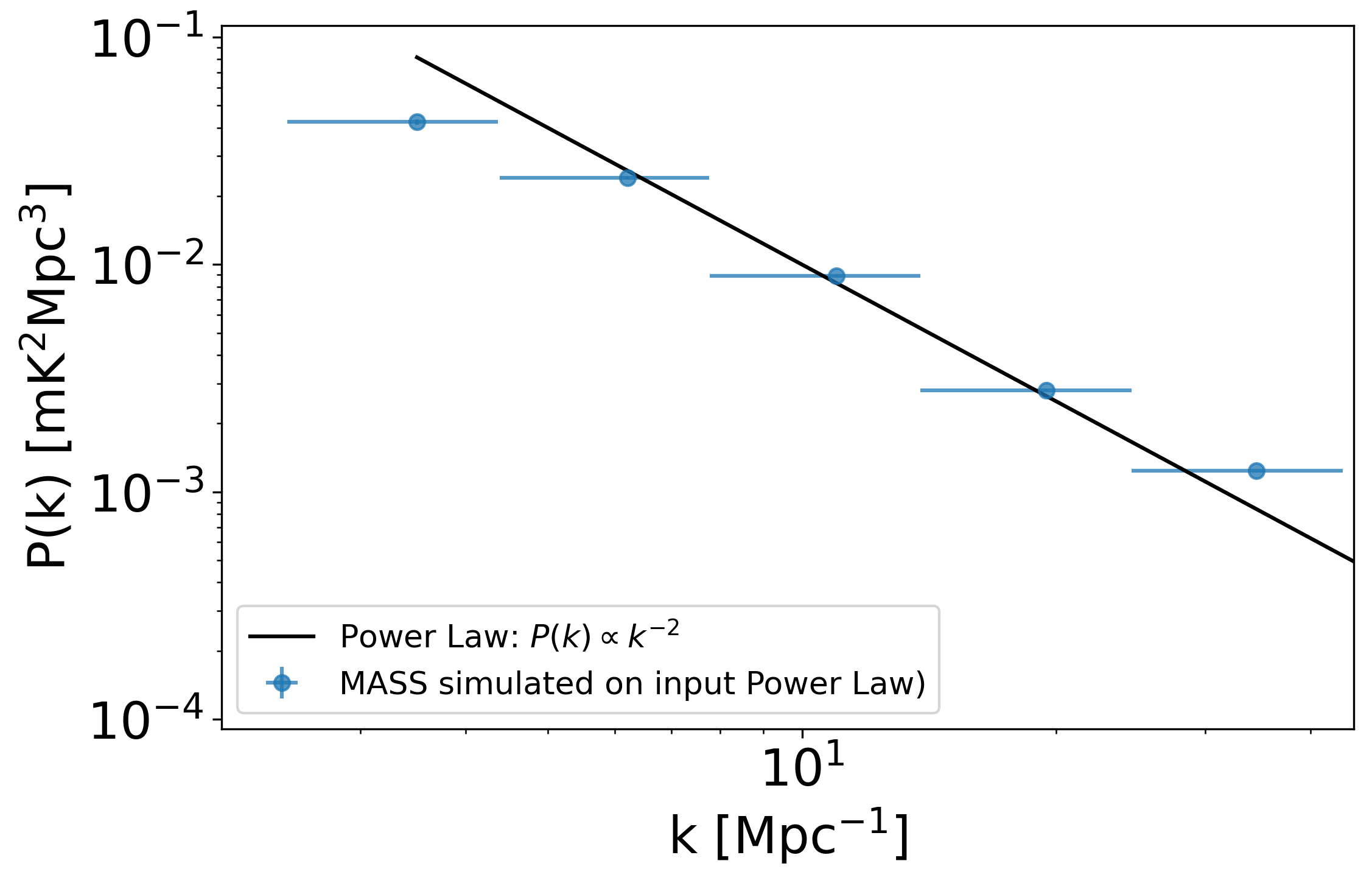}
    \caption{Comparison between an input power law \ion{H}{I} power spectrum and a single realization of a sky model generated using this power law as input to generate MASS simulated visibilities. The difference between low and high $k$ is due to the variance in this single realization, as it is randomly generated. Regardless, the two are in good agreement across the $k$ range shown.}
    \label{fig:mass_theory_input_comparison}
\end{figure}

Before doing so, we test the MASS pipeline on a realisation of the sky generated using a simple $P(k) = k^{-2}$ power law model as input. Visibilities are generated on this sky model and then put through the end-to-end pipeline and power spectrum estimator discussed in Section \ref{section:delay_pspec}. The result of this is then compared to the theoretical $k^{-2}$ power law as shown in Figure \ref{fig:mass_theory_input_comparison}. We find good agreement between the two power spectra. However, since this realisation is randomly generated, the variance will cause it to deviate from the theoretical model, which is what is observed at $k \sim 3$ Mpc$^{-1}$ and $k \gtrsim 30$ Mpc$^{-1}$. Thus, we confirm that MASS can accurately simulate the \ion{H}{I} given an input.
\par

\subsection{\texorpdfstring{Detected \ion{H}{I}}{HI} galaxies power spectrum comparison}
\label{subsection:hi_pk_validation}

To ensure that the end-to-end pipeline that generates the mesh and estimates the \ion{H}{I} power spectrum on the detected \ion{H}{I} galaxies from the catalogue is performing these processes correctly, we make use of MASS to test if we can recover the \ion{H}{I} signal present in the detected \ion{H}{I} galaxies. The input configuration for the baselines corresponds to that of the entire set of data blocks used in the analysis. In other words, we make use of the combined $uv$ distribution of all the MIGHTEE visibility data to simulate with MASS. Since there is no thermal noise added to these particular runs of the simulations, as long as the $uv$ coverage is comparable to the full $uv$ coverage of the combined data blocks, we expect to be able to recover the signal with our estimator. Once the visibilities are generated using MASS, we apply the same procedures as is done to the MIGHTEE visibility data for gridding, forming the 3D power spectrum, and estimating the 1D power spectrum with its error bars. The same thermal noise weights used for the MIGHTEE data are applied as well in order to sample the 3D power spectrum in the same way. However, since these simulations only use \ion{H}{I} signal to generate visibilities, the resulting 1D power spectra have negligible error bars. Like in the previous simulation, the main source of errors will be the catalogue \ion{H}{I} mass. To simplify the analysis, we do not include this error here since it will not impact the validation tests.
\par

\begin{figure}
    \centering
    \includegraphics[width=\columnwidth]{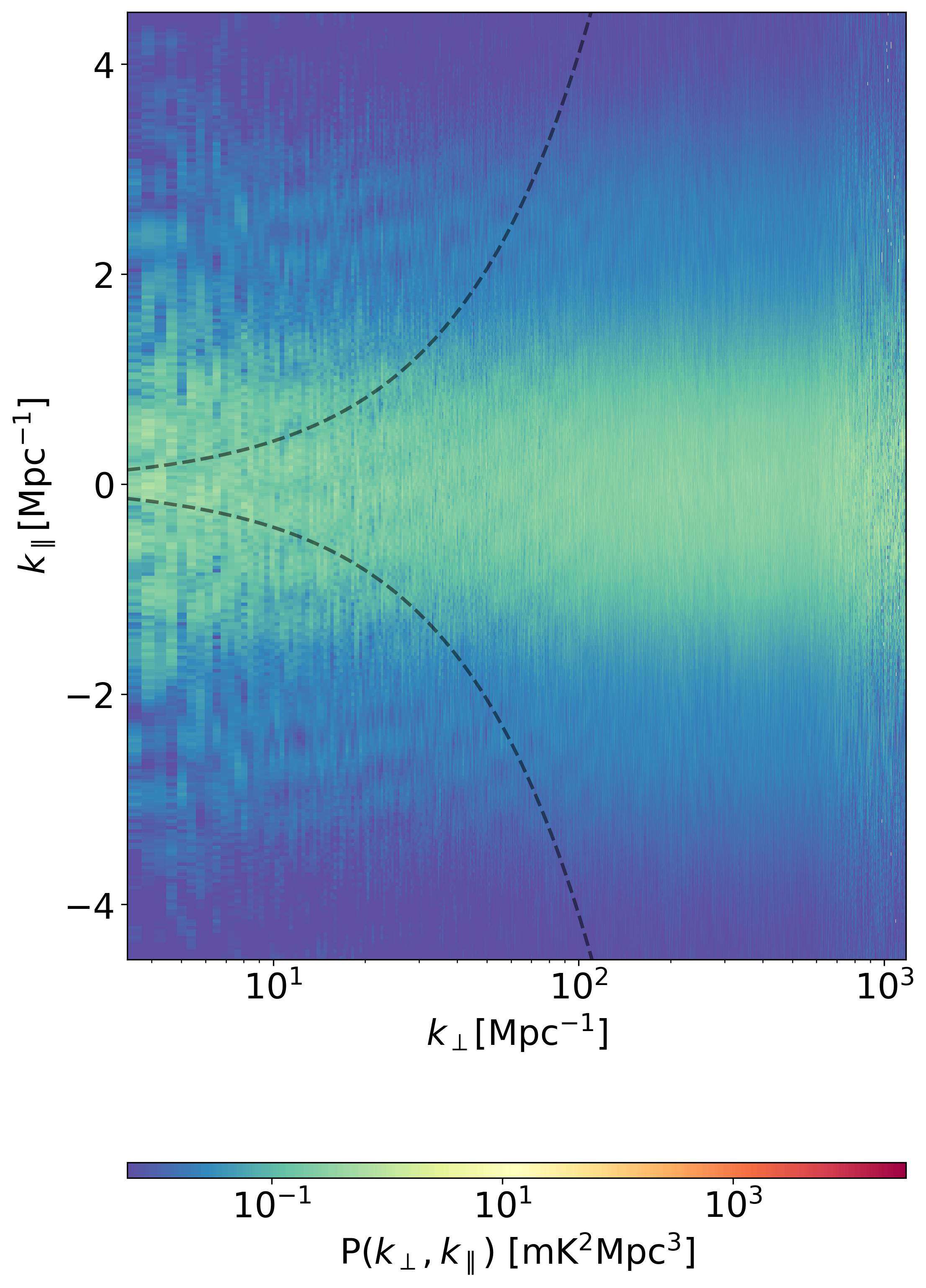}
    \caption{The cylindrical power spectrum estimated on the gridded delay-transformed visibilities generated using MASS. It uses the same information from the \ion{H}{I} detections to produce this power spectrum.}
    \label{fig:p2d_mass_boxcar}
\end{figure}

The output cylindrical power spectrum is shown in Figure \ref{fig:p2d_mass_boxcar}. We can clearly see the \ion{H}{I} signal distributed over $(k_{\perp}, k_{\parallel})$. A lot of the power is close to the foreground wedge which indicates that flagging around that region might lead to signal loss when averaging to the 1D power spectrum

\begin{figure}
    \centering
    \includegraphics[width=\columnwidth]{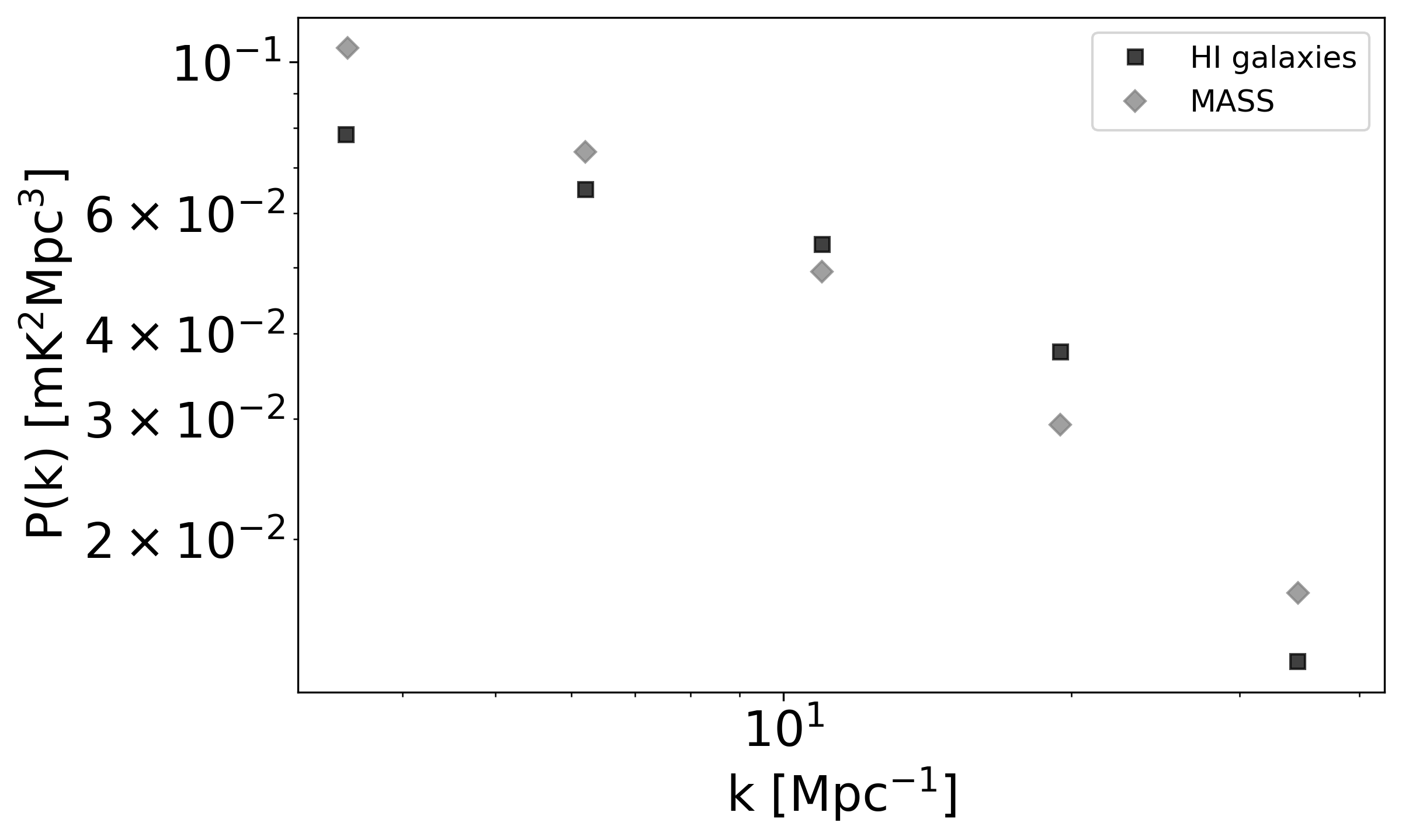}
    \caption{Comparison of the 1D power spectra generated using the pipeline described in Section \ref{section:hi_gal_pspec} and using MASS. For each, there is no thermal noise present, and therefore, we expect the two results to be consistent. The deviations between the two are likely due to how many modes are averaged in each $k$ bin to obtain the estimates, as well as differences in pre-processing steps, which introduce residual errors.}
    \label{fig:p1d_hi_gal_mass_comparison}
\end{figure}

Figure \ref{fig:p1d_hi_gal_mass_comparison} shows the spherically averaged 1D power spectrum in comparison to the one from the \ion{H}{I} galaxy pipeline described in Section \ref{section:hi_gal_pspec}. The estimates are consistent up until $k \sim 42 \ \text{Mpc}^{-1}$. This $k$ limit corresponds to the resolution set initially to the \ion{H}{I} temperature cube, $k_{\text{max}} = \pi / \Delta r_{\text{pix}}$, where $\Delta r_{\text{pix}}$ is the pixel size of the cube that we set to 0.05 Mpc/h. For the comparison between the visibility power spectrum results in both auto- and cross-correlation, it is sufficient that the comparison is consistent below $k \sim 20 \ \text{Mpc}^{-1}$, since the power spectrum estimated on the visibility data is noise-dominated beyond this $k$ and we nonetheless do not expect to extract any meaningful information from the power spectrum at $k \gtrsim 20 \ \text{Mpc}^{-1}$, as this will correspond to scales comparable to the physical sizes of galaxies. Overall, the consistency between the two results is an indication that the visibility gridding, averaging, and power spectrum estimation pipeline has a comparable output to the \ion{H}{I} galaxy power spectrum produced on the \ion{H}{I} detections.

\subsection{Cross-correlating the MIGHTEE Gridded Visibilities with MASS} \label{subsection:vis_cross_mass}

\begin{figure}
    \centering
    \includegraphics[width=\columnwidth]{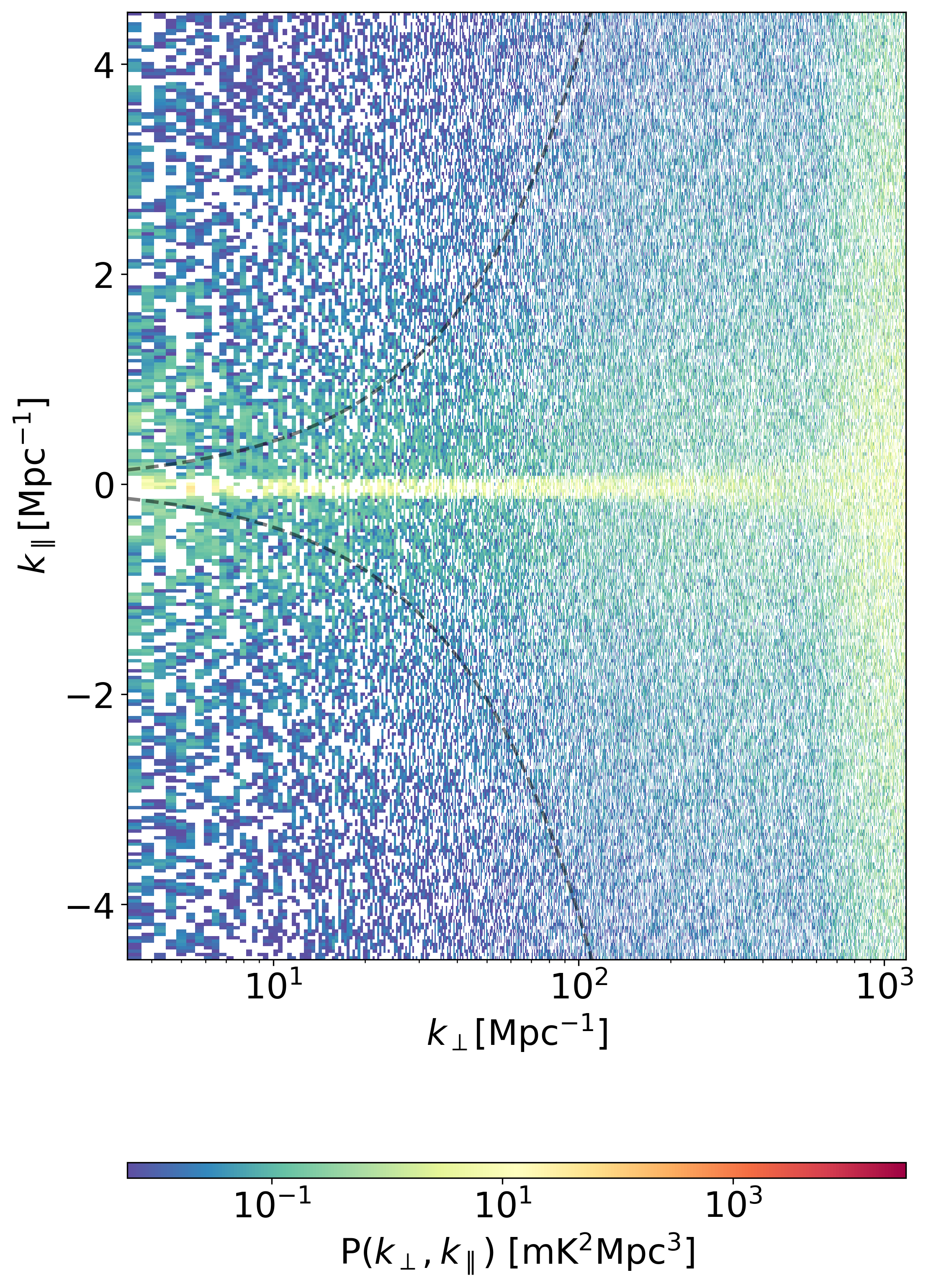}
    \caption[The cylindrical power spectrum of vis $\times$ MASS]{The cylindrical power spectrum of vis $\times$ MASS, without any power spectrum flagging applied. Also shown is the horizon limit, $k_{\parallel}^{\text{horizon}}(k_{\perp})$ as a dashed grey line for $\pm k_{\parallel}$. Clearly seen here is the removal of the foreground wedge and contamination present in the MIGHTEE data. The \ion{H}{I} present in both the visibility data and MASS is seen at low $k_{\parallel}$, and spread across $k_{\perp}$ in a similar manner to that seen in Figure \ref{fig:p2d_mass_boxcar}.}
    \label{fig:p2d_mass_cross_vis}
\end{figure}

Another important way to test our pipeline, and the validity of our power spectrum estimates, is to do a cross-correlation between the delay-transformed gridded visibilities from the MIGHTEE data and those output from MASS. To do so, we use:

\begin{equation}
\begin{split}
 P_{\text{vis} \times \text{MASS}}(\boldsymbol{u}, \tau) = & \ \frac{V_{\text{vis}}(\boldsymbol{u}, \tau) V^*_{\text{MASS}}(\boldsymbol{u}, \tau) + V^*_{\text{vis}}(\boldsymbol{u}, \tau) V_{\text{MASS}}(\boldsymbol{u}, \tau)}{2} \\
 & \ \times \left[ \left( \frac{\lambda^2}{2 k_{\text{B}}} \right)^2 \frac{A_{\text{e}}}{\lambda^2 B} \frac{r_{\nu}^2 \Delta r_{\nu}}{B}\right],
\label{eq:mass_cross_vis}
\end{split}
\end{equation}

where $V_{\text{vis}}$ is the combined, gridded, averaged, and delay-transformed data from MIGHTEE and $V_{\text{MASS}}$ is the delay-transformed gridded visibilities created on the MASS output using the \ion{H}{I} detections from MIGHTEE as the input model with flux profiles assigned using a boxcar along frequency. The power spectrum estimators for cylindrical and spherical averaging in Equation \ref{eq:pk_1d_estimator} and Equation \ref{eq:pk_2d_estimator}, respectively, are applied to this, applying the weights, $w = 1/\sigma_{P_{\text{TN}}}^2$, derived from the thermal noise simulations.
\par

The cylindrical power spectrum of this cross-correlation is shown in Figure \ref{fig:p2d_mass_cross_vis}. No flagging at the 3D power spectrum level has been applied here. However, we can see that both noise, foregrounds and systematics have been mitigated in the cross-correlation, as expected. For instance, even inside the wedge, around $k_{\parallel} = 0$, most power amplitude pixels are removed. Comparing this to \ref{fig:p2d_mass_boxcar}, we can identify the region where the \ion{H}{I} power is dominating. After the horizon cut, most of the signal will still be close to the wedge and at low $k_{\perp}$. Note, however, that we do not expect a perfect correlation between the data visibility \ion{H}{I} signal and the MASS-generated \ion{H}{I} visibilities. Small differences in the phases of the complex numbers in  Equation \ref{eq:mass_cross_vis} will inevitably lead to decorrelation. In practice, besides numerical issues, there are a few possibilities for a mismatch in the two \ion{H}{I} signals: 1) there could be more \ion{H}{I} than that detected in the high signal to noise galaxies; 2) we might be resolving the low redshift galaxies (assumed to be point sources in the MASS simulation); 3) the \ion{H}{I} line profile might be different (we are assuming a Boxcar distribution using the $W_{50}$ values) and 4) there could be a mismatch between the galaxy redshifts (which translates to a mismatch in the frequencies).
\par

\begin{figure}
    \centering
    \includegraphics[width=\columnwidth]{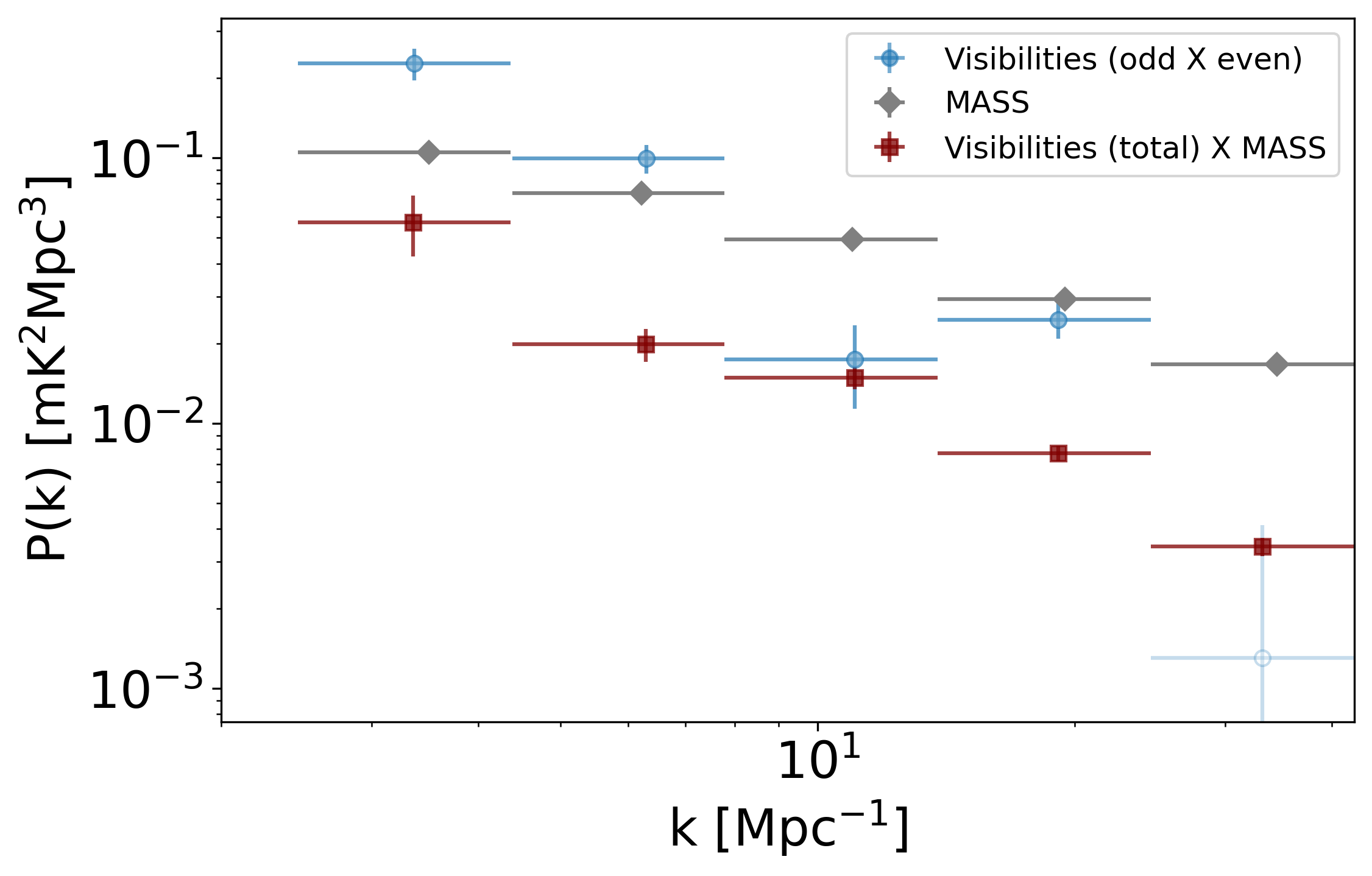}
    \caption{The 1D power spectrum of vis $\times$ MASS compared to the power spectrum of the MASS-generated gridded visibilities, and the cross-power (odd $\times$ even) Stokes I of the MIGHTEE visibility data after applying the 5$\sigma$ cut.}
    \label{fig:p1d_mass_cross_vis_result}
\end{figure}

In Figure \ref{fig:p1d_mass_cross_vis_result} we compare the auto power spectrum of the MASS visibilities to the power spectrum from our original MIGHTEE Stokes I visibilities (cross power of odd $\times$ even), after the 5$\sigma$ flagging. Included as well is the cross-power from  Equation \ref{eq:mass_cross_vis}, between MASS and the total (odd $+$ even) Stokes I visibilities (without any 5$\sigma$ flagging). The power of the MASS autocorrelation is quite close to our visibility cross-power, indicating that most of the \ion{H}{I} should be in the detected galaxies, and our 5$\sigma$ flagging is removing most of the contamination without incurring significant signal loss. On the other hand, the cross between MASS and visibility data shows a significant drop, as expected, due to a probable phase mismatch. We tested slightly moving the galaxy redshifts in a random way without any noticeable change in this cross power. The most likely culprit, at least at low $k_{\perp}$, should be point 3) above (different line profiles). Many actual \ion{H}{I} galaxy profiles would be closer to either Gaussian-shaped or exhibit a double-horn feature, meaning that there will be cases where the shapes and extents of the galaxies along frequency are quite different. Another difference, which can be noticeable but has a smaller impact, would be the beam correction. The beam correction for MASS outputs is calculated at the higher end of the bandwidth, i.e., at $\sim$ 1392 MHz, while for the visibility data, the beam correction has been calculated at the centre of the band, $\sim$ 1362 MHz. This is an approximately 30 MHz shift between beam correction calculations for each case.
\par

To further confirm that the measured cross-correlation signal between MASS and data visibilities is real, we have run another test, where we randomly shuffled the \ion{H}{I} detections along the line-of-sight and used this as input in the MASS pipeline to generate visibilities with which to cross-correlate the MIGHTEE data. The procedure is essentially the same as that described in Section \ref{subsection:vis_cross_mass}. If we indeed see a valid signal in this cross-correlation (see Figure \ref{fig:p1d_mass_cross_vis_result}), a reshuffling of the positions of galaxies would result in a decorrelation, thus making the result of this new cross-correlation consistent with zero.
\par

\begin{figure}
    \centering
    \includegraphics[width=\columnwidth]{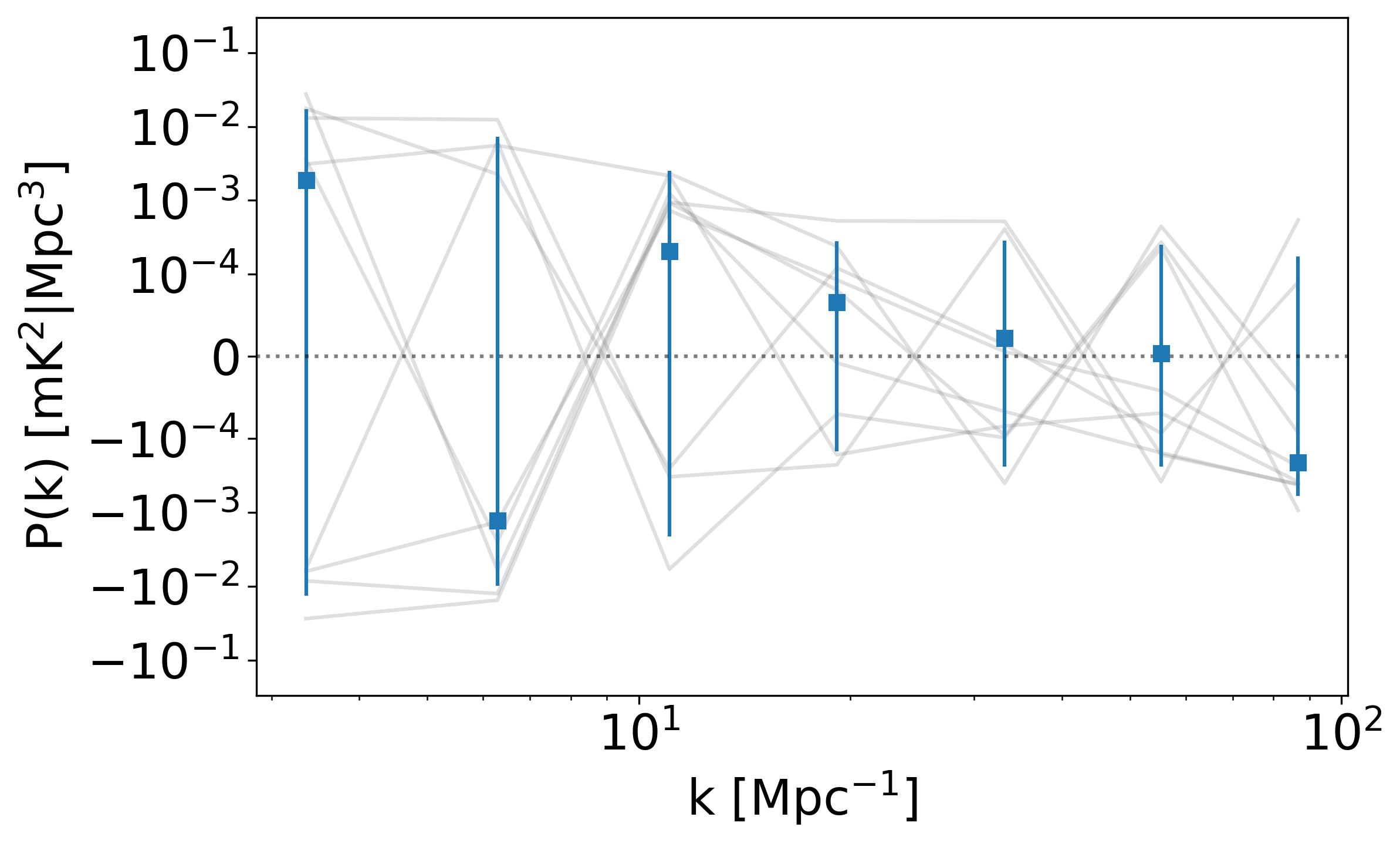}
    \caption{The mean and standard error over 9 realisations of the 1D power spectrum formed from the visibilities cross-correlated with the shuffled MASS realisations. Also shown are the individual cross power spectra for each realisation in grey.}
    \label{fig:null_test_mass_cross_vis}
\end{figure}

The new galaxy positions are completely random along each of the axes (R.A., Dec., $z$) of the volume that contains these galaxies. Due to computational limitations, this is only done for 9 realisations of this random placement of galaxies. The resulting 9 measurement sets are then gridded using the visibility pipeline. These gridded outputs are then crossed with the gridded MIGHTEE visibility data using Equation \ref{eq:mass_cross_vis}. The subsequent power spectra estimated on these are averaged and shown in Figure \ref{fig:null_test_mass_cross_vis} with the standard deviation of the 1D power spectrum realisations used as the error in each $k$ bin. Indeed, we find that with a selection of randomly shuffled \ion{H}{I} galaxy sky models, there is negligible correlation with the visibility data when compared to cross-correlating the actual MIGHTEE detected \ion{H}{I} galaxies with the visibility data.
\par

\subsection{Signal loss}
\label{subsec:sigma_flag_vis_tests}

We now study the impact of the 3D gridded visibility power spectrum flagging on signal loss with the aid of the MASS simulation. We start by considering two more cases: 9$\sigma$ and 3$\sigma$, and compare them to the standard 5$\sigma$ flagging we use for the main results in this study. The 9$\sigma$ should still remove significant contamination while mitigating some of the signal loss that inevitably results from having to remove increasingly more $\boldsymbol{k}$ modes from the 3D power spectrum cube. The 3$\sigma$ case is the least conservative cut that can be considered before considerable portions of the modes available in the $k$ bins are flagged away, resulting in larger signal loss in our power spectrum estimates.

\begin{figure*}
    \centering
    \includegraphics[width=\columnwidth]{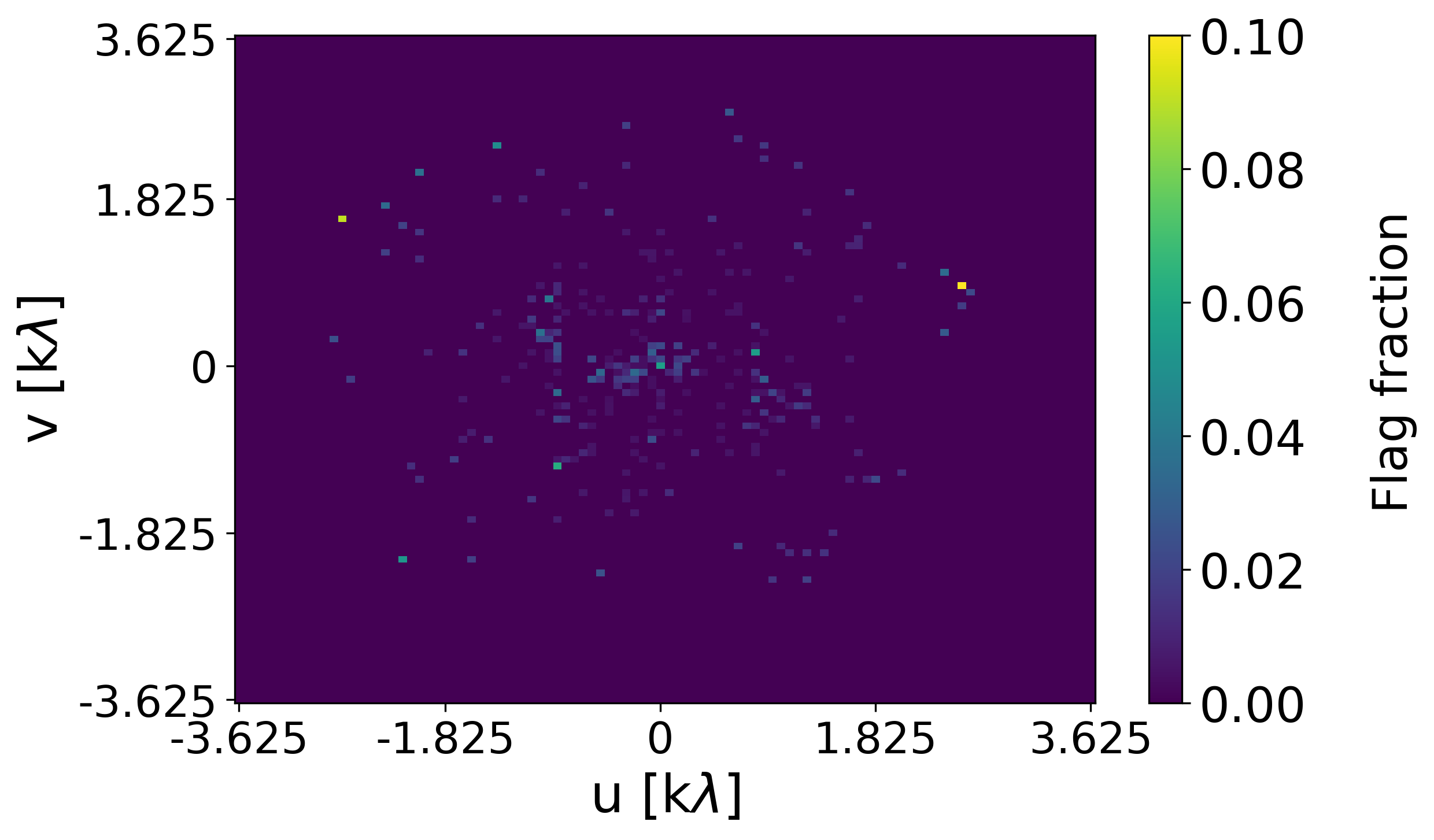}
    \includegraphics[width=\columnwidth]{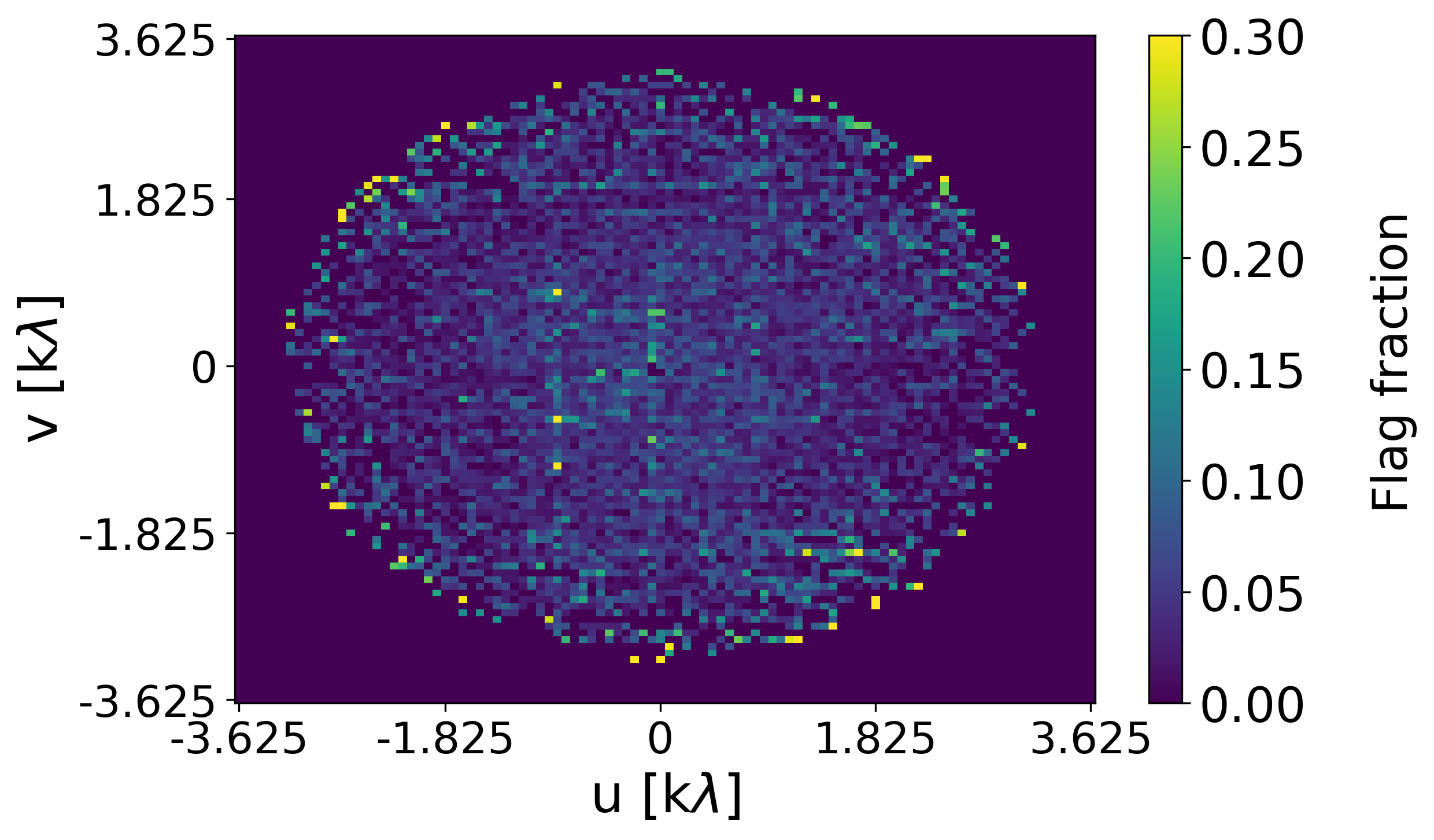}
    \caption{The flagging fraction when we apply 9$\sigma$ (left) and  3$\sigma$ (right) level flags to the 3D visibility power spectrum cube within the estimation window. In the 3$\sigma$ case, the flag fraction is a lot more pronounced throughout the $uv$ grid as the flagging is more aggressive compared to the 9$\sigma$ case. Even so, the localized contamination in the centre of the $uv$ distribution is still flagged on close inspection of this 9$\sigma$ case.}
    \label{fig:sigma_flag_frac_comparison}
\end{figure*}

Figure \ref{fig:sigma_flag_frac_comparison} shows the flagging fraction projected on the $uv$ plane for the 9 and 3$\sigma$ cases. As in the case of 5$\sigma$ shown in Figure \ref{fig:5-sigma_flag_frac}, the highest flagging fractions are confined within a region $\sim \pm \ 1.825 \ k\lambda$. Beyond this, most of the flagged points on the $uv$ grid will be thermal noise, and therefore appear random in this centralised portion of the $uv$ distribution shown. This is justified when looking at the flagging fraction for the 9$\sigma$ case, which, similarly to the 5$\sigma$ case, removes the bright contamination that appears to have stripe-like features, but has a reduced flagging fraction beyond $\sim \pm \ 1.825 \ k\lambda$ since there is less thermal noise being removed by this level of flagging as compared to 3$\sigma$. Despite this, even at the 9$\sigma$ level, the centrally-confined contamination, which appears localised in the cylindrical power spectra, is flagged out by this criterion. On the other hand, the 3$\sigma$ case removes the modes similarly to that seen at the 5$\sigma$ level, but has a noticeably higher flag fraction across the $uv$ distribution. On average, there is about a 10-15\% of $uv$ modes flagged in comparison to $\sim 5\%$ for the 5$\sigma$ case.
\par

\begin{figure*}
    \centering
    \includegraphics[width=0.667\columnwidth]{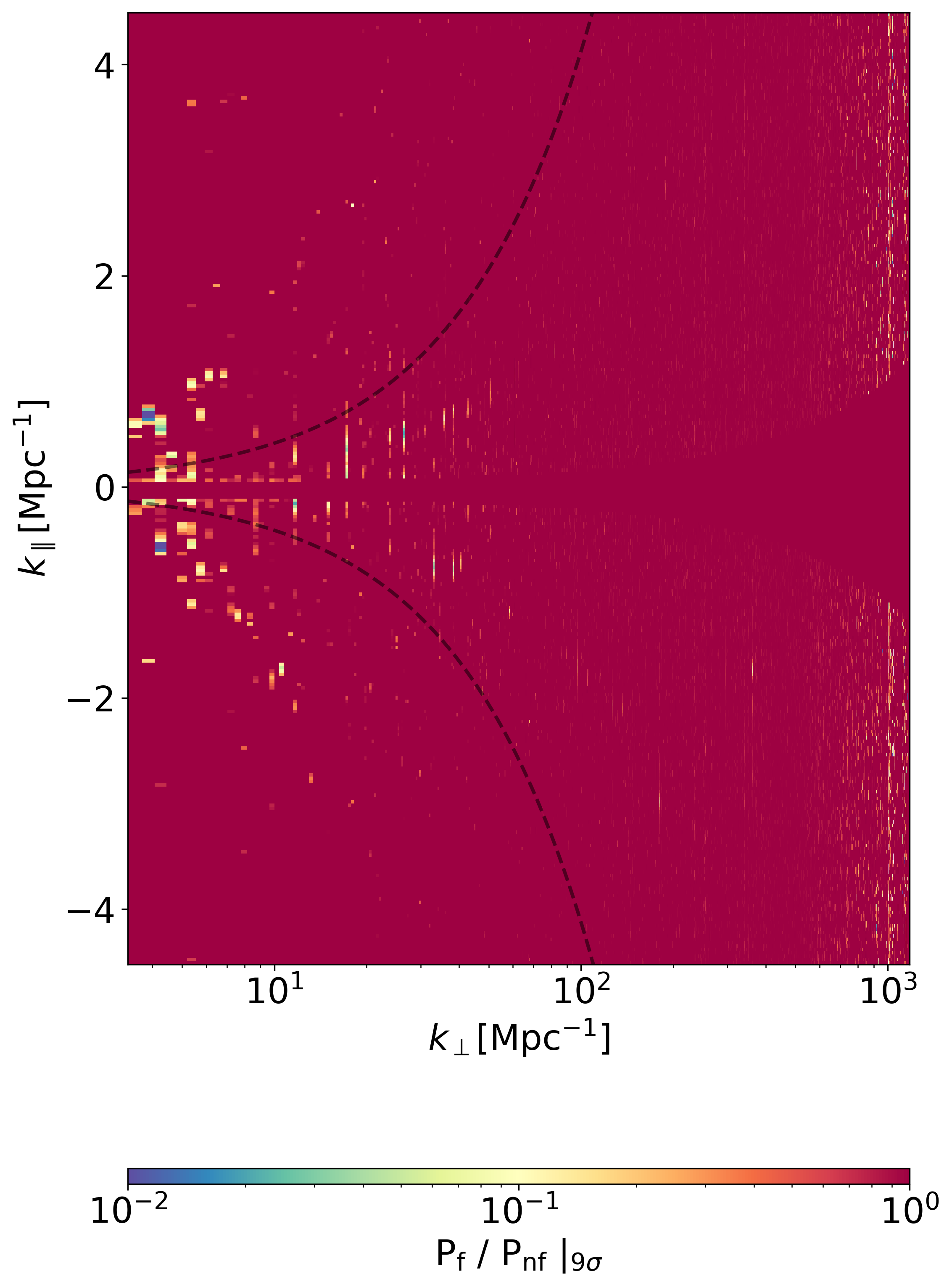}
    \includegraphics[width=0.667\columnwidth]{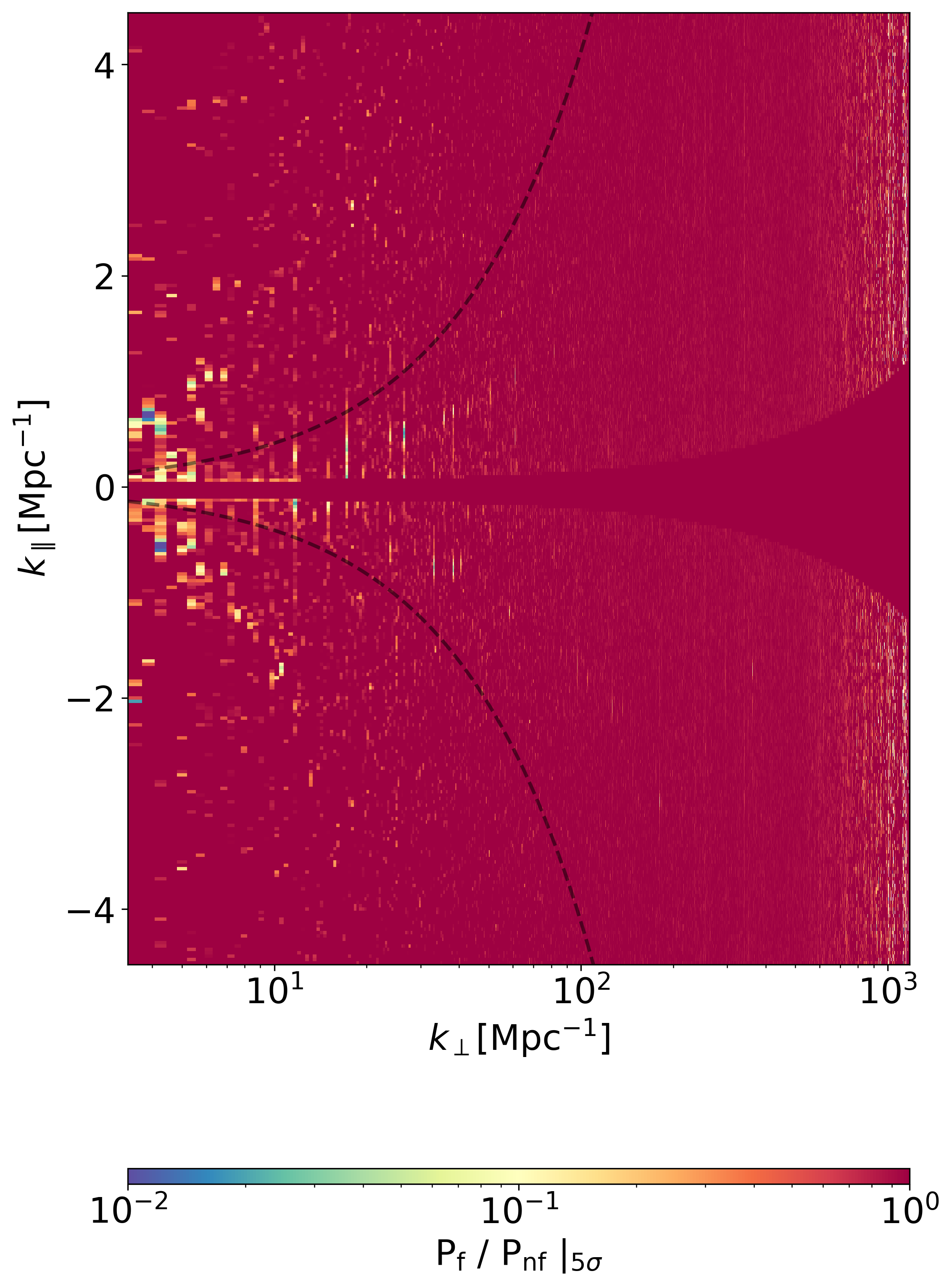}
    \includegraphics[width=0.667\columnwidth]{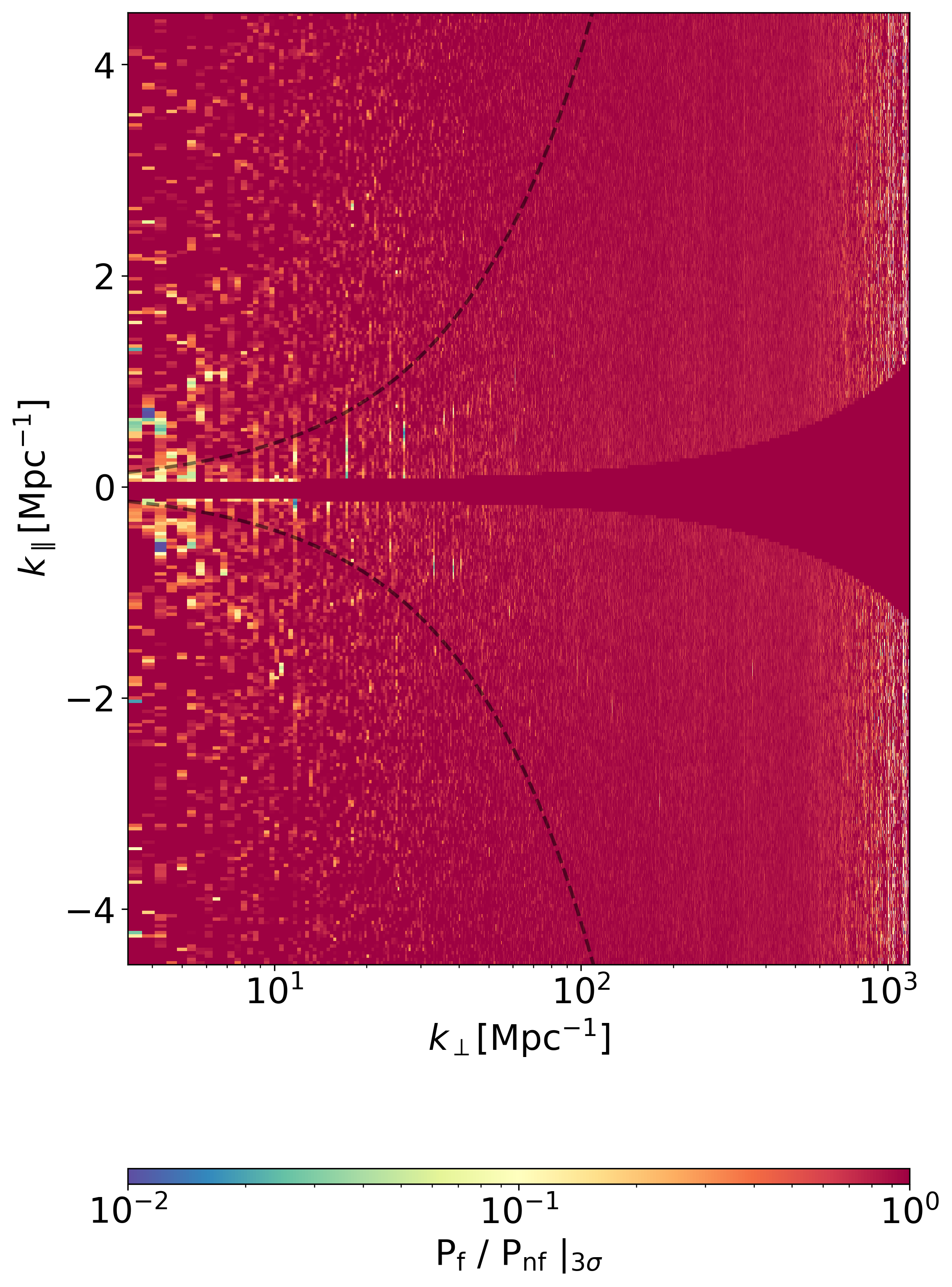}    
    \caption{The ratio of the cylindrical power after performing 9$\sigma$ (left), 5$\sigma$ (middle), and 3$\sigma$ (right) flagging to the cylindrical power spectrum without any flagging applied. The smaller foreground-dominated region inside the wedge is not flagged for any of the cases since we want to assess the level of spillover and contamination near and around this region itself. The ratios shown here help demonstrate the level of flagging done within the estimation window for each case and along $k_{\perp}$ as the thermal noise level increases, which is particularly pronounced for 3$\sigma$, as we expect this to flag a larger portion of noise-dominated $\boldsymbol{k}$ pixels in comparison to the other cases.}
    \label{fig:sigma_flag_comp_pk_2d_ratio}
\end{figure*}

\begin{figure}
    \centering
    \includegraphics[width=\columnwidth, trim=8pt 10pt 0pt 0pt, clip]{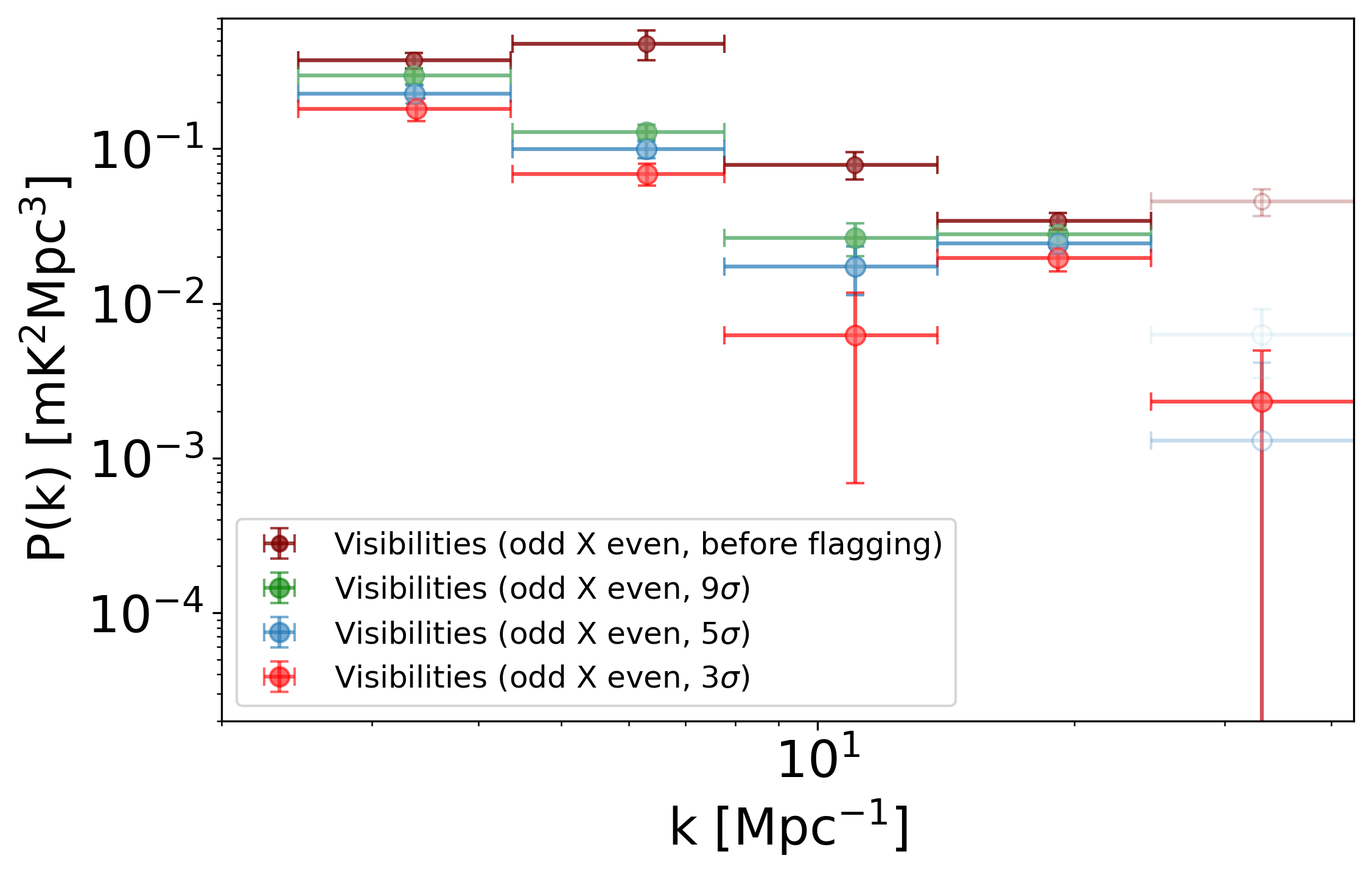}
    \caption{The \ion{H}{I} power spectrum estimates for the cases of 3, 5, and 9$\sigma$ flagging levels compared to the case where no $\sigma$ flagging has been performed. Negative amplitudes are shown as open, shaded circles with the associated error bars. Beyond $k \sim 20$ Mpc$^{-1}$, thermal noise is expected to dominate.}
    \label{fig:sigma_flag_comp_pk_1d}
\end{figure}

Figure \ref{fig:sigma_flag_comp_pk_2d_ratio} shows the ratios of the 2D cylindrical auto power spectra for the different $\sigma$ flag cases with respect to the no flagging case. We remind again that flagging is done at the 3D power spectra level. We see that the 9$\sigma$ level already picks up the main contaminants. As the flagging increases, there will be more pixels with the power reduced as expected. However, this could simply be due to noise outliers or even stronger \ion{H}{I} power. Note that flagging noise outliers should not bias the cross-power between the "even" and "odd" datasets. For 3$\sigma$, the flagging becomes more aggressive, but still is mainly restricted to removing the strong contamination at $k_{\perp} \lesssim 10 \ \text{Mpc}^{-1}$ and $k_{\parallel} \lesssim 1 \ \text{Mpc}^{-1}$. These results also translate to the 1D power spectrum estimates for each case as seen in Figure \ref{fig:sigma_flag_comp_pk_1d} (odd $\times$ even power spectra). We can see that there is a clear drop from the case with no contamination flagging to the 9$\sigma$ case, especially for the second and third bins. Interestingly, the drop from 9$\sigma$ to 5$\sigma$ is much smaller, indicating that most of the contamination is already picked up by the 9$\sigma$ flagging. At the same time, we see a continuous drop in power from 9$\sigma$ to 3$\sigma$. There could be two reasons for this: either we are still removing contamination (that correlates between the odd and even datasets), or we have signal loss.
\par

\begin{figure}
    \centering
    \includegraphics[width=\columnwidth]{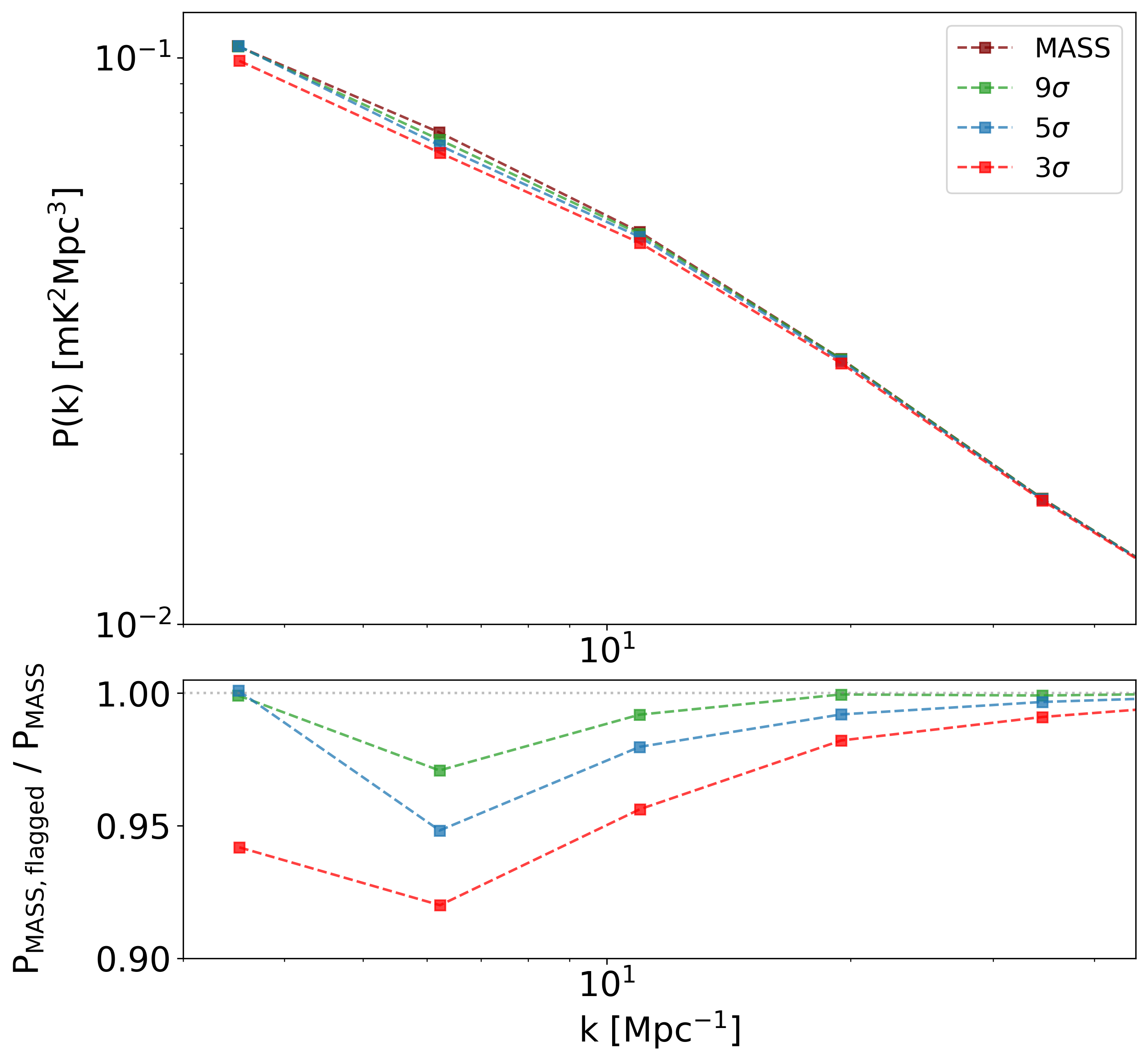}
    \caption[The top panel shows the 1D \ion{H}{I} power spectrum estimated on the MASS-generated visibilities using the \ion{H}{I} galaxy detections as input, compared to the 9, 5 and 3 $\sigma$ cut MASS power spectra cases. The bottom panel shows the ratios of each of these $\sigma$ cut cases to the case where no $\sigma$ cut is applied]{The top panel shows the 1D \ion{H}{I} power spectrum estimated on the MASS-generated visibilities using the \ion{H}{I} galaxy detections as input compared to the 9, 5 and 3 $\sigma$ cut power spectra cases. The bottom panel shows the ratios of each of these $\sigma$ cut cases to the case where no $\sigma$ cut is applied.}
    \label{fig:mass_sim_sigma_cut_comparison}
\end{figure}

Applying the same 3D ($\boldsymbol{k}_{\perp}, k_{\parallel}$) cube masks to the MASS simulation should help us understand this better. Figure \ref{fig:mass_sim_sigma_cut_comparison} shows the results for the 9$\sigma$, 5$\sigma$ and 3$\sigma$ cases. The top panel shows the 1D power spectra estimated on the MASS-generated, gridded visibilities, while the bottom panel shows the ratio of each $\sigma$ cut case to the case where no cut is applied. Although there is a consistent signal drop, this is smaller than what is observed for the original signal in Figure \ref{fig:sigma_flag_comp_pk_1d}. A similarly small drop in the signal was observed in Section \ref{subsection:pk_higal_additional_tools} where the $5\sigma$ flags were tested \ion{H}{I} galaxies placed in a cosmological volume. This is probably an indication that there is more power in the original \ion{H}{I} signal closer to the wedge, where most of the flagging is being applied (Figure \ref{fig:sigma_flag_comp_pk_2d_ratio}). This should not be a complete surprise, as we have seen that the cross-correlation of the original visibility data with the MASS output shows some decorrelation. Therefore, a final test would be to check the impact of the flags on the cross-correlation between the data and MASS visibilities.
\par

\begin{figure}
    \centering
    \includegraphics[width=\columnwidth]{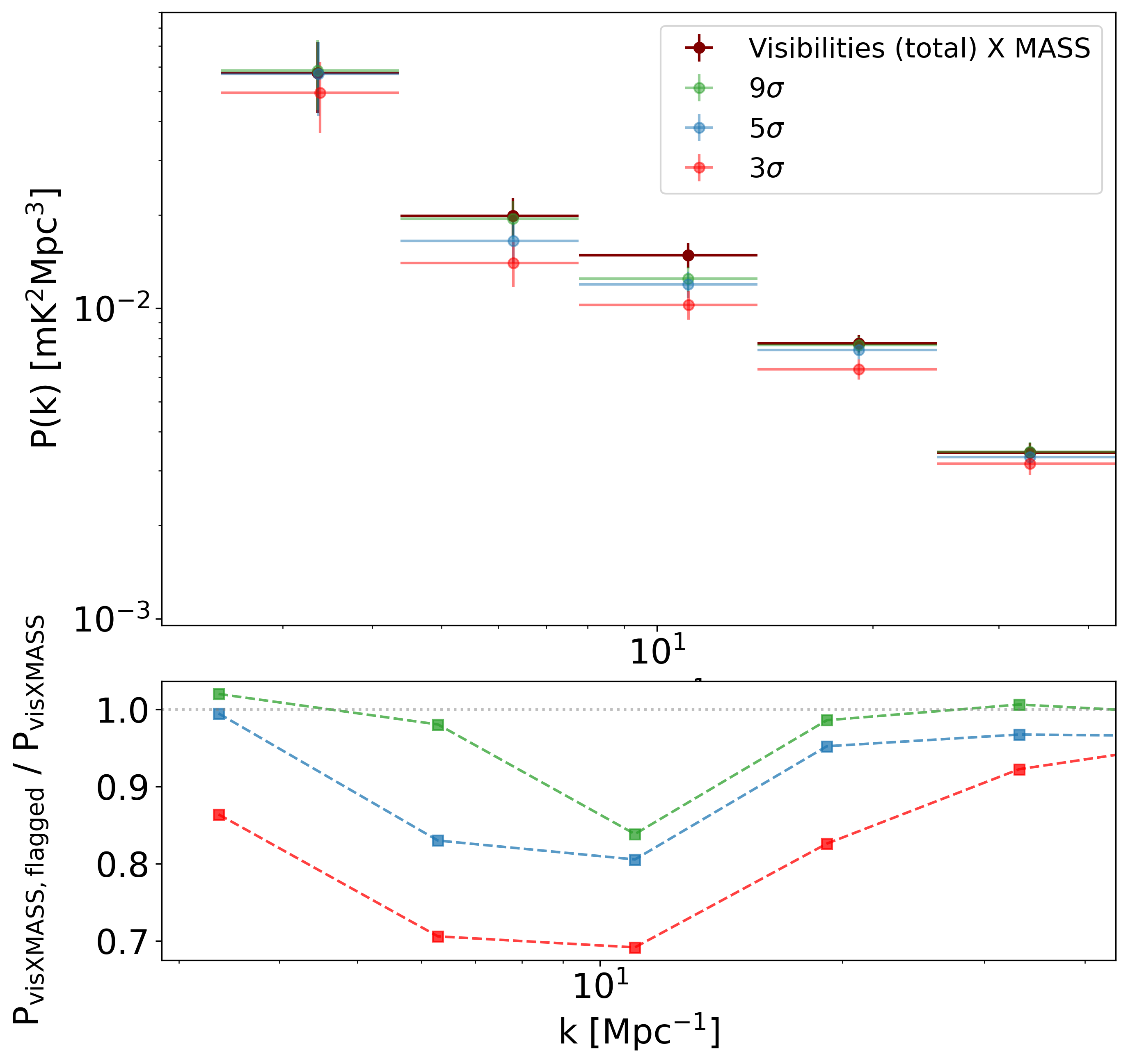}
    \caption{The vis $\times$ MASS power spectrum compared to the results obtained from applying 9, 5, and 3 $\sigma$ cuts shown in the top panel. The bottom panel shows the ratio of these $\sigma$ cut cases to the case where no $\sigma$ flagging is applied.}
    \label{fig:vis_cross_mass_sigma_cut}
\end{figure}

The cross-correlation with the MASS simulation derived from detected galaxies should be free of contamination. If the drop in power in the data as we go from 9$\sigma$ to 3 $\sigma$ flagging is only due to the removal of contaminants, we should not see that effect in the cross-correlation since the systematics will not correlate. Figure \ref{fig:vis_cross_mass_sigma_cut} shows the impact of flagging on this cross-power spectrum. Note that the same flags are applied to the delay transformed visibility cube before calculating the cross-power. Clearly we see a drop in power larger than in Figure \ref{fig:mass_sim_sigma_cut_comparison}. This is probably an indication that more \ion{H}{I} is being removed in the original visibility dataset than what is seen in the MASS simulation. The bottom panel of Figure \ref{fig:vis_cross_mass_sigma_cut} shows a signal loss below 20\% for the 5 $\sigma$ case, dropping to 30\% in the 3 $\sigma$ case. Therefore, although the level of contamination after the sigma flagging is expected to be low, some care must be taken with signal loss.

\section{Conclusions}
\label{section:conclusion}

In this work, we have investigated the feasibility of measuring the \ion{H}{I} intensity mapping signal at low redshifts ($z \simeq 0.04$) using interferometric observations from the central pointing of the MIGHTEE COSMOS field. With $\sim$17.5~hours of MIGHTEE data and a delay-spectrum-based power spectrum estimator, we have measured the \ion{H}{I} power spectrum on sub-megaparsec scales over the frequency range 1332 - 1392~MHz. This analysis extends previous MeerKAT interferometric intensity mapping studies to a regime at lower redshifts where individual \ion{H}{I} galaxy detections are available within the same dataset, enabling a direct and self-consistent validation of the methodology.
\par

As in previous MeerKAT intensity mapping analyses, we identify low-level broadband contamination in the power spectrum, primarily localised at low $k_\perp$ and close to the foreground wedge. This contamination is visible in both auto- and cross-power spectra at the cylindrical power spectrum level and is consistent with residual systematics rather than thermal noise. Following the approach adopted in earlier work \citep{Paul2023}, we mitigate this effect by flagging strong outliers directly at the three-dimensional power spectrum (or delay transformed visibility) level using detailed thermal noise simulations. The fraction of flagged voxels is small, typically in the range $\sim$ 2 - 5\%, and therefore expected to have a limited impact on the cosmological signal. After applying a conservative $5\sigma$ cut, the one-dimensional power spectrum obtained from cross-correlating visibilities split in time exhibits a signal that appears free from detectable contamination. The total signal-to-noise ratio of this cross-power spectrum measurement is approximately 13.
\par

A key objective of this study has been to validate the full end-to-end power spectrum pipeline using the detected \ion{H}{I} galaxies in the MIGHTEE COSMOS field. This validation was carried out in two complementary ways. First, the measured \ion{H}{I} masses were directly distributed into a comoving intensity field, from which a power spectrum was estimated using the same gridding and averaging procedures. Second, the detected galaxies were used as inputs to the MeerKAT All-Sky Simulator (MASS) to generate synthetic visibilities, which were then processed through the full visibility-based pipeline. In both cases, the recovered power spectra are consistent with the visibility-based measurements up to $k \sim 20~\mathrm{Mpc}^{-1}$, beyond which the measurements become noise-dominated and no meaningful constraints are expected. This agreement provides a strong internal consistency check on the gridding, delay transform, weighting, and power spectrum estimation steps of the pipeline.
\par

We further tested the astrophysical origin of the measured signal through cross-correlations between the observed visibilities and MASS-generated visibilities. A statistically significant correlation is observed when using the true galaxy distribution, while randomised galaxy realisations yield null results. These tests demonstrate that the measured signal is associated with the \ion{H}{I} galaxy population rather than residual systematics or noise fluctuations. Differences between the auto- and cross-correlations involving simulated visibilities are consistent with expected effects arising from simplified assumptions in the simulations, such as idealised line profiles and beam modelling. However, the simulations indicate that a non-negligible fraction of the \ion{H}{I} power lies close to the wedge, implying that aggressive flagging can lead to measurable signal suppression. This highlights the need for careful, simulation-calibrated treatments of foreground mitigation when interpreting one-dimensional power spectra, particularly at low redshift where the \ion{H}{I} signal is strongly anisotropic in $(k_{\parallel}, k_{\perp})$ space.
\par

Overall, this work demonstrates that interferometric \ion{H}{I} intensity mapping with MeerKAT can recover a power spectrum that is consistent, within uncertainties, with that inferred from directly-detected \ion{H}{I} galaxies in the same field over the range of scales where the measurement is signal-dominated. While the present dataset does not allow for a precision cosmological measurement, the combination of internal null tests, contamination mitigation, and end-to-end validation provides strong evidence that the recovered signal is robust and astrophysical in origin.
\par


\section*{Acknowledgements}

JT and MGS acknowledge support from the South African Radio Astronomy Observatory and National Research Foundation (Grant No.\ 84156). SC acknowledges financial support from the South African National Research Foundation (Grant No. 84156) and the Inter-University Institute for Data Intensive Astronomy (IDIA). IDIA is a partnership of the University of Cape Town, the University of Pretoria and the University of the Western Cape. IDIA is registered on the Research Organization Registry with ROR ID 01edhwb26, and on Open Funder Registry with funder ID 100031500. The authors acknowledge the use of the ilifu cloud computing facility – www.ilifu.ac.za, a partnership between the University of Cape Town, the University of the Western Cape, Stellenbosch University, Sol Plaatje University and the Cape Peninsula University of Technology. The ilifu facility is supported by contributions from IDIA, the Computational Biology division at UCT, and the Data Intensive Research Initiative of South Africa (DIRISA).


\section*{Data Availability}

The data and simulation outputs underlying this article will be shared upon reasonable requests.


\bibliographystyle{mnras}
\bibliography{library}

@article{Amiri_2023,
doi = {10.3847/1538-4357/acb13f},
url = {https://doi.org/10.3847/1538-4357/acb13f},
year = {2023},
month = {apr},
publisher = {The American Astronomical Society},
volume = {947},
number = {1},
pages = {16},
author = {{CHIME Collaboration} and Amiri, Mandana and Bandura, Kevin and Chen, Tianyue and Deng, Meiling and Dobbs, Matt and Fandino, Mateus and Foreman, Simon and Halpern, Mark and Hill, Alex S. and Hinshaw, Gary and Höfer, Carolin and Kania, Joseph and Landecker, T. L. and MacEachern, Joshua and Masui, Kiyoshi and Mena-Parra, Juan and Milutinovic, Nikola and Mirhosseini, Arash and Newburgh, Laura and Ordog, Anna and Pen, Ue-Li and Pinsonneault-Marotte, Tristan and Polzin, Ava and Reda, Alex and Renard, Andre and Shaw, J. Richard and Siegel, Seth R. and Singh, Saurabh and Vanderlinde, Keith and Wang, Haochen and Wiebe, Donald V. and Wulf, Dallas},
title = {Detection of Cosmological 21 cm Emission with the Canadian Hydrogen Intensity Mapping Experiment},
journal = {The Astrophysical Journal}
}

@ARTICLE{2015MNRAS.448.1922S,
       author = {{Serra}, Paolo and {Westmeier}, Tobias and {Giese}, Nadine and {Jurek}, Russell and {Fl{\"o}er}, Lars and {Popping}, Attila and {Winkel}, Benjamin and {van der Hulst}, Thijs and {Meyer}, Martin and {Koribalski}, B{\"a}rbel S. and {Staveley-Smith}, Lister and {Courtois}, H{\'e}l{\`e}ne},
        title = "{SOFIA: a flexible source finder for 3D spectral line data}",
      journal = {\mnras},
         year = 2015,
        month = apr,
       volume = {448},
       number = {2},
        pages = {1922-1929},
          doi = {10.1093/mnras/stv079},
archivePrefix = {arXiv},
       eprint = {1501.03906},
 primaryClass = {astro-ph.IM},
       adsurl = {https://ui.adsabs.harvard.edu/abs/2015MNRAS.448.1922S}
}

@article{10.1093/mnras/stab3064,
    author = {Spinelli, Marta and Carucci, Isabella P and Cunnington, Steven and Harper, Stuart E and Irfan, Melis O and Fonseca, José and Pourtsidou, Alkistis and Wolz, Laura},
    title = {SKAO H i intensity mapping: blind foreground subtraction challenge},
    journal = {Monthly Notices of the Royal Astronomical Society},
    volume = {509},
    number = {2},
    pages = {2048-2074},
    year = {2021},
    month = {10},
    issn = {0035-8711},
    doi = {10.1093/mnras/stab3064},
    url = {https://doi.org/10.1093/mnras/stab3064},
    eprint = {https://academic.oup.com/mnras/article-pdf/509/2/2048/41170415/stab3064.pdf},
}

@article{10.1093/mnras/stw1380,
    author = {Barry, N. and Hazelton, B. and Sullivan, I. and Morales, M. F. and Pober, J. C.},
    title = {Calibration requirements for detecting the 21 cm epoch of reionization power spectrum and implications for the SKA},
    journal = {Monthly Notices of the Royal Astronomical Society},
    volume = {461},
    number = {3},
    pages = {3135-3144},
    year = {2016},
    month = {06},
    issn = {0035-8711},
    doi = {10.1093/mnras/stw1380},
    url = {https://doi.org/10.1093/mnras/stw1380},
    eprint = {https://academic.oup.com/mnras/article-pdf/461/3/3135/8106921/stw1380.pdf},
}

@INPROCEEDINGS{2016SPIE.9906E..5XN,
       author = {{Newburgh}, L.~B. and {Bandura}, K. and {Bucher}, M.~A. and {Chang}, T.-C. and {Chiang}, H.~C. and {Cliche}, J.~F. and {Dav{\'e}}, R. and {Dobbs}, M. and {Clarkson}, C. and {Ganga}, K.~M. and {Gogo}, T. and {Gumba}, A. and {Gupta}, N. and {Hilton}, M. and {Johnstone}, B. and {Karastergiou}, A. and {Kunz}, M. and {Lokhorst}, D. and {Maartens}, R. and {Macpherson}, S. and {Mdlalose}, M. and {Moodley}, K. and {Ngwenya}, L. and {Parra}, J.~M. and {Peterson}, J. and {Recnik}, O. and {Saliwanchik}, B. and {Santos}, M.~G. and {Sievers}, J.~L. and {Smirnov}, O. and {Stronkhorst}, P. and {Taylor}, R. and {Vanderlinde}, K. and {Van Vuuren}, G. and {Weltman}, A. and {Witzemann}, A.},
        title = "{HIRAX: a probe of dark energy and radio transients}",
    booktitle = {Ground-based and Airborne Telescopes VI},
         year = 2016,
       editor = {{Hall}, Helen J. and {Gilmozzi}, Roberto and {Marshall}, Heather K.},
       series = {Society of Photo-Optical Instrumentation Engineers (SPIE) Conference Series},
       volume = {9906},
        month = aug,
          eid = {99065X},
        pages = {99065X},
          doi = {10.1117/12.2234286},
archivePrefix = {arXiv},
       eprint = {1607.02059},
 primaryClass = {astro-ph.IM},
       adsurl = {https://ui.adsabs.harvard.edu/abs/2016SPIE.9906E..5XN}
}

@ARTICLE{2022ApJS..261...29C,
       author = {{CHIME Collaboration} and {Amiri}, Mandana and {Bandura}, Kevin and {Boskovic}, Anja and {Chen}, Tianyue and {Cliche}, Jean-Fran{\c{c}}ois and {Deng}, Meiling and {Denman}, Nolan and {Dobbs}, Matt and {Fandino}, Mateus and {Foreman}, Simon and {Halpern}, Mark and {Hanna}, David and {Hill}, Alex S. and {Hinshaw}, Gary and {H{\"o}fer}, Carolin and {Kania}, Joseph and {Klages}, Peter and {Landecker}, T.~L. and {MacEachern}, Joshua and {Masui}, Kiyoshi and {Mena-Parra}, Juan and {Milutinovic}, Nikola and {Mirhosseini}, Arash and {Newburgh}, Laura and {Nitsche}, Rick and {Ordog}, Anna and {Pen}, Ue-Li and {Pinsonneault-Marotte}, Tristan and {Polzin}, Ava and {Reda}, Alex and {Renard}, Andre and {Shaw}, J. Richard and {Siegel}, Seth R. and {Singh}, Saurabh and {Smegal}, Rick and {Tretyakov}, Ian and {van Gassen}, Kwinten and {Vanderlinde}, Keith and {Wang}, Haochen and {Wiebe}, Donald V. and {Willis}, James S. and {Wulf}, Dallas},
        title = "{An Overview of CHIME, the Canadian Hydrogen Intensity Mapping Experiment}",
      journal = {\apjs},
         year = 2022,
        month = aug,
       volume = {261},
       number = {2},
          eid = {29},
        pages = {29},
          doi = {10.3847/1538-4365/ac6fd9},
archivePrefix = {arXiv},
       eprint = {2201.07869},
 primaryClass = {astro-ph.IM},
       adsurl = {https://ui.adsabs.harvard.edu/abs/2022ApJS..261...29C}
}

@INPROCEEDINGS{2015aska.confE..19S,
       author = {{Santos}, M. and {Bull}, P. and {Alonso}, D. and {Camera}, S. and {Ferreira}, P. and {Bernardi}, G. and {Maartens}, R. and {Viel}, M. and {Villaescusa-Navarro}, F. and {Abdalla}, F.~B. and {Jarvis}, M. and {Metcalf}, R.~B. and {Pourtsidou}, A. and {Wolz}, L.},
        title = "{Cosmology from a SKA HI intensity mapping survey}",
    booktitle = {Advancing Astrophysics with the Square Kilometre Array (AASKA14)},
         year = 2015,
        month = apr,
          eid = {19},
        pages = {19},
          doi = {10.22323/1.215.0019},
archivePrefix = {arXiv},
       eprint = {1501.03989},
 primaryClass = {astro-ph.CO},
       adsurl = {https://ui.adsabs.harvard.edu/abs/2015aska.confE..19S}
}

@ARTICLE{2025MNRAS.537.3632M,
       author = {{MeerKLASS Collaboration} and {Barberi-Squarotti}, Matilde and {Bernal}, Jos{\'e} L. and {Bull}, Philip and {Camera}, Stefano and {Carucci}, Isabella P. and {Chen}, Zhaoting and {Cunnington}, Steven and {Engelbrecht}, Brandon N. and {Fonseca}, Jos{\'e} and {Grainge}, Keith and {Irfan}, Melis O. and {Li}, Yichao and {Mazumder}, Aishrila and {Paul}, Sourabh and {Pourtsidou}, Alkistis and {Santos}, Mario G. and {Spinelli}, Marta and {Wang}, Jingying and {Witzemann}, Amadeus and {Wolz}, Laura},
        title = "{MeerKLASS L-band deep-field intensity maps: entering the H I dominated regime}",
      journal = {\mnras},
         year = 2025,
        month = mar,
       volume = {537},
       number = {4},
        pages = {3632-3661},
          doi = {10.1093/mnras/staf195},
archivePrefix = {arXiv},
       eprint = {2407.21626},
 primaryClass = {astro-ph.CO},
       adsurl = {https://ui.adsabs.harvard.edu/abs/2025MNRAS.537.3632M}
}

@article{10.1093/mnras/stx1388,
    author = {Wolz, L. and Blake, C. and Wyithe, J. S. B.},
    title = {Determining the H i content of galaxies via intensity mapping cross-correlations},
    journal = {Monthly Notices of the Royal Astronomical Society},
    volume = {470},
    number = {3},
    pages = {3220-3226},
    year = {2017},
    month = {06},
    issn = {0035-8711},
    doi = {10.1093/mnras/stx1388},
    url = {https://doi.org/10.1093/mnras/stx1388},
    eprint = {https://academic.oup.com/mnras/article-pdf/470/3/3220/18548205/stx1388.pdf},
}

@ARTICLE{2015MNRAS.447.1610M,
       author = {{Maddox}, Natasha and {Hess}, Kelley M. and {Obreschkow}, Danail and {Jarvis}, M.~J. and {Blyth}, S.-L.},
        title = "{Variation of galactic cold gas reservoirs with stellar mass}",
      journal = {\mnras},
         year = 2015,
        month = feb,
       volume = {447},
       number = {2},
        pages = {1610-1617},
          doi = {10.1093/mnras/stu2532},
archivePrefix = {arXiv},
       eprint = {1412.0852},
 primaryClass = {astro-ph.GA},
       adsurl = {https://ui.adsabs.harvard.edu/abs/2015MNRAS.447.1610M}
}

@ARTICLE{2012ApJ...756..113H,
       author = {{Huang}, Shan and {Haynes}, Martha P. and {Giovanelli}, Riccardo and {Brinchmann}, Jarle},
        title = "{The Arecibo Legacy Fast ALFA Survey: The Galaxy Population Detected by ALFALFA}",
      journal = {\apj},
         year = 2012,
        month = sep,
       volume = {756},
       number = {2},
          eid = {113},
        pages = {113},
          doi = {10.1088/0004-637X/756/2/113},
archivePrefix = {arXiv},
       eprint = {1207.0523},
 primaryClass = {astro-ph.CO},
       adsurl = {https://ui.adsabs.harvard.edu/abs/2012ApJ...756..113H}
}

@ARTICLE{2006PhR...433..181F,
       author = {{Furlanetto}, Steven R. and {Oh}, S. Peng and {Briggs}, Frank H.},
        title = "{Cosmology at low frequencies: The 21 cm transition and the high-redshift Universe}",
      journal = {\physrep},
         year = 2006,
        month = oct,
       volume = {433},
       number = {4-6},
        pages = {181-301},
          doi = {10.1016/j.physrep.2006.08.002},
archivePrefix = {arXiv},
       eprint = {astro-ph/0608032},
 primaryClass = {astro-ph},
       adsurl = {https://ui.adsabs.harvard.edu/abs/2006PhR...433..181F}
}

@ARTICLE{2006PhRvD..74h3517S,
       author = {{Santos}, M{\'a}rio G. and {Cooray}, Asantha},
        title = "{Cosmological and astrophysical parameter measurements with 21-cm anisotropies during the era of reionization}",
      journal = {\prd},
         year = 2006,
        month = oct,
       volume = {74},
       number = {8},
          eid = {083517},
        pages = {083517},
          doi = {10.1103/PhysRevD.74.083517},
archivePrefix = {arXiv},
       eprint = {astro-ph/0605677},
 primaryClass = {astro-ph},
       adsurl = {https://ui.adsabs.harvard.edu/abs/2006PhRvD..74h3517S}
}

@ARTICLE{2021MNRAS.508.1195P,
       author = {{Ponomareva}, Anastasia A. and {Mulaudzi}, Wanga and {Maddox}, Natasha and {Frank}, Bradley S. and {Jarvis}, Matt J. and {Di Teodoro}, Enrico M. and {Glowacki}, Marcin and {Kraan-Korteweg}, Ren{\'e}e C. and {Oosterloo}, Tom A. and {Adams}, Elizabeth A.~K. and {Pan}, Hengxing and {Prandoni}, Isabella and {Rajohnson}, Sambatriniaina H.~A. and {Sinigaglia}, Francesco and {Adams}, Nathan J. and {Heywood}, Ian and {Bowler}, Rebecca A.~A. and {Hatfield}, Peter W. and {Collier}, Jordan D. and {Sekhar}, Srikrishna},
        title = "{MIGHTEE-H I: the baryonic Tully-Fisher relation over the last billion years}",
      journal = {\mnras},
         year = 2021,
        month = nov,
       volume = {508},
       number = {1},
        pages = {1195-1205},
          doi = {10.1093/mnras/stab2654},
archivePrefix = {arXiv},
       eprint = {2109.04992},
 primaryClass = {astro-ph.GA},
       adsurl = {https://ui.adsabs.harvard.edu/abs/2021MNRAS.508.1195P}
}

@book{Dodelson2003,
   author = {Scott Dodelson},
   city = {San Diego, CA},
   publisher = {Academic Press},
   title = {Modern cosmology},
   url = {https://cds.cern.ch/record/1282338},
   year = {2003},
}

@article{Jarvis2016,
   author = {Matt J. Jarvis and A. R. Taylor and I. Agudo and James R. Allison and R. P. Deane and B. Frank and N. Gupta and I. Heywood and N. Maddox and K. McAlpine and Mario G. Santos and A. M.M. Scaife and M. Vaccari and J. T.L. Zwart and E. Adams and D. J. Bacon and A. J. Baker and Bruce A. Bassett and P. N. Best and R. Beswick and S. Blyth and Michael L. Brown and M. Brüggen and M. Cluver and S. Colafranceso and G. Cotter and C. Cress and R. Davé and C. Ferrari and M. J. Hardcastle and C. Hale and I. Harrison and P. W. Hatfield and H. R. Klöckner and S. Kolwa and E. Malefahlo and T. Marubini and T. Mauch and K. Moodley and R. Morganti and R. Norris and J. A. Peters and I. Prandoni and M. Prescott and S. Oliver and N. Oozeer and H. J.A. Röttgering and N. Seymour and C. Simpson and O. Smirnov and D. J.B. Smith and K. Spekkens and J. Stil and C. Tasse and K. van der Heyden and I. H. Whittam and W. L. WIlliams},
   issn = {18248039},
   journal = {Proceedings of Science},
   pages = {25-27},
   title = {The MeerKAT international GHz tiered extragalactic exploration (MIGHTEE) survey},
   year = {2016},
}

@article{Morales2004,
   author = {Miguel F Morales and Jacqueline Hewitt},
   doi = {10.1086/424437},
   issn = {0004-637X},
   issue = {1},
   journal = {The Astrophysical Journal},
   pages = {7-18},
   title = {Toward Epoch of Reionization Measurements with Wide‐Field Radio Observations},
   volume = {615},
   year = {2004},
}

@article{Morales2005,
   author = {Miguel F. Morales},
   doi = {10.1086/426730},
   issn = {0004-637X},
   issue = {2},
   journal = {The Astrophysical Journal},
   pages = {678-683},
   title = {Power Spectrum Sensitivity and the Design of Epoch of Reionization Observatories},
   volume = {619},
   year = {2005},
}

@article{Pober2013,
   author = {Jonathan C. Pober and Aaron R. Parsons and James E. Aguirre and Zaki Ali and Richard F. Bradley and Chris L. Carilli and Dave Deboer and Matthew Dexter and Nicole E. Gugliucci and Daniel C. Jacobs and Patricia J. Klima and Dave MacMahon and Jason Manley and David F. Moore and Irina I. Stefan and William P. Walbrugh},
   doi = {10.1088/2041-8205/768/2/L36},
   issn = {20418205},
   issue = {2},
   journal = {Astrophysical Journal Letters},
   title = {Opening the 21 cm Epoch of reionization window: Measurements of foreground isolation with paper},
   volume = {768},
   year = {2013},
}

@article{Parsons2012a,
   author = {Aaron Parsons and Jonathan Pober and Matthew McQuinn and Daniel Jacobs and James Aguirre},
   doi = {10.1088/0004-637X/753/1/81},
   issn = {15384357},
   issue = {1},
   journal = {Astrophysical Journal},
   title = {A sensitivity and array-configuration study for measuring the power spectrum of 21cm emission from reionization},
   volume = {753},
   year = {2012},
}

@article{Parsons2012,
   author = {Aaron R. Parsons and Jonathan C. Pober and James E. Aguirre and Christopher L. Carilli and Daniel C. Jacobs and David F. Moore},
   doi = {10.1088/0004-637X/756/2/165},
   issn = {15384357},
   issue = {2},
   journal = {Astrophysical Journal},
   title = {A per-baseline, delay-spectrum technique for accessing the 21cm cosmic reionization signature},
   volume = {756},
   year = {2012},
}

@article{Thyagarajan2013,
   author = {Nithyanandan Thyagarajan and N. Udaya Shankar and Ravi Subrahmanyan and Wayne Arcus and Gianni Bernardi and Judd D. Bowman and Frank Briggs and John D. Bunton and Roger J. Cappallo and Brian E. Corey and Ludi Desouza and David Emrich and Bryan M. Gaensler and Robert F. Goeke and Lincoln J. Greenhill and Bryna J. Hazelton and David Herne and Jacqueline N. Hewitt and Melanie Johnston-Hollitt and David L. Kaplan and Justin C. Kasper and Barton B. Kincaid and Ronald Koenig and Eric Kratzenberg and Colin J. Lonsdale and Mervyn J. Lynch and S. Russell McWhirter and Daniel A. Mitchell and Miguel F. Morales and Edward H. Morgan and Divya Oberoi and Stephen M. Ord and Joseph Pathikulangara and Ronald A. Remillard and Alan E.E. Rogers and D. Anish Roshi and Joseph E. Salah and Robert J. Sault and K. S. Srivani and Jamie B. Stevens and Prabu Thiagaraj and Steven J. Tingay and Randall B. Wayth and Mark Waterson and Rachel L. Webster and Alan R. Whitney and Andrew J. Williams and Christopher L. Williams and J. Stuart B. Wyithe},
   doi = {10.1088/0004-637X/776/1/6},
   issn = {15384357},
   issue = {1},
   journal = {Astrophysical Journal},
   title = {A study of fundamental limitations to statistical detection of redshifted H i from the epoch of reionization},
   volume = {776},
   year = {2013},
}

@article{Morales2019,
   author = {Miguel F. Morales and Adam Beardsley and Jonathan Pober and Nichole Barry and Bryna Hazelton and Daniel Jacobs and Ian Sullivan},
   doi = {10.1093/mnras/sty2844},
   issn = {13652966},
   issue = {2},
   journal = {Monthly Notices of the Royal Astronomical Society},
   pages = {2207-2216},
   title = {Understanding the diversity of 21 cm cosmology analyses},
   volume = {483},
   year = {2019},
}

@article{Bull2015,
   author = {Philip Bull and Pedro G Ferreira and Prina Patel and Mário G. Santos},
   doi = {10.1088/0004-637X/803/1/21},
   issn = {15384357},
   issue = {1},
   journal = {Astrophysical Journal},
   pages = {1-33},
   title = {Late-time cosmology with 21 cm intensity mapping experiments},
   volume = {803},
   year = {2015},
}

@article{Kovetz2017,
   author = {Ely D Kovetz and Marco P Viero and Adam Lidz and Laura Newburgh and Mubdi Rahman and Eric Switzer and Marc Kamionkowski and James Aguirre and Marcelo Alvarez and James Bock and J. Richard Bond and Goeffry Bower and C. Matt Bradford and Patrick C. Breysse and Philip Bull and Tzu-Ching Chang and Yun-Ting Cheng and Dongwoo Chung and Kieran Cleary and Asantha Corray and Abigail Crites and Rupert Croft and Olivier Doré and Michael Eastwood and Andrea Ferrara and José Fonseca and Daniel Jacobs and Garrett K. Keating and Guilaine Lagache and Gunjan Lakhlani and Adrian Liu and Kavilan Moodley and Norm Murray and Aurélie Pénin and Gergö Popping and Anthony Pullen and Dominik Reichers and Shun Saito and Ben Saliwanchik and Mario Santos and Rachel Somerville and Gordon Stacey and George Stein and Francesco Villaescusa-Navarro and Eli Visbal and Amanda Weltman and Laura Wolz and Micheal Zemcov},
   issue = {June},
   journal = {arXiv e-prints},
   title = {Line-Intensity Mapping: 2017 Status Report},
   url = {http://arxiv.org/abs/1709.09066},
   year = {2017},
}

@article{Liu2014,
   author = {Adrian Liu and Aaron R. Parsons and Cathryn M. Trott},
   doi = {10.1103/PhysRevD.90.023018},
   issn = {15502368},
   issue = {2},
   journal = {Physical Review D - Particles, Fields, Gravitation and Cosmology},
   title = {Epoch of reionization window. I. Mathematical formalism},
   volume = {90},
   year = {2014},
}

@article{Villaescusa-Navarro2014,
   author = {Francisco Villaescusa-Navarro and Matteo Viel and Kanan K. Datta and T. Roy Choudhury},
   doi = {10.1088/1475-7516/2014/09/050},
   issn = {14757516},
   issue = {9},
   journal = {Journal of Cosmology and Astroparticle Physics},
   title = {Modeling the neutral hydrogen distribution in the post-reionization Universe: Intensity mapping},
   volume = {2014},
   year = {2014},
}

@article{Liu2020,
   author = {Adrian Liu and J. Richard Shaw},
   doi = {10.1088/1538-3873/ab5bfd},
   issn = {00046280},
   issue = {1012},
   journal = {Publications of the Astronomical Society of the Pacific},
   month = {6},
   publisher = {Institute of Physics Publishing},
   title = {Data analysis for precision 21cm cosmology},
   volume = {132},
   year = {2020},
}

@article{Colombi2009,
       author = {{Colombi}, St{\'e}phane and {Jaffe}, Andrew and {Novikov}, Dmitri and {Pichon}, Christophe},
        title = "{Accurate estimators of power spectra in N-body simulations}",
      journal = {\mnras},
         year = 2009,
        month = feb,
       volume = {393},
       number = {2},
        pages = {511-526},
          doi = {10.1111/j.1365-2966.2008.14176.x},
archivePrefix = {arXiv},
       eprint = {0811.0313},
 primaryClass = {astro-ph},
       adsurl = {https://ui.adsabs.harvard.edu/abs/2009MNRAS.393..511C}
}

@article{Switzer2013,
   author = {E. R. Switzer and K. W. Masui and K. Bandura and L. M. Calin and T. C. Chang and X. L. Chen and Y. C. Li and Y. W. Liao and A. Natarajan and U. L. Pen and J. B. Peterson and J. R. Shaw and T. C. Voytek},
   doi = {10.1093/mnrasl/slt074},
   issn = {17453925},
   issue = {1},
   journal = {Monthly Notices of the Royal Astronomical Society: Letters},
   title = {Determination of z ~ 0.8 neutral hydrogen fluctuations using the 21 cm intensity mapping autocorrelation},
   volume = {434},
   year = {2013},
}

@article{Thyagarajan2015,
   author = {Nithyanandan Thyagarajan and Daniel C. Jacobs and Judd D. Bowman and N. Barry and A. P. Beardsley and G. Bernardi and F. Briggs and R. J. Cappallo and P. Carroll and B. E. Corey and A. De Oliveira-Costa and Joshua S. Dillon and D. Emrich and A. Ewall-Wice and L. Feng and R. Goeke and L. J. Greenhill and B. J. Hazelton and J. N. Hewitt and N. Hurley-Walker and M. Johnston-Hollitt and D. L. Kaplan and J. C. Kasper and Han Seek Kim and P. Kittiwisit and E. Kratzenberg and E. Lenc and J. Line and A. Loeb and C. J. Lonsdale and M. J. Lynch and B. McKinley and S. R. McWhirter and D. A. Mitchell and M. F. Morales and E. Morgan and A. R. Neben and D. Oberoi and A. R. Offringa and S. M. Ord and Sourabh Paul and B. Pindor and J. C. Pober and T. Prabu and P. Procopio and J. Riding and A. E.E. Rogers and A. Roshi and N. Udaya Shankar and Shiv K. Sethi and K. S. Srivani and R. Subrahmanyan and I. S. Sullivan and M. Tegmark and S. J. Tingay and C. M. Trott and M. Waterson and R. B. Wayth and R. L. Webster and A. R. Whitney and A. Williams and C. L. Williams and C. Wu and J. S.B. Wyithe},
   doi = {10.1088/0004-637X/804/1/14},
   issn = {15384357},
   issue = {1},
   journal = {Astrophysical Journal},
   title = {Foregrounds in wide-field Redshifted 21 cm power spectra},
   volume = {804},
   year = {2015},
}

@article{Ewen1951,
   author = {H.I. Ewen and E.M. Purcell},
   doi = {https://doi.org/10.1038/168356a0},
   journal = {Nature},
   title = {Observation of a Line in the Galactic Radio Spectrum: Radiation from Galactic Hydrogen at 1420 Mc./sec.},
   volume = {168},
   year = {1951},
}

@article{Hulst1945,
   author = {H.C. van de Hulst},
   issue = {11},
   journal = {Ned. Tijdschr. Natuurk.},
   pages = {210-221},
   title = {Radiogolven uit het Wereldruim},
   year = {1945},
}

@article{Chang2010,
   author = {T. Chang and U. Pen and K. et al. Bandura},
   doi = {https://doi.org/10.1038/nature09187},
   issue = {466},
   journal = {Nature},
   pages = {463-465},
   title = {An intensity map of hydrogen 21-cm emission at redshift z ≈ 0.8},
   year = {2010},
}

@article{Spinelli2020,
   author = {Marta Spinelli and Anna Zoldan and Gabriella de Lucia and Lizhi Xie and Matteo Viel},
   doi = {10.1093/mnras/staa604},
   issn = {13652966},
   issue = {4},
   journal = {Monthly Notices of the Royal Astronomical Society},
   pages = {5434-5455},
   title = {The atomic hydrogen content of the post-reionization Universe},
   volume = {493},
   year = {2021},
}

@article{Deboer2017,
   author = {David R. Deboer and Aaron R. Parsons and James E. Aguirre and Paul Alexander and Zaki S. Ali and Adam P. Beardsley and Gianni Bernardi and Judd D. Bowman and Richard F. Bradley and Chris L. Carilli and Carina Cheng and Eloy De Lera Acedo and Joshua S. Dillon and Aaron Ewall-Wice and Gcobisa Fadana and Nicolas Fagnoni and Randall Fritz and Steve R. Furlanetto and Brian Glendenning and Bradley Greig and Jasper Grobbelaar and Bryna J. Hazelton and Jacqueline N. Hewitt and Jack Hickish and Daniel C. Jacobs and Austin Julius and Maccalvin Kariseb and Saul A. Kohn and Telalo Lekalake and Adrian Liu and Anita Loots and David Macmahon and Lourence Malan and Cresshim Malgas and Matthys Maree and Zachary Martinot and Nathan Mathison and Eunice Matsetela and Andrei Mesinger and Miguel F. Morales and Abraham R. Neben and Nipanjana Patra and Samantha Pieterse and Jonathan C. Pober and Nima Razavi-Ghods and Jon Ringuette and James Robnett and Kathryn Rosie and Raddwine Sell and Craig Smith and Angelo Syce and Max Tegmark and Nithyanandan Thyagarajan and Peter K.G. Williams and Haoxuan Zheng},
   doi = {10.1088/1538-3873/129/974/045001},
   issn = {00046280},
   issue = {974},
   journal = {Publications of the Astronomical Society of the Pacific},
   title = {Hydrogen epoch of reionization array (HERA)},
   volume = {129},
   year = {2017},
}

@inproceedings{Santos2015b,
   author = {Mario Santos and Phil Bull and David Alonso and Stefano Camera and Pedro Ferreira and Gianni Bernardi and Roy Maartens and Matteo Viel and Francisco Villaescusa-Navarro and Filipe Abdalla and Matt Jarvis and R Metcalf and Alkistis Pourtsidou and Laura Wolz},
   doi = {10.22323/1.215.0019},
   booktitle = {Advancing Astrophysics with the Square Kilometre Array},
   pages = {19},
   title = {Cosmology from a SKA HI intensity mapping survey},
   year = {2015},
}

@article{Bharadwaj2001,
   author = {Somnath Bharadwaj and Biman B. Nath and Shiv K. Sethi},
   doi = {10.1007/BF02933588},
   issn = {02506335},
   issue = {1},
   journal = {Journal of Astrophysics and Astronomy},
   pages = {21-34},
   title = {Using HI to probe large scale structures at z ∼ 3},
   volume = {22},
   year = {2001},
}

@article{Battye2004,
   author = {Richard A. Battye and Rod D. Davies and Jochen Weller},
   doi = {10.1111/j.1365-2966.2004.08416.x},
   issn = {00358711},
   issue = {4},
   journal = {Monthly Notices of the Royal Astronomical Society},
   pages = {1339-1347},
   title = {Neutral hydrogen surveys for high-redshift galaxy clusters and protoclusters},
   volume = {355},
   year = {2004},
}

@article{Bharadwaj2001a,
   author = {Somnath Bharadwaj and Shiv K. Sethi},
   doi = {10.1007/BF02702273},
   issn = {02506335},
   issue = {4},
   journal = {Journal of Astrophysics and Astronomy},
   pages = {293-307},
   title = {HI fluctuations at large redshifts: I-visibility correlation},
   volume = {22},
   year = {2001},
}

@article{Chang2008,
   author = {Tzu Ching Chang and Ue Li Pen and Jeffrey B. Peterson and Patrick McDonald},
   doi = {10.1103/PhysRevLett.100.091303},
   issn = {00319007},
   issue = {9},
   journal = {Physical Review Letters},
   pages = {1-5},
   title = {Baryon acoustic oscillation intensity mapping of dark energy},
   volume = {100},
   year = {2008},
}

@inproceedings{2018AAS...23123107B,
   author = {Andrew J Baker and Sarah Blyth and Benne W Holwerda and LADUMA Team},
   booktitle = {American Astronomical Society Meeting Abstracts \#231},
   month = {1},
   pages = {231.07},
   title = {LADUMA: Looking At the Distant Universe with the MeerKAT Array},
   volume = {231},
   year = {2018},
}

@article{Paul:2016blh,
   author = {Sourabh Paul and Shiv K. Sethi and Miguel F. Morales and K. S. Dwarkanath and N. Udaya Shankar and Ravi Subrahmanyan and N. Barry and A. P. Beardsley and Judd D. Bowman and F. Briggs and P. Carroll and A. de Oliveira-Costa and Joshua S. Dillon and A. Ewall-Wice and L. Feng and L. J. Greenhill and B. M. Gaensler and B. J. Hazelton and J. N. Hewitt and N. Hurley-Walker and D. J. Jacobs and Han-Seek Kim and P. Kittiwisit and E. Lenc and J. Line and A. Loeb and B. McKinley and D. A. Mitchell and A. R. Neben and A. R. Offringa and B. Pindor and J. C. Pober and P. Procopio and J. Riding and I. S. Sullivan and M. Tegmark and Nithyanandan Thyagarajan and S. J. Tingay and C. M. Trott and R. B. Wayth and R. L. Webster and J. S. B. Wyithe and Roger Cappallo and M. Johnston-Hollitt and D. L. Kaplan and C. J. Lonsdale and S. R. McWhirter and E. Morgan and D. Oberoi and S. M. Ord and T. Prabu and K. S. Srivani and A. Williams and C. L. Williams},
   doi = {10.3847/1538-4357/833/2/213},
   issue = {2},
   journal = {Astrophysical Journal},
   pages = {213},
   title = {Delay Spectrum with Phase-Tracking Arrays: Extracting the HI power spectrum from the Epoch of Reionization},
   volume = {833},
   year = {2016},
}

@article{wyithe2008mnras,
   author = {J Stuart B Wyithe and Abraham Loeb},
   doi = {10.1111/j.1365-2966.2009.15019.x},
   issn = {0035-8711},
   issue = {4},
   journal = {Monthly Notices of the Royal Astronomical Society},
   pages = {1926-1934},
   title = {The 21-cm power spectrum after reionization},
   volume = {397},
   url = {https://doi.org/10.1111/j.1365-2966.2009.15019.x},
   year = {2009},
}

@article{Pritchard:2011xb,
   author = {Jonathan R Pritchard and Abraham Loeb},
   doi = {10.1088/0034-4885/75/8/086901},
   journal = {Rept. Prog. Phys.},
   pages = {86901},
   title = {21-cm cosmology},
   volume = {75},
   year = {2012},
}

@ARTICLE{2024MNRAS.534...76H,
       author = {{Heywood}, I. and {Ponomareva}, A.~A. and {Maddox}, N. and {Jarvis}, M.~J. and {Frank}, B.~S. and {Adams}, E.~A.~K. and {Baes}, M. and {Bianchetti}, A. and {Collier}, J.~D. and {Deane}, R.~P. and {Glowacki}, M. and {Jung}, S.~L. and {Pan}, H. and {Rajohnson}, S.~H.~A. and {Rodighiero}, G. and {Ruffa}, I. and {Santos}, M.~G. and {Sinigaglia}, F. and {Vaccari}, M.},
        title = "{MIGHTEE-H I: deep spectral line observations of the COSMOS field}",
      journal = {\mnras},
         year = 2024,
        month = oct,
       volume = {534},
       number = {1},
        pages = {76-96},
          doi = {10.1093/mnras/stae2081},
archivePrefix = {arXiv},
       eprint = {2409.17713},
 primaryClass = {astro-ph.GA},
       adsurl = {https://ui.adsabs.harvard.edu/abs/2024MNRAS.534...76H}
}

@article{Chen2021,
   author = {Zhaoting Chen and Laura Wolz and Marta Spinelli and Steven G Murray},
   doi = {10.1093/mnras/stab386},
   issn = {0035-8711},
   issue = {4},
   journal = {Monthly Notices of the Royal Astronomical Society},
   pages = {5259-5276},
   title = { Extracting H i astrophysics from interferometric intensity mapping },
   volume = {502},
   year = {2021},
}

@article{Anderson2018,
   author = {C. J. Anderson and N. J. Luciw and Y. C. Li and C. Y. Kuo and J. Yadav and K. W. Masui and T. C. Chang and X. Chen and N. Oppermann and Y. W. Liao and U. L. Pen and D. C. Price and L. Staveley-Smith and E. R. Switzer and P. T. Timbie and L. Wolz},
   doi = {10.1093/mnras/sty346},
   issn = {13652966},
   issue = {3},
   journal = {Monthly Notices of the Royal Astronomical Society},
   pages = {3382-3392},
   title = {Low-amplitude clustering in low-redshift 21-cm intensity maps cross-correlated with 2dF galaxy densities},
   volume = {476},
   year = {2018},
}

@article{Wolz2016,
   author = {L. Wolz and C Tonini and C Blake and J. S.B. Wyithe},
   doi = {10.1093/mnras/stw535},
   issn = {13652966},
   issue = {3},
   journal = {Monthly Notices of the Royal Astronomical Society},
   pages = {3399-3410},
   title = {Intensity mapping cross-correlations: Connecting the largest scales to galaxy evolution},
   volume = {458},
   year = {2016},
}

@article{Peacock2003,
   author = {J. A. Peacock},
   doi = {10.1098/rsta.2003.1288},
   issn = {1364503X},
   issue = {1812},
   journal = {Philosophical Transactions of the Royal Society A: Mathematical, Physical and Engineering Sciences},
   pages = {2479-2495},
   pmid = {14667313},
   title = {Large-scale structure and matter in the Universe},
   volume = {361},
   year = {2003},
}

@article{Jing2005,
   author = {Y. P. Jing},
   doi = {10.1086/427087},
   issn = {0004-637X},
   issue = {2},
   journal = {The Astrophysical Journal},
   pages = {559-563},
   title = {Correcting for the Alias Effect When Measuring the Power Spectrum Using a Fast Fourier Transform},
   volume = {620},
   year = {2005},
}

@article{Cui2008,
   author = {Weiguang Cui and Lei Liu and Xiaohu Yang and Yu Wang and Longlong Feng and Volker Springel},
   doi = {10.1086/592079},
   issn = {0004-637X},
   issue = {2},
   journal = {The Astrophysical Journal},
   pages = {738-744},
   title = {An Ideal Mass Assignment Scheme for Measuring the Power Spectrum with Fast Fourier Transforms},
   volume = {687},
   year = {2008},
}

@article{TheHERACollaboration2021a,
   author = {{HERA Collaboration} and Zara Abdurashidova and James E. Aguirre and Paul Alexander and Zaki Ali and Yanga Balfour and Rennan Barkana and Adam Beardsley and Gianni Bernardi and Tashalee Billings and Judd Bowman and Richard Bradley and Phillip Bull and Jacob Burba and Steven Carey and Christopher Carilli and Carina Cheng and David DeBoer and Matthew Dexter and Eloy de Lera Acedo and Joshua Dillon and John Ely and Aaron Ewall-Wice and Nicolas Fagnoni and Anastasia Fialkov and Randall Fritz and Steven Furlanetto and Kingsley Gale-Sides and Brian Glendenning and Deepthi Gorthi and Bradley Greig and Jasper Grobbelaar and Ziyaad Halday and Bryna Hazelton and Stefan Heimersheim and Jacqueline Hewitt and Jack Hickish and Daniel Jacobs and Austin Julius and Nicholas Kern and Joshua Kerrigan and Piyanat Kittiwisit and Saul Kohn and Matthew Kolopanis and Adam Lanman and Paul La Plante and Telalo Lekalake and David Lewis and Adrian Liu and Yin-Zhe Ma and David MacMahon and Lourence Malan and Cresshim Malgas and Matthys Maree and Zachary Martinot and Eunice Matsetela and Andrei Mesinger and Jordan Mirocha and Mathakane Molewa and Miguel Morales and Tshegofalang Mosiane and Julian Munoz and Steven Murray and Abraham Neben and Bojan Nikolic and Chuneeta Devi Nunhokee and Aaron Parsons and Nipanjana Patra and Samantha Pieterse and Jonathan Pober and Yuxiang Qin and N. Razavi-Ghods and Itamar Reis and Jon Ringuette and James Robnett and Kathryn Rosie and Mario Santos and Sudipta Sikder and Peter Sims and Craig Smith and Angelo Syce and Nithyanandan Thyagarajan and Peter Williams and Haoxuan Zheng},
   title = {HERA Phase I Limits on the Cosmic 21-cm Signal: Constraints on Astrophysics and Cosmology During the Epoch of Reionization},
   journal = {\apj},
   year = {2022},
   month = {feb},
   volume = {925},
   number = {2},
   eid = {221},
   pages = {221},
   url = {http://arxiv.org/abs/2108.07282}
}

@article{Rajohnson2022,
   author = {Sambatriniaina H.A. Rajohnson and Bradley S Frank and Anastasia A. Ponomare and Natasha Maddox and Renée C Kraan-Korteweg and Matt J Jarvis and Elizabeth A.K. Adams and Tom Oosterloo and Maarten Baes and Kristine Spekkens and Nathan J Adams and Marcin Glowacki and Sushma Kurapati and Isabella Prandoni and Ian Heywood and Jordan D Collier and Srikrishna Sekhar and Russ Taylor},
   doi = {10.1093/mnras/stac693},
   issn = {13652966},
   issue = {2},
   journal = {Monthly Notices of the Royal Astronomical Society},
   pages = {2697-2706},
   title = {MIGHTEE-H i: the H i size-mass relation over the last billion years},
   volume = {512},
   year = {2022},
}

@article{Chen2022,
   author = {Zhaoting Chen and Laura Wolz and Richard Battye},
   issue = {May},
   journal = {Monthly Notices of the Royal Astronomical Society},
   pages = {1-20},
   title = {Towards Optimal Foreground Mitigation Strategies for Interferometric HI Intensity Mapping in the Low-Redshift Universe},
   volume = {20},
   url = {http://arxiv.org/abs/2205.07776},
   year = {2022},
}

@article{McQuinn2006a,
   author = {Matthew McQuinn and Oliver Zahn and Matias Zaldarriaga and Lars Hernquist and Steven R. Furlanetto},
   doi = {10.1086/505167},
   issn = {0004-637X},
   issue = {2},
   journal = {The Astrophysical Journal},
   pages = {815-834},
   title = {Cosmological Parameter Estimation Using 21 cm Radiation from the Epoch of Reionization},
   volume = {653},
   year = {2006},
}

@article{2024MNRAS.528.5586C,
       author = {{Cunnington}, Steven and {Wolz}, Laura},
        title = "{Accurate Fourier-space statistics for line intensity mapping: Cartesian grid sampling without aliased power}",
      journal = {\mnras},
         year = 2024,
        month = mar,
       volume = {528},
       number = {4},
        pages = {5586-5600},
          doi = {10.1093/mnras/stae333},
archivePrefix = {arXiv},
       eprint = {2312.07289},
 primaryClass = {astro-ph.CO},
       adsurl = {https://ui.adsabs.harvard.edu/abs/2024MNRAS.528.5586C}
}

@article{Tudorache2022,
   author = {Madalina N. Tudorache and M. J. Jarvis and I. Heywood and A. A. Ponomareva and N. Maddox and B. S. Frank and N. J. Adams and R. A.A. Bowler and I. H. Whittam and M. Baes and H. Pan and S. H.A. Rajohnson and F. Sinigaglia and K. Spekkens},
   doi = {10.1093/mnras/stac996},
   issn = {13652966},
   issue = {2},
   journal = {Monthly Notices of the Royal Astronomical Society},
   pages = {2168-2177},
   title = {MIGHTEE - Hi. The relation between the Hi gas in galaxies and the cosmic web},
   volume = {513},
   year = {2022},
}

@article{Li2024,
   author = {Zhixing Li and Laura Wolz and Hong Guo and Steven Cunnington and Yi Mao},
   issue = {3},
   journal = {Monthly Notices of the Royal Astronomical Society},
   pages = {1-15},
   title = {Modeling the Nonlinear Power Spectrum in Low-redshift HI Intensity Mapping},
   volume = {534},
   url = {https://doi.org/10.1093/mnras/stae2182},
   year = {2024},
}

@ARTICLE{2017PASA...34...52M,
       author = {{Meyer}, Martin and {Robotham}, Aaron and {Obreschkow}, Danail and {Westmeier}, Tobias and {Duffy}, Alan R. and {Staveley-Smith}, Lister},
        title = "{Tracing HI Beyond the Local Universe}",
      journal = {\pasa},
         year = 2017,
        month = nov,
       volume = {34},
        pages = {52},
          doi = {10.1017/pasa.2017.31},
archivePrefix = {arXiv},
       eprint = {1705.04210},
 primaryClass = {astro-ph.CO},
       adsurl = {https://ui.adsabs.harvard.edu/abs/2017PASA...34...52M}
}

@ARTICLE{Paul2023,
       author = {{Paul}, Sourabh and {Santos}, Mario G. and {Chen}, Zhaoting and {Wolz}, Laura},
        title = "{A first detection of neutral hydrogen intensity mapping on Mpc scales at $z\approx 0.32$ and $z\approx 0.44$}",
      journal = {arXiv e-prints},
         year = 2023,
        month = jan,
          eid = {arXiv:2301.11943},
        pages = {arXiv:2301.11943},
          doi = {10.48550/arXiv.2301.11943},
archivePrefix = {arXiv},
       eprint = {2301.11943},
 primaryClass = {astro-ph.CO},
       adsurl = {https://ui.adsabs.harvard.edu/abs/2023arXiv230111943P}
}

@article{Masui2013,
   author = {K. W. Masui and E. R. Switzer and N. Banavar and K. Bandura and C. Blake and L. M. Calin and T. C. Chang and X. Chen and Y. C. Li and Y. W. Liao and A. Natarajan and U. L. Pen and J. B. Peterson and J. R. Shaw and T. C. Voytek},
   doi = {10.1088/2041-8205/763/1/L20},
   issn = {20418205},
   issue = {1},
   journal = {Astrophysical Journal Letters},
   month = {1},
   title = {Measurement of 21 cm brightness fluctuations at z ∼ 0.8 in cross-correlation},
   volume = {763},
   year = {2013},
}

@article{Cunnington2022,
   author = {Steven Cunnington and Yichao Li and Mario G. Santos and Jingying Wang and Isabella P. Carucci and Melis O. Irfan and Alkistis Pourtsidou and Marta Spinelli and Laura Wolz and Paula S. Soares and Chris Blake and Philip Bull and Brandon Engelbrecht and José Fonseca and Keith Grainge and Yin-Zhe Ma},
   doi = {10.1093/mnras/stac3060},
   month = {6},
   journal = {Monthly Notices of the Royal Astronomical Society},
   title = {HI intensity mapping with MeerKAT: power spectrum detection in cross-correlation with WiggleZ galaxies},
   url = {http://arxiv.org/abs/2206.01579 http://dx.doi.org/10.1093/mnras/stac3060},
   year = {2022},
}

@ARTICLE{Maddox2021,
       author = {{Maddox}, N. and {Frank}, B.~S. and {Ponomareva}, A.~A. and {Jarvis}, M.~J. and {Adams}, E.~A.~K. and {Dav{\'e}}, R. and {Oosterloo}, T.~A. and {Santos}, M.~G. and {Blyth}, S.~L. and {Glowacki}, M. and {Kraan-Korteweg}, R.~C. and {Mulaudzi}, W. and {Namumba}, B. and {Prandoni}, I. and {Rajohnson}, S.~H.~A. and {Spekkens}, K. and {Adams}, N.~J. and {Bowler}, R.~A.~A. and {Collier}, J.~D. and {Heywood}, I. and {Sekhar}, S. and {Taylor}, A.~R.},
        title = "{MIGHTEE-HI: The H I emission project of the MeerKAT MIGHTEE survey}",
      journal = {\aap},
         year = 2021,
        month = feb,
       volume = {646},
          eid = {A35},
        pages = {A35},
          doi = {10.1051/0004-6361/202039655},
archivePrefix = {arXiv},
       eprint = {2011.09470},
 primaryClass = {astro-ph.GA},
       adsurl = {https://ui.adsabs.harvard.edu/abs/2021A&A...646A..35M}
}

@article{Wolz2021,
   author = {Laura Wolz and Alkistis Pourtsidou and Kiyoshi W. Masui and Tzu-Ching Chang and Julian E. Bautista and Eva-Maria Mueller and Santiago Avila and David Bacon and Will J. Percival and Steven Cunnington and Chris Anderson and Xuelei Chen and Jean-Paul Kneib and Yi-Chao Li and Yu-Wei Liao and Ue-Li Pen and Jeffrey B. Peterson and Graziano Rossi and Donald P. Schneider and Jaswant Yadav and Gong-Bo Zhao},
   doi = {10.1093/mnras/stab3621},
   month = {2},
   journal = {Monthly Notices of the Royal Astronomical Society},
   title = {HI constraints from the cross-correlation of eBOSS galaxies and Green Bank Telescope intensity maps},
   url = {http://arxiv.org/abs/2102.04946 http://dx.doi.org/10.1093/mnras/stab3621},
   year = {2021},
}

@article{Pan2022,
   author = {Hengxing Pan and Matt J. Jarvis and Mario G. Santos and Natasha Maddox and Bradley S. Frank and Anastasia A. Ponomareva and Isabella Prandoni and Sushma Kurapati and Maarten Baes and Pavel E. Mancera Piña and Giulia Rodighiero and Martin J. Meyer and Romeel Davé and Gauri Sharma and Sambatriniaina H. A. Rajohnson and Nathan J. Adams and Rebecca A. A. Bowler and Francesco Sinigaglia and Thijs van der Hulst and Peter W. Hatfield and Srikrishna Sekhar and Jordan D. Collier},
   doi = {10.1093/mnras/stad2343},
   month = {10},
   journal = {Monthly Notices of the Royal Astronomical Society},
   title = {MIGHTEE-HI: The HI mass-stellar mass relation over the last billion years},
   url = {http://arxiv.org/abs/2210.04651 http://dx.doi.org/10.1093/mnras/stad2343},
   year = {2022},
}

@article{Hand2018,
   author = {Nick Hand and Yu Feng and Florian Beutler and Yin Li and Chirag Modi and Uroš Seljak and Zachary Slepian},
   doi = {10.3847/1538-3881/aadae0},
   issn = {0004-6256},
   issue = {4},
   journal = {The Astronomical Journal},
   month = {10},
   pages = {160},
   publisher = {American Astronomical Society},
   title = {nbodykit: An Open-source, Massively Parallel Toolkit for Large-scale Structure},
   volume = {156},
   year = {2018},
}

@article{Ponomareva2023,
   author = {Anastasia A. Ponomareva and Matt J. Jarvis and Hengxing Pan and Natasha Maddox and Michael G. Jones and Bradley S. Frank and Sambatriniaina H. A. Rajohnson and Wanga Mulaudzi and Martin Meyer and Elizabeth A. K. Adams and Maarten Baes and Kelley M. Hess and Sushma Kurapati and Isabella Prandoni and Francesco Sinigaglia and Kristine Spekkens and Madalina Tudorache and Ian Heywood and Jordan D. Collier and Srikrishna Sekhar},
   doi = {10.1093/mnras/stad1249},
   month = {4},
   journal = {Monthly Notices of the Royal Astronomical Society},
   title = {MIGHTEE-HI: The first MeerKAT HI mass function from an untargeted interferometric survey},
   url = {http://arxiv.org/abs/2304.13051 http://dx.doi.org/10.1093/mnras/stad1249},
   year = {2023},
}

@article{Mauch2020,
   author = {T. Mauch and W. D. Cotton and J. J. Condon and A. M. Matthews and T. D. Abbott and R. M. Adam and M. A. Aldera and K. M. B. Asad and E. F. Bauermeister and T. G. H. Bennett and H. Bester and D. H. Botha and L. R. S. Brederode and Z. B. Brits and S. J. Buchner and J. P. Burger and F. Camilo and J. M. Chalmers and T. Cheetham and D. de Villiers and M. S. de Villiers and M. A. Dikgale-Mahlakoana and L. J. du Toit and S. W. P. Esterhuyse and G. Fadana and B. L. Fanaroff and S. Fataar and S. February and B. S. Frank and R. R. G. Gamatham and M. Geyer and S. Goedhart and S. Gounden and S. C. Gumede and I. Heywood and M. J. Hlakola and J. M. G. Horrell and B. Hugo and A. R. Isaacson and G. I. G. Józsa and J. L. Jonas and R. P. M. Julie and F. B. Kapp and V. A. Kasper and J. S. Kenyon and P. P. A. Kotzé and N. Kriek and H. Kriel and T. W. Kusel and R. Lehmensiek and A. Loots and R. T. Lord and B. M. Lunsky and K. Madisa and L. G. Magnus and J. P. L. Main and J. A. Malan and J. R. Manley and S. J. Marais and A. Martens and B. Merry and R. Millenaar and N. Mnyandu and I. P. T. Moeng and O. J. Mokone and T. E. Monama and M. C. Mphego and W. S. New and B. Ngcebetsha and K. J. Ngoasheng and M. T. O. Ockards and N. Oozeer and A. J. Otto and A. A. Patel and A. Peens-Hough and S. J. Perkins and A. J. T. Ramaila and Z. R. Ramudzuli and R. Renil and L. L. Richter and A. Robyntjies and S. Salie and C. T. G. Schollar and L. C. Schwardt and M. Serylak and R. Siebrits and S. K. Sirothia and O. M. Smirnov and L. Sofeya and G. Stone and B. Taljaard and C. Tasse and I. P. Theron and A. J. Tiplady and O. Toruvanda and S. N. Twum and T. J. van Balla and A. van der Byl and C. van der Merwe and V. Van Tonder and B. H. Wallace and M. G. Welz and L. P. Williams and B. Xaia},
   doi = {10.3847/1538-4357/ab5d2d},
   issn = {15384357},
   issue = {2},
   journal = {The Astrophysical Journal},
   month = {1},
   pages = {61},
   publisher = {American Astronomical Society},
   title = {The 1.28 GHz MeerKAT DEEP2 Image},
   volume = {888},
   year = {2020},
}

@article{Datta2010,
   author = {A. Datta and J. D. Bowman and C. L. Carilli},
   doi = {10.1088/0004-637X/724/1/526},
   issn = {15384357},
   issue = {1},
   journal = {Astrophysical Journal},
   month = {11},
   pages = {526-538},
   publisher = {Institute of Physics Publishing},
   title = {Bright source subtraction requirements for redshifted 21cm measurements},
   volume = {724},
   year = {2010},
}

@ARTICLE{2025MNRAS.541..476M,
       author = {{Mazumder}, Aishrila and {Wolz}, Laura and {Chen}, Zhaoting and {Paul}, Sourabh and {Santos}, Mario G. and {Jarvis}, Matt and {Townsend}, Junaid and {Sekhar}, Srikrishna and {Taylor}, Russ},
        title = "{HI intensity mapping with the MIGHTEE Survey: first results of the HI power spectrum}",
      journal = {\mnras},
         year = 2025,
        month = jul,
       volume = {541},
       number = {1},
        pages = {476-493},
          doi = {10.1093/mnras/staf975},
archivePrefix = {arXiv},
       eprint = {2501.17564},
 primaryClass = {astro-ph.CO},
       adsurl = {https://ui.adsabs.harvard.edu/abs/2025MNRAS.541..476M}
}

@inproceedings{2002ASPC..278...81C,
       author = {{Campbell}, Donald B.},
        title = {Measurement in Radio Astronomy},
    booktitle = {Single-Dish Radio Astronomy: Techniques and Applications},
         year = 2002,
       editor = {{Stanimirovic}, Snezana and {Altschuler}, Daniel and {Goldsmith}, Paul and {Salter}, Chris},
       series = {Astronomical Society of the Pacific Conference Series},
       volume = {278},
        month = dec,
        pages = {81-90},
       adsurl = {https://ui.adsabs.harvard.edu/abs/2002ASPC..278...81C}
}

@article{2024MNRAS.528.2511T,
       author = {{Taylor}, A.~R. and {Sekhar}, S. and {Heino}, L. and {Scaife}, A.~M.~M. and {Stil}, J. and {Bowles}, M. and {Jarvis}, M. and {Heywood}, I. and {Collier}, J.~D.},
        title = "{MIGHTEE polarization early science fields: the deep polarized sky}",
      journal = {\mnras},
         year = 2024,
        month = feb,
       volume = {528},
       number = {2},
        pages = {2511-2522},
          doi = {10.1093/mnras/stae169},
archivePrefix = {arXiv},
       eprint = {2312.13230},
 primaryClass = {astro-ph.GA},
       adsurl = {https://ui.adsabs.harvard.edu/abs/2024MNRAS.528.2511T}
}

@article{Chen2023,
   author = {Zhaoting Chen and Emma Chapman and Laura Wolz and Aishrila Mazumder},
   doi = {10.1093/mnras/stad2102},
   issn = {13652966},
   issue = {3},
   journal = {Monthly Notices of the Royal Astronomical Society},
   month = {9},
   pages = {3724-3740},
   publisher = {Oxford University Press},
   title = {Detecting the H i power spectrum in the post-reionization Universe with SKA-Low},
   volume = {524},
   year = {2023},
}

@article{Parsons2014,
   author = {Aaron R. Parsons and Adrian Liu and James E. Aguirre and Zaki S. Ali and Richard F. Bradley and Chris L. Carilli and David R. Deboer and Matthew R. Dexter and Nicole E. Gugliucci and Daniel C. Jacobs and Pat Klima and David H.E. Macmahon and Jason R. Manley and David F. Moore and Jonathan C. Pober and Irina I. Stefan and William P. Walbrugh},
   doi = {10.1088/0004-637X/788/2/106},
   issn = {15384357},
   issue = {2},
   journal = {Astrophysical Journal},
   month = {6},
   publisher = {Institute of Physics Publishing},
   title = {New limits on 21 cm epoch of reionization from paper-32 consistent with an x-ray heated intergalactic medium at z = 7.7},
   volume = {788},
   year = {2014},
}

@article{Paul2021,
   author = {Sourabh Paul and Mario G. Santos and Junaid Townsend and Matt J. Jarvis and Natasha Maddox and Jordan D. Collier and Bradley S. Frank and Russ Taylor},
   doi = {10.1093/mnras/stab1089},
   issn = {13652966},
   issue = {2},
   journal = {Monthly Notices of the Royal Astronomical Society},
   month = {8},
   pages = {2039-2050},
   publisher = {Oxford University Press},
   title = {H i intensity mapping with the MIGHTEE survey: Power spectrum estimates},
   volume = {505},
   year = {2021},
}

@article{Planck_18_params,
   author = {{Planck Collaboration} and N. Aghanim and Y. Akrami and M. Ashdown and J. Aumont and C. Baccigalupi and M. Ballardini and A. J. Banday and R. B. Barreiro and N. Bartolo and S. Basak and R. Battye and K. Benabed and J. P. Bernard and M. Bersanelli and P. Bielewicz and J. J. Bock and J. R. Bond and J. Borrill and F. R. Bouchet and F. Boulanger and M. Bucher and C. Burigana and R. C. Butler and E. Calabrese and J. F. Cardoso and J. Carron and A. Challinor and H. C. Chiang and J. Chluba and L. P.L. Colombo and C. Combet and D. Contreras and B. P. Crill and F. Cuttaia and P. De Bernardis and G. De Zotti and J. Delabrouille and J. M. Delouis and E. Di Valentino and J. M. Diego and O. Doré and M. Douspis and A. Ducout and X. Dupac and S. Dusini and G. Efstathiou and F. Elsner and T. A. Enßlin and H. K. Eriksen and Y. Fantaye and M. Farhang and J. Fergusson and R. Fernandez-Cobos and F. Finelli and F. Forastieri and M. Frailis and A. A. Fraisse and E. Franceschi and A. Frolov and S. Galeotta and S. Galli and K. Ganga and R. T. Génova-Santos and M. Gerbino and T. Ghosh and J. González-Nuevo and K. M. Górski and S. Gratton and A. Gruppuso and J. E. Gudmundsson and J. Hamann and W. Handley and F. K. Hansen and D. Herranz and S. R. Hildebrandt and E. Hivon and Z. Huang and A. H. Jaffe and W. C. Jones and A. Karakci and E. Keihänen and R. Keskitalo and K. Kiiveri and J. Kim and T. S. Kisner and L. Knox and N. Krachmalnicoff and M. Kunz and H. Kurki-Suonio and G. Lagache and J. M. Lamarre and A. Lasenby and M. Lattanzi and C. R. Lawrence and M. Le Jeune and P. Lemos and J. Lesgourgues and F. Levrier and A. Lewis and M. Liguori and P. B. Lilje and M. Lilley and V. Lindholm and M. López-Caniego and P. M. Lubin and Y. Z. Ma and J. F. Maciás-Pérez and G. Maggio and D. Maino and N. Mandolesi and A. Mangilli and A. Marcos-Caballero and M. Maris and P. G. Martin and M. Martinelli and E. Martínez-González and S. Matarrese and N. Mauri and J. D. McEwen and P. R. Meinhold and A. Melchiorri and A. Mennella and M. Migliaccio and M. Millea and S. Mitra and M. A. Miville-Deschênes and D. Molinari and L. Montier and G. Morgante and A. Moss and P. Natoli and H. U. Nørgaard-Nielsen and L. Pagano and D. Paoletti and B. Partridge and G. Patanchon and H. V. Peiris and F. Perrotta and V. Pettorino and F. Piacentini and L. Polastri and G. Polenta and J. L. Puget and J. P. Rachen and M. Reinecke and M. Remazeilles and A. Renzi and G. Rocha and C. Rosset and G. Roudier and J. A. Rubiño-Martín and B. Ruiz-Granados and L. Salvati and M. Sandri and M. Savelainen and D. Scott and E. P.S. Shellard and C. Sirignano and G. Sirri and L. D. Spencer and R. Sunyaev and A. S. Suur-Uski and J. A. Tauber and D. Tavagnacco and M. Tenti and L. Toffolatti and M. Tomasi and T. Trombetti and L. Valenziano and J. Valiviita and B. Van Tent and L. Vibert and P. Vielva and F. Villa and N. Vittorio and B. D. Wandelt and I. K. Wehus and M. White and S. D.M. White and A. Zacchei and A. Zonca},
   doi = {10.1051/0004-6361/201833910},
   issn = {14320746},
   journal = {Astronomy and Astrophysics},
   month = {9},
   publisher = {EDP Sciences},
   title = {Planck 2018 results: VI. Cosmological parameters},
   volume = {641},
   year = {2020},
}

@article{Dave_2019,
   author = {Romeel Davé and Daniel Anglés-Alcázar and Desika Narayanan and Qi Li and Mika H. Rafieferantsoa and Sarah Appleby},
   doi = {10.1093/mnras/stz937},
   issn = {13652966},
   issue = {2},
   journal = {Monthly Notices of the Royal Astronomical Society},
   month = {6},
   pages = {2827-2849},
   publisher = {Oxford University Press},
   title = {SIMBA: Cosmological simulations with black hole growth and feedback},
   volume = {486},
   year = {2019},
}

@ARTICLE{2025MNRAS.536.2187H,
       author = {{Hale}, C.~L. and {Heywood}, I. and {Jarvis}, M.~J. and {Whittam}, I.~H. and {Best}, P.~N. and {An}, Fangxia and {Bowler}, R.~A.~A. and {Harrison}, I. and {Matthews}, A. and {Smith}, D.~J.~B. and {Taylor}, A.~R. and {Vaccari}, M.},
        title = "{MIGHTEE: the continuum survey Data Release 1}",
      journal = {\mnras},
         year = 2025,
        month = jan,
       volume = {536},
       number = {3},
        pages = {2187-2211},
          doi = {10.1093/mnras/stae2528},
archivePrefix = {arXiv},
       eprint = {2411.04958},
 primaryClass = {astro-ph.GA},
       adsurl = {https://ui.adsabs.harvard.edu/abs/2025MNRAS.536.2187H}
}

@article{Heywood2021,
   author = {I. Heywood and M. J. Jarvis and C. L. Hale and I. H. Whittam and H. L. Bester and B. Hugo and J. S. Kenyon and M. Prescott and O. M. Smirnov and C. Tasse and J. M. Afonso and P. N. Best and J. D. Collier and R. P. Deane and B. S. Frank and M. J. Hardcastle and K. Knowles and N. Maddox and E. J. Murphy and I. Prandoni and S. M. Randriamampandry and M. G. Santos and S. Sekhar and F. Tabatabaei and A. R. Taylor and K. Thorat},
   doi = {10.1093/mnras/stab3021},
   month = {10},
   journal = {Monthly Notices of the Royal Astronomical Society},
   title = {MIGHTEE: total intensity radio continuum imaging and the COSMOS / XMM-LSS Early Science fields},
   url = {http://arxiv.org/abs/2110.00347 http://dx.doi.org/10.1093/mnras/stab3021},
   year = {2021},
}



\appendix


\section{The \texorpdfstring{\ion{H}{I}}{HI} shot noise power spectrum on a set of \texorpdfstring{\ion{H}{I}}{HI} galaxies}
\label{appendix:A}

The derivation of the model of the expected \ion{H}{I} shot noise power spectrum is showcased here. This model is useful as it allows us to directly compare the results from the \ion{H}{I} galaxy catalogue to a model depending only on the \ion{H}{I} mass in the catalogue and the average \ion{H}{I} temperature in the given cosmological volume, as will be seen below.
\par

\noindent
The derivation is based on the formalism outlined in \citep{Wolz2016} and extended in \citep{Chen2022}. Start by defining the Fourier convention:

\begin{equation} \label{eq:ft}
    \tilde{f}(\textbf{k}) = \int \frac{d^3\textbf{x}}{V} \ f(\textbf{x}) \ e^{i \textbf{k} \cdot \textbf{x} },
\end{equation}

\noindent
and

\begin{equation} \label{eq:ift}
    f(\textbf{k}) = \frac{V}{(2\pi)^3} \int  d^3\textbf{k} \ \tilde{f}(\textbf{k}) \ e^{-i \textbf{k} \cdot \textbf{x} },
\end{equation}

\noindent
with

\begin{equation} \label{eq:dirac_delta}
    \int \frac{d^3 \textbf{x}}{V} \delta_{\text{D}}^3 (\textbf{x} - \textbf{x}^{\prime}) f(\textbf{x}) = f(\textbf{x}^{\prime}),
\end{equation}

\noindent
where $\delta_{\text{D}}^3$ is the Dirac delta function in three dimensions. The \ion{H}{I} temperature at a given position, $\textbf{x}$, can be expressed as

\begin{equation} \label{eq:thi_c_rho_hi}
    T_{\ion{H}{I}}(\textbf{x}) = C_{\ion{H}{I}} \ \rho_{\ion{H}{I}} (\textbf{x}),
\end{equation}

\noindent
with the \ion{H}{I} density is given by

\begin{equation} \label{eq:rho_hi}
    \rho_{\ion{H}{I}} = \frac{1}{V} \sum_{\text{i}} M_{\ion{H}{I}}^{\text{i}} \delta_{\text{D}}^3 (\textbf{x} - \textbf{x}^{\prime}),
\end{equation}

\noindent
where $i$ iterates over the \ion{H}{I} sources in the given cosmological volume, $V$. $C_{\ion{H}{I}}$ is a multiplicative term that depends on redshift, $z$, which converts this \ion{H}{I} density to temperature, and is given in \cite{Chen2022} as

\begin{equation} \label{eq:c_hi}
    C_{\ion{H}{I}} (z) = \frac{3 A_{12} h_{\text{P}} c^3 (1+z)^2}{32 \pi \ m_{\text{H}} k_{\text{B}} \nu^2_{21} H(z)}.
\end{equation}

\noindent
In \ref{eq:rho_hi}, $\rho_{\ion{H}{I}}$ is explicitly expressed in terms of the \ion{H}{I} mass, $M_{\ion{H}{I}}$, in the considered spatial volume, $V$. Here it is assumed that all the \ion{H}{I} is contained in discrete sources over a relatively small spatial volume, i.e., that the \ion{H}{I} content in the considered spatial volume is contained entirely within galaxies. Further, the \ion{H}{I} temperature can also be expressed as 

\begin{equation} \label{eq:t_hi_ito_mhi}
\begin{split}
    T_{\ion{H}{I}} (\textbf{x}) & = C_{\ion{H}{I}} \rho_{\ion{H}{I}} (\textbf{x}) \\
    & = \frac{C_{\ion{H}{I}}}{V} \sum_i M_{\ion{H}{I}}^i \delta^3_{\text{D}} (\textbf{x} - \textbf{x}^{\prime}),
\end{split}
\end{equation}

\noindent
so that the Fourier transform becomes

\begin{equation} \label{eq:t_hi_ito_mhi_ft}
\begin{split}
    \tilde{T} (\textbf{k}) & = \int \frac{d^3\textbf{x}}{V} T_{\ion{H}{I}}(\textbf{x}) e^{-i \textbf{k} \cdot \textbf{x}} \\
    & = \frac{C_{\ion{H}{I}}}{V} \int \frac{d^3\textbf{x}}{V} \sum_i M_{\ion{H}{I}}^i \delta^3_{\text{D}} (\textbf{x} - \textbf{x}^{\prime}) e^{-i \textbf{k} \cdot \textbf{x}} \\
    & = \frac{C_{\ion{H}{I}}}{V} \int \frac{d^3\textbf{x}}{V} \delta^3_{\text{D}} (\textbf{x} - \textbf{x}^{\prime}) \left[ \sum_i M_{\ion{H}{I}}^i e^{-i \textbf{k} \cdot \textbf{x}}\right] \\
    & = \frac{C_{\ion{H}{I}}}{V} \sum_i M_{\ion{H}{I}}^i e^{-i \textbf{k} \cdot \textbf{x}_i}.
\end{split}
\end{equation}

\noindent
With the form given for the Fourier-transformed \ion{H}{I} temperature in Equation \ref{eq:t_hi_ito_mhi_ft}, the \ion{H}{I} power spectrum can be written out in the form

\begin{equation} \label{eq:full_pk_mhi}
\begin{split}
    P(\textbf{k}) & = V \ \tilde{T}(\textbf{k}) \ \tilde{T}^* (\textbf{k}) \\
    & = V \frac{C_{\ion{H}{I}}^2}{V^2} \sum_i M_{\ion{H}{I}}^i e^{-i\textbf{k} \cdot \textbf{x}_i} \sum_j M_{\ion{H}{I}}^j e^{-i\textbf{k} \cdot \textbf{x}_j} \\
    & = V \frac{C_{\ion{H}{I}}^2}{V^2} \sum_{ij} M_{\ion{H}{I}}^i M_{\ion{H}{I}}^j e^{-i \textbf{k} \cdot (\textbf{x}_i - \textbf{x}_j)}.
\end{split}
\end{equation}

\noindent
Now consider the conditions for Equation \ref{eq:full_pk_mhi} when the shot noise term dominates. Equation \ref{eq:full_pk_mhi} is essentially a weighted average of galaxy pair counts done over $ij$ for each galaxy pair. This equation is valid for all $\textbf{k}$. However, in this case, we are looking for the limit in which $P(\textbf{k})$ becomes constant for $\textbf{k}$, i.e., the shot noise limit of the \ion{H}{I} power spectrum expression. The shot noise originates from the discrete, noise-like nature of the distribution of \ion{H}{I} sources.  Hence, we are looking at the self-pairing component of Equation \ref{eq:full_pk_mhi}, i.e., when $i = j$. We, therefore have:

\begin{equation} \label{eq:pkhi_full_shot}
\begin{split}
    P_{\ion{H}{I}}^{\text{SN}} & = P_{\ion{H}{I}}(\textbf{k}) |_{i=j} \\
    & = \frac{C_{\ion{H}{I}}^2}{V^2} V \sum_i (M^i_{\ion{H}{I}})^2 \\
    & = \frac{C_{\ion{H}{I}}^2}{V} \sum_i (M_{\ion{H}{I}}^i)^2.
\end{split}    
\end{equation}

\noindent
The shot noise can be written in terms of the average \ion{H}{I} temperature, given by

\begin{equation} \label{eq:t_hi_avg}
\begin{split}
    \Bar{T}_{\ion{H}{I}} & = \int\frac{d^3 \textbf{x}}{V} \ T_{\ion{H}{I}}(\textbf{x}) \\
    & = \int \frac{d^3\textbf{x}}{V} \frac{C_{\ion{H}{I}}}{V} \sum_i M_{\ion{H}{I}}^i \delta_{\text{D}}^3 (\textbf{x} - \textbf{x}^{\prime}) \\
    & = \frac{C_{\ion{H}{I}}}{V} \sum_i M_{\ion{H}{I}}^i \int \frac{d^3 \textbf{x}}{V} \delta_{\text{D}}^3 (\textbf{x} - \textbf{x}^{\prime}) \\
    & = \frac{C_{\ion{H}{I}}}{V} \sum_i M_{\ion{H}{I}}^i.
\end{split}
\end{equation}

Therefore,

\begin{equation} \label{eq:t_hi_avg_squared}
\begin{split}
    \Bar{T}^2 = \frac{C_{\ion{H}{I}}^2}{V^2} \left( \sum_i M_{\ion{H}{I}}^i \right)^2,
\end{split}
\end{equation}

\noindent
which implies that

\begin{equation} \label{eq:c_hi_squared}
    C_{\ion{H}{I}}^2 = \Bar{T}^2 V^2 / \ \left( \sum_i M_{\ion{H}{I}}^i \right)^2.
\end{equation}

\noindent
Substituting Equation \ref{eq:c_hi_squared} into \ref{eq:pkhi_full_shot}, the shot noise can therefore be rewritten as

\begin{equation} \label{eq:pk_sn_ito_t_hi_avg}
\begin{split}
    P_{\text{SN}} & = \frac{C_{\ion{H}{I}}^2}{V} \sum_i (M_{\ion{H}{I}}^i)^2 \\
    & = \frac{\Bar{T}_{\ion{H}{I}}^2 V^2}{V} \sum_i \left( M_{\ion{H}{I}}^i \right)^2 / \ \left( \sum_i M_{\ion{H}{I}}^i \right)^2 \\
    & = \Bar{T}_{\ion{H}{I}}^2 V \sum_i \left( M_{\ion{H}{I}}^i \right)^2 / \ \left( \sum_i M_{\ion{H}{I}}^i \right)^2.
\end{split}
\end{equation}

\noindent
In other words, the shot noise power spectrum will then depend on the average \ion{H}{I} temperature in a given cosmological volume, $V$, and the total \ion{H}{I} mass in said volume. To calculate the shot noise, it is sufficient to know the masses of the galaxies in the cosmological volume, as the average \ion{H}{I} temperature can also be computed from this. 
\par

\noindent
Lastly, it is easy to see that in the case of all the galaxies having the same \ion{H}{I} mass, the shot noise reduces to

\begin{equation} \label{eq:pk_sn_same_mhi}
\begin{split}
    P_{\text{SN}} & = \Bar{T}_{\ion{H}{I}}^2 V \frac{N_{\text{g}}}{N_{\text{g}}^2} \\
    & = \Bar{T}_{\ion{H}{I}}^2 \frac{V}{N_{\text{g}}} \\
    & = \frac{\Bar{T}_{\ion{H}{I}}^2}{\Bar{n}_{\text{g}}},
\end{split}
\end{equation}

\noindent
which is a close analogue to the shot noise found in the galaxy power spectrum, $P_{g}$. In that case, the shot noise is usually obtained by dividing \ref{eq:pk_sn_same_mhi} by $T^2$, resulting in $P_{\text{SN}, \ g} (\textbf{k}) \sim 1/\Bar{n}_{\text{g}}$.

\bsp
\label{lastpage}
\end{document}